\documentclass{article}

\usepackage{natbib}
\usepackage{cancel}
\usepackage{soul}
\usepackage[english]{babel}
\bibpunct{(}{)}{;}{a}{,}{,}

\usepackage{fullpage}

\usepackage{amsmath,amsthm, amssymb}
\usepackage{bm}
\usepackage{bbm} 
\usepackage{graphicx}
\usepackage[colorinlistoftodos]{todonotes}
\usepackage[colorlinks=true, allcolors=blue]{hyperref}
\usepackage{enumitem}
\usepackage{algorithm,algpseudocode}
\usepackage{multirow}
\usepackage{comment}

\newtheorem{thm}{Theorem}
\newtheorem{cor}{Corollary}[thm]

\newtheorem{Def}{Definition}

\usepackage{setspace}
\usepackage{float} 
\usepackage{caption}
\usepackage{subcaption}
\usepackage[export]{adjustbox} 

\usepackage{booktabs}
\usepackage{longtable}
\usepackage{array}
\usepackage{multirow}
\usepackage{wrapfig}
\usepackage{float}
\usepackage{colortbl}
\usepackage{pdflscape}
\usepackage{tabu}
\usepackage{threeparttable}
\usepackage{threeparttablex}
\usepackage[normalem]{ulem}
\usepackage{makecell}
\usepackage{xcolor}

\newtheorem{proposition}{Proposition}

\theoremstyle{definition}

\newtheorem{assumption}{Assumption}

\usepackage{scalerel,stackengine}
\stackMath
\newcommand\reallywidehat[1]{%
\savestack{\tmpbox}{\stretchto{%
  \scaleto{%
    \scalerel*[\widthof{\ensuremath{#1}}]{\kern-.6pt\bigwedge\kern-.6pt}%
    {\rule[-\textheight/2]{1ex}{\textheight}}
  }{\textheight}%
}{0.5ex}}%
\stackon[1pt]{#1}{\tmpbox}%
}
\definecolor{ashgrey}{rgb}{0.7, 0.75, 0.71}
\definecolor{aurometalsaurus}{rgb}{0.43, 0.5, 0.5}
\definecolor{battleshipgrey}{rgb}{0.52, 0.52, 0.51}
\definecolor{alizarin}{rgb}{0.82, 0.1, 0.26}
\newcommand{\ed}[1]{\color{black}#1 \color{black}}

\title{Copula-based Sensitivity Analysis for Multi-Treatment Causal Inference with Unobserved Confounding
\protect\thanks{Jiajing Zheng is a PhD candidate in the Department of Statistics and Applied Probability at the University of Caliornia Santa Barbara (\href{mailto:jzheng@pstat.ucsb.edu}{jzheng@pstat.ucsb.edu}). Alexander D'Amour is a Research Scientist at Google Research, Cambridge, MA (\href{mailto:alexdamour@google.com}{alexdamour@google.com}). Alexander M. Franks is an Assistant Professor of Statistics at the University of California, Santa Barbara (\href{mailto:afranks@pstat.ucsb.edu}{afranks@pstat.ucsb.edu}). We thank Steve Yadlowksy, Victor Veitch, and Avi Feller for thoughtful comments and discussion.}}

\usepackage{graphicx,scalerel}

\author{Jiajing Zheng\\UCSB \and  Alexander D'Amour \\Google Research \and Alexander Franks\\UCSB}
\date{\today}

\begin{document}
\maketitle


%
%


\def\-{\text{-}}
\def\Bern{\text{Bern}}

\def\PATE{$\text{PATE}_{\Delta t}$ }
\def\Bias{$\text{Bias}_{\Delta t}$ }

\newcommand\sbullet[1][.5]{\mathbin{\ThisStyle{\vcenter{\hbox{%
  \scalebox{#1}{$\SavedStyle\bullet$}}}}}%
}

\newcommand\cmnt[2]{\;
{\textcolor{red}{[{\em #1 --- #2}] \;}
}}
\newcommand\jiajing[1]{\cmnt{#1}{Jiajing}}
\newcommand\damour[1]{\cmnt{#1}{D'Amour}}
\newcommand\franks[1]{\cmnt{#1}{AF}}

\thispagestyle{empty}
\pagenumbering{gobble}

\begin{abstract}

Recent work has focused on the potential and pitfalls of causal identification in observational studies with multiple simultaneous treatments. Building on previous work, we show that even if the conditional distribution of unmeasured confounders given treatments were known exactly, the causal effects would not in general be identifiable, although they may be partially identified.  Given these results, we propose a sensitivity analysis method for characterizing the effects of potential unmeasured confounding, tailored to the multiple treatment setting, that can be used to characterize a range of causal effects that are compatible with the observed data. Our method is based on a copula factorization of the joint distribution of outcomes, treatments, and confounders, and can be layered on top of arbitrary observed data models. We propose a practical implementation of this approach making use of the Gaussian copula, and establish conditions under which causal effects can be bounded. We also describe approaches for reasoning about effects, including calibrating sensitivity parameters, quantifying robustness of effect estimates, and selecting models that are most consistent with prior hypotheses.

\vspace{2em}
\end{abstract}

\clearpage
\pagenumbering{arabic}

\doublespacing
\section{Introduction}

Although it is well-established that treatment effects are not generally identifiable in the presence of unobserved confounding, some have focused on whether this challenge can be mitigated when there are multiple simultaneous treatments \citep{wang2018blessings}. Intuitively, dependence among multivariate treatments could provide information about latent confounders, which could in turn be leveraged to facilitate causal inference and identification.  This intuition has motivated latent variable approaches such as ``the deconfounder'', a much discussed approach for estimating causal effects for multiple treatments \citep{wang2018blessings}.

Unfortunately, it was shown that this strategy has limited practical applicability for point identification and estimation of causal effects.
For example, \citet{d2019comment} and \citet{damour2019aistats} note that causal effects are not nonparametrically identifiable in this setting, even when the distribution of latent confounders can be identified.
Likewise, \citet{ogburn2019comment} and \citet{ogburn2020counterexamples} provide several additional counterexamples and detailed rebuttals to previous theoretical results, while \citet{grimmer2020ive} argue that the approach cannot consistently outperform na\"ive regression, even when stringent assumptions are met.

Nonetheless, latent variable--type strategies are used to estimate causal quantities in genomics \citep{price2006principal},  computational neuroscience, social science and medicine \citep{zhang2019medical}, and time series applications \citep{bica2020real}.
Given the practical importance of these questions, some have focused on finding stronger identifying assumptions for causal effects in the multi-treatment setting. \citet{miao2020identifying} propose identifying assumptions involving proxy or negative control variables and in settings when over half of the treatments are assumed to have a null effect (without specifying which treatments are null).  \citet{kong2019multi} consider identification in a parametric model with binary outcomes.


To date, the literature on multi-treatment causal inference has revolved around a binary question about point identification: can causal effects be identified or not?
However, a potentially more productive question is: what information about treatment effects, if any, can be gained from a latent variable model?
We propose that sensitivity analysis---which explores a range of causal effects that are consistent with the observed data in the context of a given problem---can be a useful tool to address this question.
Specifically, sensitivity analysis can show what can be gained by leveraging latent structure in a given application, even if this (usually) falls short of fully identifying the causal effect of interest.

We focus on settings in which residual dependence between treatments is presumed to be caused by unmeasured confounders.  To extend sensitivity analysis to this setting, we replace untestable assumptions about the magnitude of the dependence between each treatment and unmeasured confounders with an assumption about the  suitability of a latent variable model. Given a (partially) identifiable latent variable model linking confounders to  treatments, we are still left to specify the relationship between unmeasured confounders and the outcome.  Here, we suggest using copulas, which completely characterize this confounder-outcome dependence without affecting the model fit.  For practical analyses, we focus on characterizing this dependence with Gaussian copulas. Under the Gaussian copula specification, we establish that causal effects which are unbounded under unrestricted sensitivity models are bounded when the latent variable model for the treatments is identifiable. In this sense, we show that appropriately motivated latent variable models can sharpen causal conclusions and provide insights about the implications of specific assumptions, even if they cannot point-identify causal effects.

The paper proceeds as follows.
We begin by defining the relevant quantities and notation in Section \ref{sec:setup}.  In Section \ref{sec:gaussian} we illustrate the intuition behind our approach when we can model the treatments via a linear factor model.   In Section \ref{sec:gaussian_copula} we describe a more general framework for latent variable sensitivity analysis via a copula factorization, introducing a special case of the more general approach in which we assume confounder-outcome relationships can be characterized by a Gaussian copula.  We discuss sensitivity parameter interpretation, calibration, and measures of robustness in Section \ref{sec:cali&robust} and, finally, in Section \ref{sec:simulation} and Section \ref{sec:mouse} we demonstrate our approach in simulation and with a gene expression data set recently reanalyzed by \citet{miao2020identifying}.



\section{Preliminaries}
\label{sec:setup}

\subsection{Setup}
Let $T = (T_1, ..., T_k)$ be a random $k$-vector of treatment variables, $Y$ be a scalar random outcome of interest, and $t$ and $y$ be realizations of the respective random variables.  We let $U = (U_1, .., U_m)$ be a random $m$-vector denoting potential unobserved confounders, and $X$ denote any observed pre-treatment variables. We use the \emph{do}-calculus framework \citep{pearl2009causality} and let $f(y \mid do(t))$ denote the density of $Y$ in the population in which we have intervened to assign treatment level $t$ to all units.  In general, this is distinct from the observed outcome density, $f(y \mid t)$, which represents the density of the outcome in the subpopulation that received treatment $t$.  These two densities are the same if and only if there are no confounders \citep{vanderweele2013definition}. 

The goal of observational causal inference with multiple treatments is to quantify the effects of different treatments by comparing the intervention distribution at different levels of treatment $T$ \citep{lechner1999, lopez2017estimation}. In this work we focus on \emph{marginal contrast estimands}  \citep{franks2019flexible} under arbitrary outcome and treatment distributions.  An estimand is a ``marginal contrast'' if it can be expressed as a function of the marginal distributions of the intervention outcomes, e.g. $\tau(E[v(Y)\mid do(t_1)], E[v(Y)\mid do(t_2)])$ for some functions $v$ and $\tau$.  This  includes the vast majority of commonly used estimands.  For continuous outcomes, our primary estimand is the difference in the population average outcome for treatment $T=t_1$ and the population average outcome given treatment $T=t_2$:
\begin{equation}
    \label{eqn:theta_t1t2}
    \text{PATE}_{t_1, t_2} := E(Y \mid do(t_1)) - E(Y \mid do(t_2)).
\end{equation}
Here, $v(Y) = Y$ is the identity function and $\tau(a, b) = a-b$.  We also consider the difference in the population average outcome receiving treatment $t$ and observed population average outcome, which we denote
\begin{equation}
    \label{eqn:eta_t1}
    \text{PATE}_{t, \sbullet} := E(Y \mid do(t)) - E(Y),
\end{equation}
where $E(Y) = \int E(Y \mid t)f(t) dt$ and $\text{PATE}_{t_1, t_2} = \text{PATE}_{t_1, \sbullet} - \text{PATE}_{t_2, \sbullet}$.  By analogy with the PATE, we also define quantile treatment effect for quantile $q$, as $QTE^q$, 

\begin{equation}
    \label{eqn:qte}
    \text{QTE}^q_{t_1, t_2} := \text{quantile}_q(Y \mid do(t_1)) - \text{quantile}_q(Y \mid do(t_2)),
\end{equation}
\noindent and let $\text{MTE}_{t_1, t_2} = \text{QTE}^{1/2}_{t_1, t_2}$ be the median treatment effect. For binary outcomes, our primary estimand is the causal risk ratio between treatments $t_1$ and $t_2$
\begin{equation}
RR_{t_1, t_2} :=  P(Y=1\mid do(t_1))/P(Y=1\mid do(t_2)).
\end{equation}
\noindent where $RR_{t, \sbullet}$ is defined analogously to Equation \ref{eqn:eta_t1}, as $P(Y=1 \mid do(t))/P(Y=1)$, so that we can express $RR_{t_1, t_2} = RR_{t_1, \sbullet}/RR_{t_2, \sbullet}$. Here $v(Y) = I[Y=1]$ is the indicator function and $\tau(a, b) = a/b$.

In general, it is difficult to infer PATEs or RRs from observational data since the potential presence of unmeasured confounders, which affect both treatment and outcome, can bias naive estimates. If $U$ were to be observed, the following assumptions would be sufficient to identify the intervention distribution, and hence the treatment effect:

\begin{assumption}[Backdoor Criterion]\label{asm:unconfoundedness}
$X$ and $U$ block all backdoor paths between $T$ and $Y$ so that $f(Y=y \mid do(T)=t, X=x, U=u) = f(Y \mid T=t, X=x, U=u)$ \citep{pearl2009causality}.
\end{assumption}
\begin{assumption}[Positivity]\label{asm:positivity}
$f(T = t \mid U=u, X=x) > 0$ for all $u$ and $x$ such that $P(U=u, X=x) > 0$.
\end{assumption}
\begin{assumption}[SUTVA]\label{asm:sutva} There are no hidden versions of the treatments and there is no interference between units \citep[see][]{rubin1980comment}.

\end{assumption}

\begin{figure}	
	\centering
	\begin{subfigure}[t]{0.31\textwidth}
		\centering
\includegraphics[width=\textwidth,valign=t]{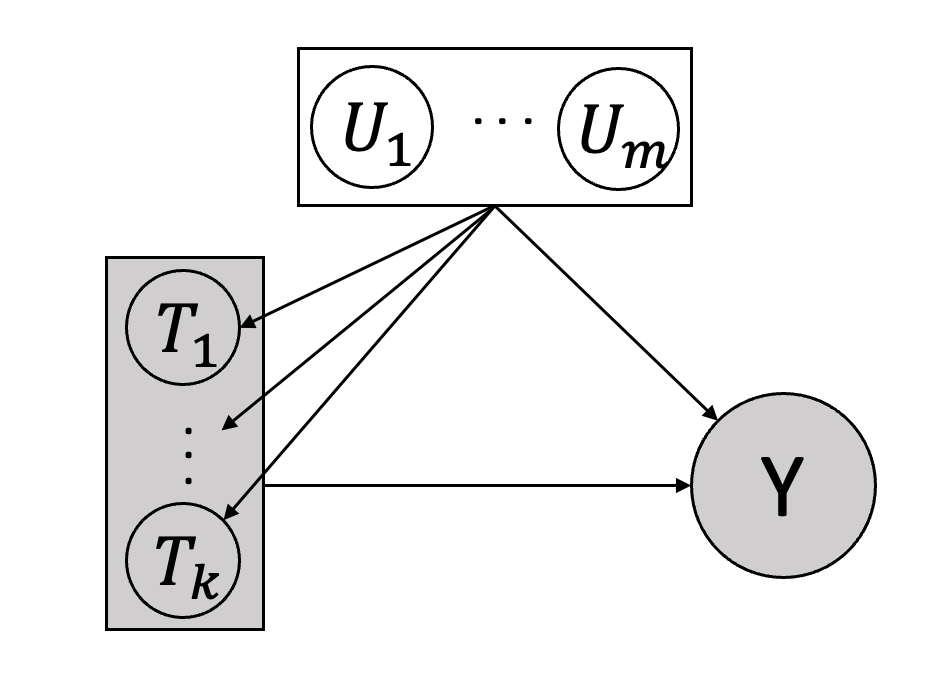}
		\vspace*{0pt}
	\end{subfigure}
	\caption{$k$-vector of treatments $T$, $m$-vector of unmeasured confounders $U$ and a scalar outcome $Y$. In the linear factor model given by Equation \ref{eqn:ppca}, the treatments are conditionally independent given the unmeasured confounders. We exclude observed covariates, $X$, from the diagram for simplicity.
    }
    \label{fig:diagram}
\end{figure}

\noindent Assumption \ref{asm:positivity}, also called the overlap condition, ensures that every observable level of the potential confounders $(U, X)$ has a positive probability of being observed with any treatment $t$, and is needed for nonparametric identification of causal effects. Assumption \ref{asm:sutva} is a standard consistency condition. The focus on this work relates to Assumption \ref{asm:unconfoundedness}. By conditioning on $U$ and $X$, we ``block" non-causal paths between the treatments and the outcome, so that any residual dependence between the treatments and the outcome must be induced by the intervention on the treatment (See Figure \ref{fig:diagram}). Under this assumption $f(Y=y \mid do(T=t)) = \int_{\mathcal X, \mathcal U} f(Y=y \mid T=t, X=x, U=u) f(X=x, U=u) dx du$.

However, since $U$ is not observed, and it is not generally true that $f(Y=y \mid do(T)=t, X=x) = f(Y=y \mid T=t, X=x)$, treatment effects are not identifiable without additional assumptions about the influence of $U$. 
In this case, a common solution is to conduct a sensitivity analysis which characterizes how the implied causal effects change under different assumptions about $U$ and its relationship to $T$ and $Y$ given $X$. 


\subsection{Sensitivity Analysis}

There is an extensive literature on assessing sensitivity to violations of unconfoundedness in single treatment models, dating back at least to the work of \citet{cornfield1959smoking} on the link between smoking and lung cancer.  Since then, a wide range of strategies have been proposed for assessing sensitivity to unobserved confounding  \citep[e.g. see][]{greenland1996basic, gastwirth1998dual, vansteelandt2006ignorance, imbens2003sensitivity, vanderweele2011bias, vanderweele2012sensitivity, robins2000sensitivity, franks2019flexible, cinelli2020making, veitch2020sense}.
The sensitivity analysis approach that we propose in this paper builds on latent confounder approaches to sensitivity analysis. These approaches assert, as in Assumption \ref{asm:unconfoundedness}, that unconfoundness would hold if only an additional latent variable $U$ were observed \citep{rosenbaum1983assessing, ROBINS_NON_1997,vansteelandt2006ignorance, daniels2008missing}.

In a typical latent confounder analysis, we posit densities $f(u \mid x)$, the marginal density of the latent confounders, $f_{\psi_T}(t \mid x, u)$, the conditional density or probability mass function (PMF) for treatment assignment given all confounders and $f_{\psi_{Y}}(y \mid x, u, t)$, the outcome density in the treatment arm $t$.   The dependence of $Y$ and $T$ on $U$  is indexed by a vector of sensitivity parameters $\psi = (\psi_Y, \psi_T)$ \citep[e.g. see][]{imbens2003sensitivity, dorie2016}.  
Practitioners can then reason about how assumptions about these parameters translate to different causal conclusions.
Often, this is done through \emph{calibration}, by determining reasonable ranges for $\psi$ using analogies about observable associations and through a \emph{robustness} assessment, by examining how strong associations with unobserved confounders must be for conclusions to change. Latent confounder models are usually parameterized so that some specific values of the sensitivity parameters $\psi$ indicate the ``no unobserved confounding'' case.  For example, we can take $\psi_T = 0$ to imply that $f_{\psi_T}(t \mid x, u) = f_{\psi_T}(t \mid x)$ and $\psi_Y = 0$ to imply that $f_{\psi_{Y}}(y \mid x, u, t) = f_{\psi_{Y}}(y \mid x, t)$.  Then, when either $\psi_T = 0$ or $\psi_Y =0$, $U$ is not a confounder \citep{vanderweele2013definition}.  Without loss of generality, we suppress conditioning on $x$ throughout the remainder of the manuscript, and comment on the role of observed covariates where appropriate.  



\subsection{Partial Identification and Copula Parameterizations}
\label{sec:partial_Identification_CopParam}

In this paper, we focus on models for which it is possible to learn the conditional confounder density $f_{\psi_T}(u\mid t)$ from a latent variable model on the multiple treatments. We then explore how the causal effects change under different assumptions about the $U$-$Y$ relationship, as governed by the sensitivity parameter $\psi_Y$.  To do so, we use a sensitivity parameterization in which the observed data densities are invariant to the choice of sensitivity parameters \citep{gustafson2018sensitivity, franks2019flexible}. We consider this to be desirable because it implies that the data offer equivalent support to the various causal conclusions admitted by the sensitivity analysis.
The model for $Y$ conditional on treatments and potential unobserved confounders can be decomposed into the observed data density and a conditional copula as 
\begin{equation}
f_\psi(y \mid u, t) = f(y \mid t)c_{\psi}(F_{Y \mid t}(y), F^{\psi_T}_{U_1 \mid t}(u_1), ..., F^{\psi_T}_{U_m \mid t}(u_m) \mid t)
\label{eqn:cop_full}
\end{equation}
\noindent where $F_{Y\mid t}$ is the CDF associated with $f(y \mid t)$ and $F_{U_i\mid t}^{\psi_T}$ are the CDFs associated with $f_i^{\psi_T}(u_i \mid t)$, the density for $i$th latent confounder given treatments. $c_{\psi}$ is the conditional copula density, defined on the $m+1$ dimensional unit hypercube and parameterized by $\psi = \{\psi_Y, \psi_T\}$, which characterizes the joint density of $Y$ and $U_1, ..., U_m$ conditional on $T=t$ after transforming the marginals to uniform random variables \citep{nelsen2007introduction}. This factorization holds for all densities (or PMFs) $f(y \mid t)$ and $f_{\psi_T}(u \mid t)$ and any number of treatments, and thus can be used to characterize the outcome-confounder dependence for any model of the observables.

With Equation \ref{eqn:cop_full}, we can express the intervention distribution, $f_{\psi}(y \mid do(t))$, in terms of the observed conditional distribution, $f(y\mid t)$, as:
\begin{align} \label{eqn:fac_y|do(t)}
        f_{\psi}(y \mid do(t)) &= f(y \mid t) \int c_{\psi}(F_{Y \mid t}(y), F^{\psi_T}_{U_1 \mid t}(u_1), ..., F^{\psi_T}_{U_m \mid t}(u_m) \mid t)
    f(u) du
\end{align}

\noindent Note that when either $c_\psi$ is the independence copula or $f(u\mid t) = f(u)$ there is no confounding and the integral on the right-hand side of Equation \ref{eqn:fac_y|do(t)} evaluates to $1$  so that the intervention density is identical to the conditional outcome density, as expected.

\ed{In cases where we can learn $f_{\psi_T}(u\mid t)$ from a latent variable model, we show that the copula integrand in Equation \ref{eqn:fac_y|do(t)} can be decomposed further, yielding some pieces that are identified as part of $f_{\psi_T}(u \mid t)$, and some fully unidentified pieces governing the $U$-$Y$ dependence which are governed by a parameter we denote as $\psi_Y$ (see Proposition \ref{prop:vine_cop}). Thus, in these cases, it suffices explore how causal conclusions change under different assumptions about $\psi_Y$, which governs the conditional copula, $c_{\psi}$ once $\psi_T$ is known.  Because this general parameterization can be complex, we focus primarily on the setting in which $c_{\psi}$ is a Gaussian copula which characterizes monotone dependences between unmeasured confounders and the outcome.  For the Gaussian copula, we show that the causal effects are partially identified, with the most extreme outcomes achieved when there is perfect dependence between the outcome and unmeasured confounders given the treatment.}

Our results contribute to an extensive literature on partial identification \citep{manski2003partial, gustafson2015bayesian}, and in particular, approaches to partial identification involving copulas \citep{tamer2010partial}.  For partially identified parameters, a common approach is to consider the worst-case bounds under a set of weaker assumptions and show how additional assumptions can further sharpen inferences \citep{manski2003partial, manski2008identification}. Partial identification results have been established in causal settings  with instrumental variables \citep{swanson2018partial, flores2013partial}, causal inference with noisy covariate data \citep{guo2022partial}, and for estimation of individual treatment effects (ITEs). A key result from the copula literature, due to Fr\'echet and Hoeffding, characterizes model-free bounds on the joint CDF of random variables as functions of the marginal CDFS. The Fr\'echet-Hoeffding bound and other related bounds have been specifically used to bound the distribution of ITEs and other functionals of the joint distribution of potential outcomes \citep[see e.g.][]{heckman1997making, fan2010sharp, firpo2019partial}.  Our formulation is fundamentally different from approaches using model-free copula bounds, since we remain focused on marginal contrasts (not joint distributions over potential outcomes) and also focus on parametric copula models for characterizing the dependence between the outcome and potential unmeasured variables.

%
%




\section{Confounding Bias in the Linear Factor Model}
\label{sec:gaussian}

Before detailing our general copula-based approach, we provide some crucial intuition about our sensitivity analysis in a simple linear Gaussian factor model. The more general approach which we will introduce in the next section uses the same ideas presented here but relaxes the requirements on the marginal distributions of the treatment and outcome by using copulas. In the linear-Gaussian model, we highlight the following results:


\begin{itemize}
    \item For causal inference with multiple treatments, we show that the magnitude of the confounding bias for $\text{PATE}_{t_1, t_2}$ is bounded.  Given standard assumptions for factor model identifiability this bound is identifiable. We characterize how the magnitude of this bound depends on the parameters of the latent confounder model and a scalar sensitivity parameter.
    \item The confounding bias depends on the treatment contrast.  We characterize which treatment contrasts lead to the largest bounds and which treatment contrasts (if any) imply identifiable effects.    
    \item For causal inference with a single treatment, for which the conditional confounder distribution is not identifiable, we cannot identify a bound for confounding bias of $\text{PATE}_{t_1, t_2}$ without additional assumptions.  
\end{itemize}

\subsection{Sensitivity Bounds in the Linear Gaussian Model}

\noindent In this section, we establish expressions for confounding bias of $\text{PATE}_{t_1,t_2}$ in terms of the parameters of a linear Gaussian model. We assume the following linear structural model:
\begin{align} 
    U & = \epsilon_u \label{eqn:confounder}\\
    T &= BU + \epsilon_{t} \label{eqn:ppca}\\
    Y &= \tau^\top  T + \gamma^\top  U + \epsilon_{y} \label{eqn:outcome}
\end{align}
\noindent where $U$ is $m$-dimensional, $T$ is $k$-dimensional and $Y$ is scalar, with $\epsilon_u, \epsilon_t$ and $\epsilon_y$ all mean-zero Gaussian, $\text{Var}(\epsilon_u) =I_m$, $\text{Var}(\epsilon_y)=\sigma^2$ and $\text{Var}(\epsilon_t) = \Lambda_{t}$ a $k \times k$ positive diagonal matrix. Further, $B \in \mathbb{R}^{k \times m}$, $\tau \in \mathbb{R}^k$, $\mathbb{\gamma} \in \mathbb{R}^{m}$. When either $B = 0$ or $\gamma = 0$, there is no confounding.  In the more general framing introduced from Equation \ref{eqn:cop_full} in the Section \ref{sec:partial_Identification_CopParam},  $\psi_t = \{B\}$ determines the copula defining the $T$--$U$ relationship and $\psi_Y = \{\gamma\}$ specifies the copula defining $Y$--$U$ dependence.

Equations \ref{eqn:confounder} and \ref{eqn:ppca} imply that the conditional distribution of the confounder can be expressed as $f_{\psi_t}( u \mid t) \sim N(\mu_{u\mid t}, \Sigma_{u \mid t})$, where

 \begin{align}
\mu_{u\mid t} &= B^\top ( B  B^\top  + \Lambda_{t})^{-1} t \label{eqn:mu_u_t}\\
\Sigma_{u\mid t} &= I - B^\top (BB^\top  + \Lambda_t)^{-1}B \label{eqn:sigma_u_t}
 \end{align}
\noindent  Under model in Equations \ref{eqn:confounder}-\ref{eqn:outcome}, the intervention distribution has density
\begin{equation}
f(y \mid do(T = t)) \sim N(\tau^\top  t, \ \sigma^2 +  \gamma^\top  \gamma).
\end{equation}
For any $t_1, t_2$, $\text{PATE}_{t_1,t_2}$ is characterized entirely by the regression coefficients $\tau$.  The observed outcome distribution can be expressed as
\begin{equation}
    f(y \mid T = t) \sim N(\tau_{\text{naive}}^\top  t, \  \sigma^2_{y\mid t}),
\end{equation}
where 
\begin{align}
\tau_{\text{naive}} &= \tau + \gamma^\top \mu_{u\mid t}\\
\sigma^2_{y\mid t} &= \sigma^2 + \gamma^\top \Sigma_{u|t}\gamma. \label{eqn:var_decomp}
\end{align}
We refer to $\tau_{\text{naive}}$ as the naive estimate since it naively neglects the effect of unobserved confounders.  Equation \ref{eqn:var_decomp} shows that the observed residual outcome variance, $\sigma^2_{y\mid t} := \text{Var}(Y\mid T)$, can be decomposed into nonconfounding variation $\sigma^2$ and confounding variation, $\gamma^\top \Sigma_{u|t}\gamma$. 

We note that the population average treatment effect and the bias of the naive estimator depends only on the difference between the treatment vectors,  $(t_1 - t_2)$.  This is the case since the population average treatment effect can be expressed as
\begin{equation}
\text{PATE}_{t_1, t_2} = \tau^\top  (t_1 - t_2)
\end{equation}

\noindent and the confounding bias, $\text{Bias}_{t_1, t_2} = (\tau_{\text{naive}} - \tau)^\top  (t_1 - t_2)$.   It is then straightforward to show that the bias is linear in the difference in the confounder means in each treatment group:
\begin{align}
\label{eqn:pate_bias}
\text{Bias}_{t_1, t_2} & = \gamma^\top  (\mu_{u|t_1} - \mu_{u | t_2}) = \gamma^\top B^\top ( B  B^\top  + \Lambda_{t})^{-1} (t_1 - t_2),
\end{align}

\noindent For the results that follow, it is useful to define the fraction of residual outcome variance explained by confounders as a key quantity:
\begin{equation}
    0 \leq R_{Y \sim U|T}^2 = \frac{{\gamma}^\top  {\Sigma}_{u|t} {\gamma}}{\sigma_{y\mid t}^2} \leq 1 \label{eqn:partial_r2} 
\end{equation}
This R-squared value can be viewed as a parameter governing the copula $c_{\psi_Y}$ in the general model given by Equation \ref{eqn:cop_full}, and plays a central role in sensitivity analysis frameworks such as \citet{cinelli2020making}.
Using $R_{Y \sim U|T}^2$, we can write a bound on the omitted variable bias of any $PATE_{t_1, t_2}$.

\begin{thm} \label{thm:linear_bound}
Suppose that the observed data is generated by model in Equations \ref{eqn:confounder}-\ref{eqn:outcome} with $\Lambda_{t|u} > 0$. 
Then,  $\forall \gamma$,
    \begin{equation} \label{eqn:ellipse_gamma}
    {\gamma}^\top  \Sigma_{u|t} {\gamma} \leq \sigma^2_{y\mid t}
    \end{equation}

\noindent For any given $t_1$, $t_2$, we have
    \begin{equation}
    \label{eq:bound,multi-T}
    \text{Bias}_{t_1, t_2}^2 \leq \sigma^2_{y|t}R_{Y \sim U|T}^2 \| \Sigma_{u|t}^{-1/2} (\mu_{u|t_1} - \mu_{u|t_2}) \|_2^2.
    \end{equation}
The bound is achieved when ${\gamma}$ is colinear with $\Sigma_{u|t}^{-1}(\mu_{u|t_1} - \mu_{u\mid t_2})$ and is maximized when all the residual outcome variance is due to unmeasured confounders, e.g. $R^2_{Y\sim U|T} = 1$.\\
\noindent Proof. See appendix \ref{sec:proof_thm_linear_bound}.
\end{thm}
\noindent This theorem states that the true causal effect lies in the interval 
\begin{equation}
\label{eqn:ig_region}
\tau_{naive}^\top (t_1 - t_2) \pm \sqrt{\sigma^2_{y|t}R_{Y \sim U|T}^2} \| \Sigma_{u|t}^{-1/2} (\mu_{u|t_1} - \mu_{u|t_2}) \|_2.    
\end{equation}

\noindent We refer to the right-hand side of Equation \ref{eq:bound,multi-T} as the ``worst-case bias'' of the naive estimator.   In particular, since $\tau_{\text{naive}}$ is the midpoint of the ignorance region, it has the minimum worst-case bias over all alternative causal effect estimators. This is consistent with \citet{grimmer2020ive} who emphasize that the deconfounder proposed by \citet{wang2018blessings} cannot outperform the naive estimator in general. 

In the following corollary, we provide additional intuition by establishing the worst-case bias over all possible treatment contrasts in the special case of the homoskedastic factor model, for which $\Lambda_{t} = \sigma^2_{t} I$:
\begin{cor}
\label{thm:anyt}
Assume $\Lambda_{t} = \sigma^2_{t} I$ and let $d_1$ be the largest singular value of $B$.  For all $t_1, t_2$ with $\parallel (t_1 - t_2) \parallel_2 = 1$, the squared bias  is bounded by
\begin{equation}
\label{eq:bound_width,multi-T}
\text{Bias}_{t_1, t_2}^2
\leq \frac{ d_1^2}{ (d_1^2 + \sigma_{t}^2)}\frac{\sigma_{y|t}^2}{\sigma_{t}^2} R_{Y \sim U|T}^2,
\end{equation}
with equality when $(t_1 - t_2) = u_1^{B}$, the first left singular vector of $B$. When $(t_1 - t_2) \in  Null(B^\top )$, the naive estimate is unbiased, that is, $PATE_{t_1, t_2} = \tau_{\text{naive}}^\top  (t_1 - t_2)$. \\
Proof: See Appendix \ref{sec:proof_thm_anyt}.
\end{cor}

\noindent The first term in Equation \ref{eq:bound_width,multi-T}, $\frac{ d_1^2}{ (d_1^2 + \sigma_{t}^2)}$, is the fraction of variance in the first principal component of the causes that can be explained by confounding.  The first principal component corresponds to the projection of treatments which is most correlated with confounders, and thus is the causal contrast with the largest ignorance region.  We illustrate and discuss some additional insights from Corollary \ref{thm:anyt} in Figure \ref{fig:gaussian_example}, Appendix A.

\subsection{Identifiability of Sensitivity Bounds}

In the previous subsection, we established bounds on the omitted variable bias, but did not characterize whether the bounds are themselves identifiable.  Now, we show that factor model identifiability assumptions (when appropriate) indeed imply that these bounds are identifiable.  Thus, multi-treatment inference can yield shaper sensitivity analyses, beyond what is possible when considering inference one treatment at a time.

Crucially, in model in Equations \ref{eqn:confounder}-\ref{eqn:outcome}, $B$ can be identified (up to rotation) under standard factor model conditions \citep{anderson1956statistical}.  The parameter $\gamma$ is not identifiable but can be considered a sensitivity vector that parameterizes the residual correlation between the m-dimensional unobserved confounder U and
the outcome Y after conditioning on the treatment vector T. We use results on the identification of $B$ to establish the following proposition.

\begin{proposition} \label{prop:equivalence}
Suppose that the observed data is generated by model in Equations \ref{eqn:confounder}-\ref{eqn:outcome}.  If B is rank $m$ and there remain two disjoint matrices of rank $m$ after deleting any row of $B$, then $\| \Sigma_{u|t}^{-1/2} (\mu_{u|t_1} - \mu_{u|t_2}) \|_2^2$ and $\frac{ d_1^2}{ (d_1^2 + \sigma_{t}^2)}$ are both identified.  \\
\noindent Proof. See appendix \ref{sec:proof_prop_equivalence}.
\end{proposition}

\noindent As a result, the bounds in both Equations \ref{eq:bound,multi-T} and \ref{eq:bound_width,multi-T} are identified given $R^2_{Y \sim U |T}$.  Further, since $R^2_{Y \sim U |T}$ is itself at most $1$, we can indeed identify an upper bound on the omitted variable bias under the assumptions in Proposition \ref{prop:equivalence}. Notably, the sufficient conditions in \ref{prop:equivalence} can only be satisfied when the number of confounders is $m \leq (k-1)/2$.
This fact can be useful for practitioners to reason about the number of treatments that suffice, relative to the number of unmeasured confounders, to bound the confounding bias. 

Proposition \ref{prop:equivalence} also establishes a key distinction between the multi-treatment and single treatment settings for sensitivity analysis. Specifically, when $k=1$ (single treatment causal inference), we cannot identify a bound on the omitted variable bias. As shown in \citet{cinelli2020making}, in the single treatment case, the squared confounding bias of the PATE can be expressed as 
\begin{equation}
\label{eqn:bias_single}
{Bias}_{t_1, t_2}^2 =  \frac{\sigma^2_{y\mid t}}{\sigma^2_{T}}\left(\frac{R^2_{T \sim U}}{1-R^2_{T \sim U}}\right)R^2_{Y \sim U | T}
\end{equation}
where $\sigma^2_T:= BB^\top  + \Lambda_{t}$ is the marginal variance of the treatment and
\begin{equation}
0 \le R^2_{T \sim U} = \frac{\sigma^2_T ||\mu_{u\mid t_1} - \mu_{u\mid t_2}||_2^2}{(t_1-t_2)^2} \le 1 
\end{equation}
is the unidentified fraction of treatment variance explained by confounders. In the single treatment case, neither $R^2_{T\sim U}$ nor $R^2_{Y \sim U \mid T}$ are identifiable, and since  $\frac{R^2_{T \sim U}}{1-R^2_{T \sim U}}$ can be arbitrarily large, the confounding bias is unbounded without additional assumptions.  As such, \citet{cinelli2020making} consider a variety of calibration and robustness criteria for reasoning about plausible magnitudes for both $R^2_{Y\sim U \mid T}$ and $R^2_{T\sim U}$.
In contrast, as we show above, in the multiple treatment case, the marginal bounds on the omitted variable bias depend on the sensitivity vector $\gamma$ only through the fraction of outcome variance explained by confounders given the treatment, $R^2_{Y \sim U|T} = \frac{\gamma^\top \Sigma_{u\mid t}\gamma}{\sigma^2_{y\mid t}}$. In later sections, we explore how domain knowledge combined with techniques for calibrating $\gamma$ and its magnitude can be used to further sharpen the set of plausible causal conclusions.

\section{Sensitivity Analysis via Copula Parameterizations}
\label{sec:gaussian_copula}

We now establish a sensitivity parameterization for multiple treatment causal inference in a more general class of models \ed{by further factorizing the copula parameterization Equation \ref{eqn:fac_y|do(t)} into identifiable and unidentifiable parts}. For this class of models, we propose an algorithm for estimating any marginal contrast estimand.  

As illustrated in the previous section, when $T$ is multivariate, we might gain information about $\psi_T$ if we are willing to make assumptions about the class of latent variable models linking the unmeasured confounders to the treatment.  
Here, we show generically how identification of the $T$-$U$ relationship translates to bounds on marginal contrast estimands.

\begin{assumption}[Latent variable model identification]
\label{asm:latent_identification}
The potential confounders are continuous and their distribution given treatments, $f\psi_{T}(u_1, ..., u_m \mid t)$, is identifiable up to rotation and scale.
\end{assumption}

\noindent \ed{
We can rewrite the $(m+1)-$dimensional copula as a product of the $m$-dimensional copula characterizing U-dependence, as determined by a latent variable model fit to the treatments, and a product of bivariate (conditional) copulas between each dimension of $U$ and $Y$.  This factorization cleanly separates the terms that are identifiable under Assumptions \ref{asm:unconfoundedness}-\ref{asm:latent_identification} from those that are not identifiable.} 

\ed{
\begin{proposition} For any ordering of the latent confounders $U = U_1, ..., U_m$,
the joint (m+1) dimensional copula between the scalar outcome and $m$ latent confounders can be re-expressed as

\begin{align}
&c_{\psi}(F_{Y \mid t}(y), F_{U_1\mid t}^{\psi_T}(u_1), ..., F^{\psi_T}_{U_m\mid t}(u_m)\mid t) =\\ 
&c_{\psi_T}(F_{U_1\mid t}^{\psi_T}(u_1), ..., F^{\psi_T}_{U_m\mid t}(u_m)\mid t)\prod_{i=1}^{m}
c_{\psi}({y,u_i; u_1, ..., u_{i-1}}, t)
\end{align}
where $c_{\psi_T}$ is the $m$-dimensional copula governing the dependence amongst latent variables, $U$ given $T=t$ and we let $c_{\psi}({y,u_i; u_1, ..., u_{i-1}}, t) := c_{\psi}(F_{Y\mid u_1, ... u_{i-1}, t}(y),F_{U_i\mid u_1, ... u_{i-1}, t} (u_i) \mid u_1, ..., u_{i-1}, t)$ is a bivariate (conditional) copula characterizing the dependence between $Y$ and $U_i$ given $i-1$ of the ``preceding'' variables. Given Assumption \ref{asm:latent_identification}, $c_{\psi_T}$ and $F^{\psi_T}_{U_i\mid t}$ are fully identified.  All bivariate copulas, $c_{\psi}({y,u_i; u_1, ..., u_{i-1}}, t)$ are unidentified without further assumptions.
\label{prop:vine_cop}

\noindent Proof: See Appendix \ref{prop:vine_cop_proof}.
\end{proposition}

\noindent Proposition \ref{prop:vine_cop} results from a decomposition of multivariate copulas into products of bivariate copulas known as D-vine decomposition \citep{czado2019analyzing}.  For us, the proposition highlights that when $\psi_T$ is identifiable from a latent variable model, we need only specify $\psi_Y$, which characterizes the $U$-$Y$ dependence, for example by parameterizing $m$ bivariate conditional copulas.  



}

\ed{Given $f(y\mid t)$ (identifiable), $f_{\psi_T}(u_1, ..., u_m \mid t)$ (identifiable by Assumption \ref{asm:latent_identification}) and $c_{\psi}({y,u_i; u_1, ..., u_{i-1}}, t)$ for all $i$ (nonidentifiable, governed by chosen sensitivity parameter $\psi_Y$)}, we can compute the expected value of any function of the outcome under the intervention distribution, $E[v(Y) \mid do(t)] = \int v(y) f(y \mid do(t)) dy$.   This can be in turn used to compute any marginal contrast estimand. Applying Equation \ref{eqn:fac_y|do(t)}, we write this intervention expectation as
\begin{equation}
\label{eqn:interven_expect}
    E[v(Y) \mid do(t)]\ =  \int v(y) w_\psi(y, t) f(y \mid t) dy,
\end{equation}
where  $w_\psi(y, t) = \int c_{\psi}(F_{Y|t}(y), F^{\psi_T}_{U|t}(u) \mid t) f_{\psi_T}(u) du$ is the importance weight associated with sampling from the observed data distribution instead of the intervention distribution. In practice, we can approximate the marginal distribution of the unobserved confounder with the mixture density $f_{\psi_T}(u) \approx \frac{1}{n} \sum_i f^{\psi_T}(u \mid t_i)$ where $t_i \in  \mathcal{T}$ is the $i$th observed treatment and $\mathcal{T}$ is the set of all observed treatment vectors.  Thus, the importance weight can be approximated as
\begin{equation}
\label{eqn:weight}
    w_\psi(y, t) \approx \frac{1}{|\mathcal{T}|}\sum_{t_i \in \mathcal{T}} \left[ \int ... \int c_{\psi}(F_{Y \mid t}(y), F_{U_1\mid t}^{\psi_T}(u_1), ..., F^{\psi_T}_{U_m\mid t}(u_m)\mid t) f_{\psi_T}(u \mid t_i) du_1, ..., du_m\right].
\end{equation}
We use this approximation to derive importance sampling algorithm for computing the expected value in Equation \ref{eqn:interven_expect} for any copula and conditional confounder distributions $f(u \mid t)$ (Appendix A, Algorithm \ref{algm:general}).  This can in turn be used to compute any marginal contrast estimand, $\tau(E[v(y)|do(t_1)], E[v(y)|do(t_2)])$.

Note that the results in this work require continuous unmeasured confounders.  We leave explorations of discrete latent variable models for future work.  Latent variable model identification is essential for the validity of our sensitivity analysis and will not hold in all settings.  However, while a complete discussion of identifiability in latent variable models is outside the scope of this work, there is a broad range of mathematical settings in which this assumption does hold.  In the previous section, we noted the classical result due to \citet{anderson1956statistical} establishing identifiablity conditions for linear factor models. \citet{allman2009identifiability} establish identifiability for many latent class models, including those with limited direct dependence between observations; and \citet{miao2020identifying} where weak sufficient conditions are given for similar identifiability in linear models. \citet{barber2022half} consider a set of criteria for establishing identifiability of direct effects in linear structural equation models with latent variables \citet{rohe2022vintage} consider identifiability in a broader class of (non-Gaussian) factor models. It is up to the practitioner to decide whether latent variable identifying assumptions are compatible with their applied setting. Finally, in Appendix \ref{sec:causal_equiv} we formalize the idea that it is sufficient to identify the latent variable density only up to invertible linear transformations (rotation and scale). To do so we introduce the notion of a ``causal equivalence class'', which establishes that the substantive causal conclusions do not depend on a particular rotation or scale for the latent confounders.


\subsection{The Gaussian Copula Sensitivity Parameterization}


\ed{A challenge with the most general copula specification is that the sensitivity parameter $\psi_Y$ can be complex, hard to reason about and interpret.  In order to bridge the gap between the interpretable sensitivity parameterization and established theory in the linear Gaussian model (Section \ref{sec:gaussian}) with the most general formulation from Proposition \ref{prop:vine_cop}, we provide additional results in an intermediate setting in which the full $m+1$ dimensional coupla, $c_\psi$, is a Gaussian copula.
}

First, we establish the class of latent variable models for the treatments, $f_{\psi_T}(u\mid t)$ by assuming the latent variables are conditionally Gaussian.  This assumption facilitates sensitivity parameter interpretation and the theoretical results that follow.

\begin{assumption}[Conditionally Gaussian latent confounders]
\label{asm:gaussian_latent}
\sloppy
The conditional distribution of latent confounders given treatments is Gaussian, 
that is $f_{\psi_T}(u\mid t) \allowbreak \sim \allowbreak N(\mu_{u\mid t}, \Sigma_{u\mid t})$ with $\psi_T = \{\mu_{u\mid t}, \Sigma_{u\mid t} : t \in \mathcal{T}\}$ 
where $\Sigma_{u|t}$ is a positive semi-definite covariance matrix. Further, the marginal means, $E[U_i]$ exist for all $i$ in $1, \allowbreak \dots, \allowbreak m$ and we assume without loss of $E[U_i] = 0$.
\end{assumption}

\noindent Assumption \ref{asm:gaussian_latent} includes the linear Gaussian case (see Equations \ref{eqn:mu_u_t}- \ref{eqn:sigma_u_t}) and is also commonly assumed in many flexible neural network-based latent variable models like variational autoencoder (VAE) models (see VAE examples in Section \ref{sec:sim_nongaussian_t} and Appendix \ref{sec:movie}). By definition, Assumption \ref{asm:gaussian_latent} implies $c_{\psi_T}(F_{U_1\mid t}^{\psi_T}(u_1), ..., F^{\psi_T}_{U_m\mid t}(u_m)\mid t)$ is a Gaussian copula.

Next, we establish the confounder-outcome relationship by assuming the form of the conditional outcome density from Proposition \ref{prop:vine_cop}.
\begin{assumption}[Gaussian outcome-confounder dependence]
\label{asm:copula}


\sloppy
For any ordering of $U_1, \allowbreak \dots, \allowbreak U_m$, the bivariate (conditional) copula relating the outcome to the $i$th latent confounder, $c_{\psi_Y}(y, u_i;\mid u_1, ... u_{i-1}, t)$ is a bivariate Gaussian copula and invariant to the specific levels $u_1, ... u_{i-1}$ for all $i = 1, ..., m$ with $\psi_Y = \{ \rho_1, ..., \rho_m\}$ where $\rho_i \in [-1, 1]$ is the scalar copula parameter governing the dependence between $Y$ and $U_i$ given $U_1, ..., U_{i-1}$ and $t$.
 
\end{assumption}

\noindent \ed{

Assumption \ref{asm:copula} is a natural choice when the conditional median of the outcome is plausibly monotone in the conditional median of the confounders. It includes the special limiting case in which the outcome is comonotone (perfect positive dependence) or countermonotone (perfect negative dependence) with confounders. Outcome-confounder monotonocity is often plausible, at least approximately, conditional on each level $T=t$.

Assumptions \ref{asm:gaussian_latent} and \ref{asm:copula} imply the full (m+1) dimensional copula between all latent confounders is also a Gaussian copula.  Further, these assumptions imply the following a particular structural equation model which we state in the following proposition.

\begin{proposition} Given Assumptions \ref{asm:gaussian_latent} and \ref{asm:copula}, the full $(m+1)$-dimensional conditional copula between the outcome and all $m$ latent confounders given treatments,\\ $c_{\psi}(F_{Y \mid t}(y), F_{U_1\mid t}^{\psi_T}(u_1), ..., F^{\psi_T}_{U_m\mid t}(u_m)\mid t)$, is a Gaussian copula. Further, any data generating process satisfying Assumptions 5 and 6 can be represented using the following structural equation model, which links $Y$  to $U$ through a scalar latent Gaussian variable $Z_Y$ with unit variance:
 \begin{align}
    U &= \mu_{u\mid t} + \Sigma_{u\mid t}^{1/2}\epsilon_U \label{eqn:u_mid_t}\\
    Z_Y &= \gamma_t^\top U + \sqrt{1 - \gamma_t^\top \Sigma_{u\mid t}\gamma_t}\epsilon_{Z_Y} \label{eqn:ytilde_x,u_gauss}\\
     Y &= F^{-1}_{Y|T}(\Phi(Z_Y -\gamma_t^\top \mu_{u\mid t}))\label{eqn:y_ytilde_relationship_gauss}
\end{align}
where $\epsilon_U \sim N(0, I_m)$, and $\epsilon_{Z_Y} \sim N(0, 1)$. $\psi_Y = \{\gamma_t\}$ is a reparameterization of the $\rho_i$ parameterization.
\label{prop:model_gcop}

\noindent Proof: See Appendix \ref{sec:proof_prop_model_gcop}.
\end{proposition}}



\noindent The full $m+1$ dimensional Gaussian copula under this model is fully determined by the correlation matrix associated with the covariance matrix
\begin{equation} 
\label{eqn:gaussian_copula}
    \text{Cov}([Z_Y, U] \mid T=t) = 
    \begin{bmatrix}
        1 & \gamma_t^\top \Sigma_{u|t}\\
    \ \Sigma_{u|t}\gamma_t & \ \Sigma_{u|t}
    \end{bmatrix}
\end{equation}
Per Assumption \ref{asm:latent_identification}, we assume that $\psi_T$ is identified up to invertible linear transformations of $U$. We can then consider $\psi_Y = \{\gamma_t\}$ the sole $m-$dimensional sensitivity vector governing the magnitude of the omitted variable bias, and can explore the range of possible causal effects for different $\gamma_t \in \mathbb{R}^m$ satisfying $\gamma^\top _t\Sigma_{u \mid t}\gamma_t \leq 1$.

In Algorithm \ref{algm:gaussian} (Appendix A), we provide a modification of Algorithm \ref{algm:general} tailored to the Gaussian copula setting.  At a high level, we compute a Monte Carlo estimate of $f(y \mid do(t))$ via the following three-step procedure: (1) draw a sample from $f(u)$, (2)  compute the conditional density of the Gaussianized outcome $f(z_y \mid u, t)$ via the Gaussian copula and (3) transform $Z_Y$ back to original outcome space via the conditional quantile function $F^{-1}_{Y\mid t}$ (see Figure \ref{fig:calibration}, Appendix A). In the following sections, we introduce some theoretical insights about our Gaussian copula approach and provide a method for calibrating the magnitude of $\gamma_t$ and reasoning about its direction.




\subsection{Bounds on the Causal Effects in Gaussian Copula Models}
\label{sec:generalizing}

Although the causal effects given any set of values $\gamma_t$ can be inferred from Algorithm \ref{algm:gaussian}, when the observed outcome distribution is non-Gaussian, we cannot necessarily express bounds on the $\text{PATE}_{t_1, t_2}$ analytically. In fact, there is not even a guarantee that the intervention density $f(y\mid do(T))$ has finite mean without additional assumptions about the outcome density.  
For quantile estimands, on the other hand, we prove that the effects are bounded. When the marginal distribution of latent confounders is assumed to be symmetric, we state these tight analytic bounds for the median treatment effect.

\begin{thm}
\label{prop:non_gaussian_bound} 
Assume model  \ref{eqn:u_mid_t} - \ref{eqn:y_ytilde_relationship_gauss} and that $\sigma_{y|t}$, $\Sigma_{u|t}$, and $\gamma_t$ can vary with $t$ and assume $\mu_{u|t_1}$ and $\mu_{u|t_2}$ are in the row space of $\Sigma_{u|t_1}$ and $\Sigma_{u|t_2}$ respectively, and $Y$ is continuous.  Then the omitted variable bias for all quantile treatment effects are bounded. If $U_i$ is marginally symmetric for all $i=1, ..., m$ then the median treatment effect, $MTE_{t_1, t_2} = \text{med}\big(Y \mid do(t_1)\big) - \text{med}\big(Y \mid do(t_2)\big)$  is in the interval $m_l \leq MTE_{t_1, t_2} \leq m_u$ where
\begin{align}
m_l &= F^{-1}_{Y|T=t_1}(\Phi(-\parallel(\Sigma_{u|t_1}^{\dagger})^{1/2}\mu_{u|t_1}\parallel_2)) - F^{-1}_{Y|T=t_2}(\Phi(\parallel(\Sigma_{u|t_2}^{\dagger})^{1/2}\mu_{u|t_2}\parallel_2))\\
m_u &= F^{-1}_{Y|T=t_1}(\Phi(\parallel(\Sigma_{u|t_1}^{\dagger})^{1/2}\mu_{u|t_1}\parallel_2)) - F^{-1}_{Y|T=t_2}(\Phi(-\parallel(\Sigma_{u|t_2}^{\dagger})^{1/2}\mu_{u|t_2}\parallel_2))
\end{align}
\noindent where $\Sigma_{u\mid t_1}^{\dagger}$ and $\Sigma_{u\mid t_2}^{\dagger}$ denote the pseudo-inverses of $\Sigma_{u \mid t_1}$ and $\Sigma_{u \mid t_2}$ respectively and $m_l$ and $m_u$ are identifiable under Assumptions \ref{asm:latent_identification} and \ref{asm:copula}.\\
\noindent Proof: See Appendix \ref{sec:prop_non_gaussian_bound}.

\end{thm}

\noindent When $Y$ is conditionally Gaussian, i.e., $F^{-1}_{Y|T=t}$ is the inverse-CDF of a Gaussian random variable for all $t$, then the mean and median are the same so that $MTE_{t_1, t_2} = \text{PATE}_{t_1, t_2}$ and thus we can use the result from Theorem  \ref{prop:non_gaussian_bound} to bound the bias of the PATE.

\begin{cor}
\label{thm:general_gaus}
Assume the model in Equations \ref{eqn:u_mid_t} - \ref{eqn:y_ytilde_relationship_gauss} where $Y$ is conditionally Gaussian given treatments, and where $\sigma_{y|t}$, $\Sigma_{u|t}$, and $\gamma_t$ can vary with $t$. If $\Sigma_{u|t_1}$ and $\Sigma_{u|t_2}$ are non-invertible, then $\text{Bias}_{t_1, t_2}$ is bounded if and only if $\mu_{u|t_1}$ and $\mu_{u|t_2}$ are in the row space of $\Sigma_{u|t_1}$ and $\Sigma_{u|t_2}$ respectively.  When bounded, 
\begin{equation}
\label{eqn:gaus_general}
\text{Bias}_{t_1, t_2}^2 \leq 
    \bigg(
     \sigma_{y|t_1}\sqrt{R_{ Y \sim U|t_1}^2} \| (\Sigma_{u|t_1}^{\dagger})^{1/2}\mu_{u|t_1}\|_2 + 
     \sigma_{y|t_2}\sqrt{R_{ Y \sim U|t_2}^2} \| (\Sigma_{u|t_2}^{\dagger})^{1/2}\mu_{u|t_2}\|_2 
    \bigg)^2,
\end{equation}
with equality when $\gamma_{t_1} \propto \Sigma_{u\mid t_1}^{\dagger}\mu_{u \mid t_1}$ and $\gamma_{t_2} \propto \Sigma_{u\mid t_2}^{\dagger}\mu_{u \mid t_1}$ and where $\Sigma_{u\mid t_1}^{\dagger}$ and $\Sigma_{u\mid t_2}^{\dagger}$ are the pseudo-inverses of $\Sigma_{u \mid t_1}$ and $\Sigma_{u \mid t_2}$.  If $\Sigma_{u\mid t_1} = \Sigma_{u\mid t_2} = \Sigma_{u|t}$ and $\gamma_t = \gamma$ is invariant to $t$ (i.e. there are no treatment-confounder interactions), and $\sigma^2_{y\mid t_1} = \sigma^2_{y \mid t_2} = \sigma^2_{y|t}$ (homoskedastic outcome model) then $\text{Bias}_{t_1, t_2}$ is bounded if and only if $\mu_{u|t_1} - \mu_{u|t_2}$ is in the row space of $\Sigma_{u|t}$ and when bounded, 
\begin{equation}
\label{eqn:gaus_homoskedastic}
  \text{Bias}_{t_1, t_2}^2 \leq \sigma_{y \mid t}^2 R_{ Y \sim U | T}^2\|(\Sigma_{u|t}^{\dagger})^{1/2}(\mu_{u|t_1} - \mu_{u|t_2}) \|_2^2.
\end{equation}
Proof: See Appendix \ref{sec:proof_thm_general_gaus}.
\end{cor}

\noindent As expected, Equation \ref{eqn:gaus_homoskedastic} takes the same form as Equation \ref{eq:bound,multi-T}, but generalizes it in two ways: first, we do not require that $\Sigma_{u\mid t}$ has the form in Equation \ref{eqn:sigma_u_t} and is only required to be non-negative definite; and second, $\mu_{u\mid t}$ can be nonlinear in $t$ (i.e. it does not need to follow Equation \ref{eqn:mu_u_t}).  As in Section \ref{sec:gaussian}, when bounded, the bias is proportional to the norm of the scaled difference in confounder means in the two treatment arms.  When there exists an m-vector, $q$, such that $\text{Var}(q^\top U \mid T=t) = 0$, then $\Sigma_{u\mid t}$ is non-invertible because there exists a projection of the confounders that is point identified.  Corollary \ref{thm:general_gaus} says that in this case, the ignorance region for the PATE is bounded if and only if $q^\top (\mu_{u| t_1} - \mu_{u| t_2}) = 0$.  In words, if a projection of the confounders can be identified, then the confounding bias is bounded if and only if the identifiable projection of the confounders has the same value in both treatment arms.
This corresponds to observations in \citet{damour2019aistats}  about violations of the positivity assumption (Assumption \ref{asm:positivity}) when confounders are ``pinpointed'' by the latent variable model.
In particular, if the ``pinpointed'' confounders do not match between $t_1$ and $t_2$, this implies that positivity has been violated.


Finally, we note that for binary outcomes, we focus primarily on the risk ratio as the estimands of interest. Interestingly, unlike the ATE and quantile treatment effects, $RR_{t, \sbullet}$ and $RR_{t_1, t_2}$ are non-monotone in the magnitude of $\gamma$.  We discuss this in more detail in Appendix \ref{sec:binary} and provide simulation results with binary outcomes in Section \ref{sec:simulation}. For simplicity, for the remainder of the paper, we focus on settings in which $\gamma_t = \gamma$ does not vary with the level of treatment.  This corresponds to a model in which there are no treatment-confounder interactions in the outcome model.

\section{Calibration and Robustness}
\label{sec:cali&robust}

Sensitivity analyses consist of two parts: first, the sensitivity model itself, which specifies a set of data-compatible causal models, indexed by sensitivity parameters; and secondly, exploratory tools for mapping external assumptions to particular causal models in this set.
We now turn to discussing the latter in the context of our proposed model.

In the sensitivity analysis literature so far, two exploratory techniques have been particularly popular in single treatment studies: \emph{calibration}, which maps sensitivity parameter values to interpretable observable or hypothetical quantities; and \emph{robustness analysis}, which characterizes the ``strength'' of confounding necessary to change the conclusion of a study.
Here, we show how to adapt these techniques to our sensitivity model in the multi-treatment setting.
In addition, we introduce a third class of tools that are particularly well-suited to the multi-treatment setting, which we call \emph{multiple contrast criteria} (MCCs).
MCCs specify aggregate properties of the treatment effects for multiple treatment contrasts that are implied by a single causal model, e.g., the L2 norm of PATEs corresponding to contrasts in each individual treatment variable in $T$. 
In many multi-treatment settings, assumptions are often expressed in terms of the aggregates---e.g., in genomics, the idea that the effect of most single nucleotide polymorphisms is small---and we show here how these can be used in conjunction with our sensitivity model to characterize candidate causal models that may be of interest in an application.

\subsection{Calibration for a Single Contrast}
\label{sec:calibration_single}

We begin by describing calibration for  $\gamma$ in our sensitivity model when the focus is on a single treatment contrast, between levels $T = t_1$ and $T = t_2$.
The goal is to develop heuristics for specifying ``reasonable'' values or ranges for $\gamma$, e.g., to derive bounds on treatment effects by specifying bounds on the strength or direction of confounding.
Following previous work in the single treatment setting, we outline how to calibrate our sensitivity parameter vector $\gamma$ in terms of a fraction of outcome variance explained by the unobserved confounder.
Recall that $\gamma$ is a vector that parameterizes the residual correlation between the $m$-dimensional unobserved confounder $U$ and the outcome $Y$ after conditioning on the treatment vector $T$.

First, we briefly review calibration in single-treatment settings.
In latent variable approaches for single treatment sensitivity analysis, the causal effect is identified given two sensitivity parameters: the fraction of outcome variance explained by unobserved confounders, $R_{Y \sim U \mid T }^{2}$, and the fraction of treatment variance explained by unobserved confounders,  $R_{T \sim U}^{2}$ \citep{cinelli2020making}. In a linear model, these two scalar quantities identify the confounding bias (Equation \ref{eqn:bias_single}).  Neither R-squared value is identifiable and thus many authors have proposed strategies for drawing analogies between these values and other observable or hypothetical quantities \citep{cinelli2020sensemakr, veitch2020sense, franks2019flexible}.   

We borrow this strategy for calibration in our setting, with some modifications.
First, in our setting there is no need to calibrate $R^2_{T \sim U}$, because we have restricted ourselves to a setting in which this is implicitly identified (Assumption \ref{asm:latent_identification})
This leaves calibration of the outcome-confounder relationship, which in our setting is more complex because it is parameterized by a vector $\gamma$\footnote{Unlike the single treatment setting, the confounder-outcome relationship cannot be sufficiently summarized in terms of a scalar $R_{Y \sim U \mid T}^{2}$. Each confounder can impact each treatment in different ways.
}.
However, we can reparameterize $\gamma$ in terms of a direction $d$ and an R-squared for interpretable calibration:
\begin{equation} \label{eqn:reparam_rho}
 \gamma = \sigma_{y\mid t}\sqrt{R_{Y \sim U | T}^2} \Sigma^{-1/2}_{u|t} d,
\end{equation}
where $d \in \mathbb{S}^{m-1}$ is an $m$-dimensional unit vector on the $(m-1)$-sphere.
We discuss strategies for calibrating both the magnitude and direction separately.\\


\noindent \textbf{Calibrating the magnitude of $\gamma$}. For Gaussian outcomes, the magnitude of $\gamma$ is characterized entirely by $R^2_{ Y \sim U | T}$, the partial fraction of outcome variance explained by $U$ given $T$. When  $R_{ Y \sim U|T}^2=0$ there is no unobserved confounding, and when $R_{ Y \sim U|T}^2=1$ all the observed residual variance in $Y$ is due to confounding factors.  In order to calibrate this magnitude, we adopt an idea proposed by \citet{cinelli2020making} for causal inference with single treatments. 

First, we consider calibration when observed covariates, $X$, are also available and consider the importance of an unmeasured confounder $U$ relative to a measured confounder (or set of confounders), $X_j$, given all other confounders $X_{-j}$.  Specifically, assume that we believe that $R^2_{Y \sim U |X_{-j}, T} \leq \kappa R^2_{Y \sim X_j |X_{-j}, T}$, where $\kappa$ is a user chosen parameter reflecting an upper bound on how much ``stronger'' $U$ might be than $X_j$. Then \citet{cinelli2020making} show that this implies 

\begin{equation}
\label{eqn:cali_benchmark}
R^2_{Y \sim U | X, T} \leq \kappa \frac{ R^2_{Y \sim X_j |X_{-j}, T}}{1- R^2_{Y \sim X_j |X_{-j}, T}.}
\end{equation}  

\noindent We use the right hand side of Equation \ref{eqn:cali_benchmark}, which is estimable given any choice of $\kappa$, to benchmark the fraction of outcome variance explained by unmeasured confounders given observed confounders and treatments.

When there are no measured confounders, we can still use the same strategy as above, by leveraging the presence of multiple treatments to calibrate $R^2_{Y\sim U |T}$.  For example, in the context of the example to come in Section \ref{sec:mouse}, where treatments are gene expression levels, we might posit that unmeasured confounders cannot explain more variation in the outcome than a set of genes $T_j$, given all other genes $T_{-j}$.  We can compute this quantity, the fraction of variation in $Y$ that can be explained by a specific treatment (or set of treatments), $T_j$,  after controlling for all other treatments $T_{-j}$ as
\begin{equation} \label{eqn:partial_R2_tildeY}
    R_{Y \sim T_j \mid T_{-j}}^2 := \frac{R_{Y \sim T}^2 - R_{Y \sim T_{-j}}^2}{1- R_{Y \sim T_{-j}}^2}.
\end{equation}
\sloppy
\noindent and then, analogously to Equation \ref{eqn:cali_benchmark}, can make the assumption that $R^2_{Y \sim U|T_{-j}} \allowdisplaybreaks \leq \allowdisplaybreaks \kappa R^2_{Y \sim T_j|T_{-j}}$.  As before, this implies the benchmark $R^2_{Y \sim U|T} \leq \kappa \frac{R^2_{Y \sim T_j|T_{-j}}}{1-R^2_{Y \sim T_j|T_{-j}}}$.

When the observed outcome is non-Gaussian, we calibrate the ``implicit $R^2$'', by considering the explained variance of the transformed outcome in the standard normal space, denoted by $Z_Y$ in Equation \ref{eqn:ytilde_x,u_gauss}.   The implicit $R^2$ of $T$ for model in Equations \ref{eqn:u_mid_t} - \ref{eqn:y_ytilde_relationship_gauss}
is defined as 
$R_{Z_Y \sim T}^2 = \frac{{\text{Var}}(E[Z_Y | T])}{{\text{Var}}(E[Z_Y | T]) + 1}$,
and the implicit partial R-squared of treatment $T_j$, $R_{Z_Y \sim T_j \mid T_{-j}}^2$, is defined analogously to Equation \ref{eqn:partial_R2_tildeY}.  As before, these estimable partial R-squared values can be used to provide a useful comparison for the partial R-squared of potential unobserved confounders, $R_{Z_Y \sim U \mid T}^2$. For more detail, see \citet{imbens2003sensitivity} and \citet{franks2019flexible} who discuss calibration with implicit R-squared values in  logistic regression models. See \citet{veitch2020sense} and \citet{cinelli2020sensemakr} propose useful graphical summaries for calibration based on these metrics in the single treatment setting. \\

\noindent \textbf{Choosing the direction of $\gamma$}. Given a magnitude, we now propose a default method for identifying the direction of $\gamma$ for a single contrast.  The dot product $d^\top \Sigma^{-1/2}_{u|t}( \mu_{u\mid t_1} - \mu_{u\mid t_2})$ corresponds to the projection of the scaled difference in confounder means onto the outcome space.  By default, we suggest using the direction which maximizes the squared bias.  As shown in Corrolary \ref{thm:general_gaus}, when $d$ is colinear with $\Sigma^{-1/2}_{u|t}( \mu_{u\mid t_1} - \mu_{u\mid t_2})$,  the confounding bias of the naive estimator for Gaussian outcomes is maximized at
\begin{equation}
\label{eqn:bias_sr}
|\text{Bias}_{t_1,t_2}| = \sigma_{y|t} \sqrt{R_{Y \sim U | T}^2}   \| \Sigma_{u|t}^{-1/2}(\mu_{u|t_1} - \mu_{u|t_2}) \|_2,
\end{equation}
Choosing the direction of the sensitivity vector in this way provides conservative bounds for each contrast of interest. For non-Gaussian outcomes or alternative estimands, there may not be an analytic solution to the direction which maximizes the bias, but  we can still compute the direction via numerical optimization. 

\subsection{Robustness for Individual Contrasts}
\label{sec:robustness}
We now turn to assessing the robustness of conclusions using our sensitivity model, extending work by \citet{cinelli2020making} and \citet{evalue} in the single treatment setting.
Specifically, we propose an extension of the robustness value (RV) within our model, which characterizes the minimum strength of confounding needed to change the sign of the treatment effect.
As in the previous section, the extension is most straightforward when considering the effect of a single treatment contrast, between levels $T = t_1$ and $T = t_2$.

To review briefly, in single treatment settings, Cinelli and Hazlett define the robustness value as the smallest value of $\text{max}(R^2_{Y\sim U \mid T}, R^2_{T \sim U})$, needed to change the sign of the effect. A robustness value close to one means the treatment effect maintains the same sign even if nearly all the observed residual variance in the outcome is due to confounding \emph{and} all the residual treatment variance is due to confounding. On the other hand, a robustness value close to zero means that even weak confounding would change the sign of the point estimate. In the multi-treatment setting, we can more precisely characterize the robustness of causal effects, subject to Assumptions~\ref{asm:latent_identification} and \ref{asm:copula}.

In single treatment analyses, the smallest value of $\text{max}(R^2_{Y\sim U \mid T}, R^2_{T \sim U})$ needed to change the sign of the treatment effect is achieved when $R^2_{Y\sim U \mid T} = R^2_{T \sim U}$.  As such, the value of the single treatment robustness value can be misleading when $R^2_{Y\sim U \mid T}$ is very different from $R^2_{T \sim U}$.  In detail, when $R_{Y \sim U \mid T}^{2} > R_{T \sim U }^{2}$, the single-treatment RV will be too conservative. Conversely, when $R_{Y \sim U \mid T}^{2} <  R_{T \sim U}^{2}$ the single-treatment RV will overestimate the robustness of the effect.  In the multiple treatment setting, Assumptions \ref{asm:latent_identification} and \ref{asm:copula} imply that for any treatment, the fraction of treatment variance due to confounding  is identifiable, which allows us to define the multi-treatment RV as the minimum value of $R_{Y \sim U \mid T}^{2}$ needed to explain away the treatment effect of interest, assuming the direction of the sensitivity vector is chosen to maximize the bias.  This allows us to more precisely characterize robustness.

When the observed outcomes are Gaussian, the robustness value can be computed in closed form.

\begin{cor}
Assume the model in Equations \ref{eqn:u_mid_t} - \ref{eqn:y_ytilde_relationship_gauss} where $Y$ is conditionally Gaussian given treatments.  Further, assume a homoskedastic outcome with no interaction between unmeasured confounders and treatments, so that $\sigma_{y|t}$, $\Sigma_{u|t}$, and $\gamma$ are invariant to the level of $t$.  Then,

\begin{equation} 
\label{eqn:rv_gaussian}
    \text{RV}_{t_1, t_2} = \text{min}\left(\frac{(\mu_{y|t_1} - \mu_{y|t_2})^2 / \sigma_{y|t}^2}{ \| \Sigma_{u|t}^{\dagger})^{1/2}(\mu_{u|t_1} - \mu_{u|t_2})\|_2^2}, \ 1\right).
\end{equation}
Proof: Immediate from Corollary \ref{thm:general_gaus}, by setting $\text{Bias}_{t_1, t_2}$ equal to the observed difference in outcomes, $\mu_{y\mid t_1} - \mu_{y\mid t_2}$.
\end{cor}

\noindent Note that because the bias is bounded, it is possible that for some treatment effects, no matter how much variance in the outcome is due to unmeasured confounding, the sign of the effect will remain the same. In this case, by convention we say anything with an RV of 1 is ``robust''. The numerator of the robustness value corresponds to the squared difference in mean outcomes under each treatment arm, in units of residual standard deviations.  Likewise, the denominator is the squared difference in confounder means in each treatment arm, in units of residual standard deviations. When the (scaled) difference in mean outcomes is larger than the (scaled) difference in unmeasured confounders, the causal effect is robust.

RV metrics for alternative estimands and/or non-Gaussian data can still be computed using the same principle.  For example, when the observed outcome is binary, the RV can be computed numerically by solving $RR_{t_1, t_2} = 1$,  which corresponds to the minimum strength of confounding needed for the observed risk ratio (RR) to equal to one.  We can view this robustness value as a multi-treatment parametric analog of the ``E-value'' proposed by \citet{evalue}.

In our setting, we can also make stronger statements about robustness than in the single treatment setting: under the latent variable model, it is possible to declare an effect robust to \emph{any} level of confounding.
In particular, when the latent variable model implies $R_{T \sim U}^{2}<1$ (i.e., we have confounder overlap), then even when $R_{Y \sim U \mid T}^{2}=1$, the ignorance region is bounded (Corollary~\ref{thm:general_gaus}).
When this ignorance region excludes zero, we declare the effect ``robust''.
This operation is consistent with the result in \citet{miao2020identifying}, showing that hypotheses of zero effect can be tested in this setting, even if the treatment effect cannot be identified.



\subsection{Multiple Contrast Criteria}


\label{sec:joint_cali}

So far, we have examined the sensitivity of causal conclusions by exploring the marginal bounds on a treatment contrast in isolation. However, the multi-treatment setting presents opportunities for exploring sensitivity models in new ways.  Here we characterize a choice of sensitivity vector $\gamma$ by concurrently considering its implications for the causal effects of multiple treatment contrasts. Thus, while the sensitivity vector $\gamma$ that gives the worst-case bias may differ across individual contrasts, here we explore criteria for selecting a single $\gamma$ which concurrently incorporates implications for multiple treatment contrasts in aggregate.  We term these ``multiple contrast criteria'' or MCCs.

Formally, for a set of treatment contrasts $\mathcal T^2 = \{(t_{1}, t_{2})_k\}_{k=1}^K$, and a candidate sensitivity vector $\gamma$, let $\textbf{PATE}_{\mathcal T^2}(\gamma)$ be the vector of PATEs implied by the causal model indexed by $\gamma$.
An MCC is a scalar summary of this treatment effect vector, which we write as $\omega(\textbf{PATE}_{\mathcal T^2}(\gamma))$.
An MCC is specified by the set of contrasts $\mathcal T^2$ and the summary function $\omega$, both of which can be chosen to meet the needs of a given analysis.

MCCs can be used in many ways, but here we consider how they can be used to search for the causal model that yields the minimum norm treatment effect vector, subject to Assumptions \ref{asm:unconfoundedness}-\ref{asm:copula} and a confounding limit $\mathcal{R}^2$.
Specifically, we take $\omega$ to be an $L_p$ norm for some $p$, and consider sensitivity vectors $\gamma_*$ that satisfy: 
\begin{align}
\label{eqn:multi_robust}
    \gamma_{*} = \underset{\gamma}{\text{argmin }}
    \omega(\textbf{PATE}_{\mathcal T^2}(\gamma))
    \text{ subject to } R^2_{Y \sim U \mid T}(\gamma) \leq \mathcal{R}^2
\end{align}
\noindent
where $R^2_{Y \sim U \mid T, X}(\gamma) = \frac{\gamma^\top \Sigma_{u \mid t}\gamma}{\sigma^2_{y\mid t}}$ is the partial fraction of outcome variance explained by confounding for sensitivity vector $\gamma$.
Causal models selected in this way are often highly interpretable, in terms of either ``worst case'' effect sizes or established prior knowledge.
For example, we can choose $\omega$ to be the $L_\infty$ norm, so that $\gamma_*$ is the sensitivity vector that minimizes the maximum absolute treatment effect across contrasts.
Alternatively, we could choose $\omega$ to be the $L_1$ or $L_2$ norm of the treatment effects to incorporate prior knowledge that might imply small ``typical'' effect sizes.
We demonstrate how this minimization approach can be used to express prior knowledge about small effects in simulated data in Section \ref{sec:sim_nongaussian_t}, and how it can be used to evaluate robustness on a real data set in Section \ref{sec:mouse}.

\section{Simulation Studies}
\label{sec:simulation}

In this section, we demonstrate our sensitivity analysis workflow in several numerical simulations.
The goal of these simulations is twofold: first, to demonstrate some of the operating characteristics of the approach in settings that are more realistic than the linear Gaussian settings we characterized analytically; and secondly, to show how exploratory tools like calibration, robustness analysis, and MCCs can be used to draw conclusions and choose interesting candidate models.

We consider two broad simulation settings.
In the first setting, we construct simulations with non-linear responses to treatment to show how the ignorance regions returned by our method can vary in different scenarios.  
In the second setting, we construct a simulation that mimics the structure of a Genome Wide Association Study (GWAS).
Here, we examine the behavior of our method when a popular approximate latent variable method---the Variational Auto Encoder (VAE)---is used to estimate the effects of latent confounders, and demonstrate how MCCs can be useful tools for using prior information to choose potentially useful causal models from the set that is compatible with the observed data.
In both subsections, we simulate data from the following generating process:
\begin{align}
    U & := \epsilon_u, \quad \epsilon_u \sim N(0, I),\label{eq:sim_first}\\
 T &:= h_T(BU + \epsilon_{t}), \quad \epsilon_{t} \sim N(0, \sigma_{t}^2 I)\\
Y &:= h_{Y|T}(g(T) + \gamma^\top  U + \epsilon_{y}), \quad \epsilon_{y} \sim N(0, \sigma^2) \label{eq:sim_last}
\end{align}
\noindent The functions $h_{Y \mid T}$ and $h_T$ are chosen according to be either the identity for Gaussian data, or an indicator function for binary data.

\subsection{Example with Non-Linear Response Functions}
\label{sec:sim_nongaussian}

We start by exploring variation in the size of ignorance regions for different contrasts in a simple simulated example with four treatments where the outcome is a nonlinear function of these treatments.  We consider two cases: first, a case where $Y$ is Gaussian with $h_{Y|T}(Z_Y) = Z_Y$; and secondly, a case where $Y$ is binary with $h_{Y|T}(Z_Y) = I_{Z_Y > 0}$. We aim to estimate the $\text{PATE}_{e_i, 0}$ for Gaussian outcome and $\text{RR}_{e_i, 0}$ for binary outcome, where $e_i$ denotes the $i$th canonical vector, i.e. the vector with a 1 in the $i$-th coordinate and 0's elsewhere.  

In both examples, we generate the data with a 1-dimensional latent confounder ($m=1$), $k=4$ treatments, $B = [2, 0.5, -0.4, 0.2]$, $\sigma_{t}^2 = 1$, $\gamma = 2.8$, $\sigma^2 = 1$, $h_T(Z_T) = Z_T$ and
$$g(T) = 3T_1 - T_2 + T_3 I_{T_3 > 0} + 0.7 T_3 I_{T_3 \leq 0} - 0.06T_4 - 4T_1^2.$$
Based on the choice of $g$, contrasts along the $j$th dimension of $T$ have effects of widely varying magnitude. Based on our choice for $B$, the worst-case confounding bias also varies significantly across contrasts.   For example, the effect of confounding is larger when estimating the treatment of $T_1$, since the first entry of $B$ has the largest magnitude, meaning $T_1$ is the feature most correlated with $U$. In order to demonstrate this in simulation, we first apply probabilistic PCA (PPCA) to estimate the distribution $f(u \mid t)$, and then model  $f(y \mid t)$ using Bayesian Additive Regression Tree (BART) with R package BART \citep{bart}.

\begin{figure}
    \begin{subfigure}[t]{0.49\textwidth}
        \centering
        \includegraphics[scale=0.49]{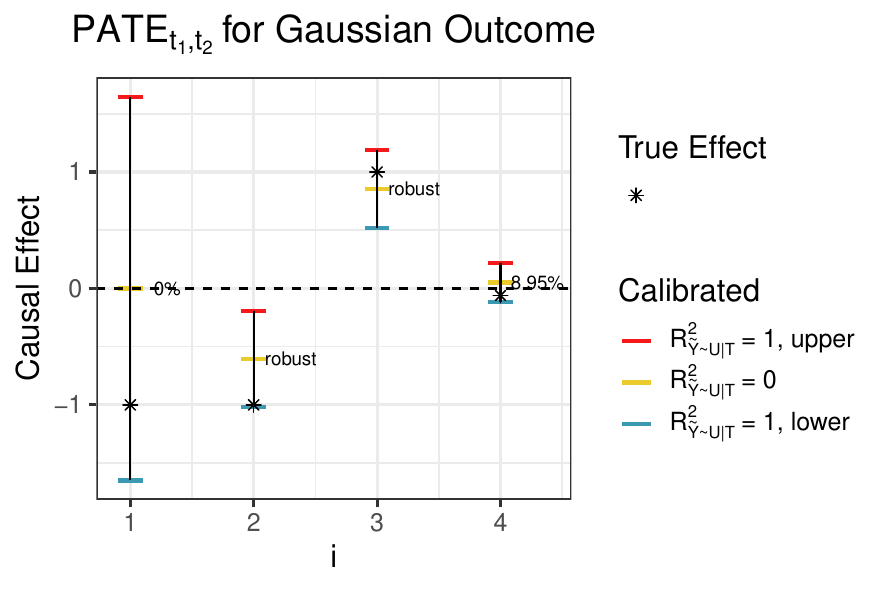}
        \caption{\label{fig:nonlinear_yt_gaussianY} Gaussian Outcome}
	\end{subfigure}
	\begin{subfigure}[t]{0.49\textwidth}
		\centering
		\includegraphics[scale=0.5]{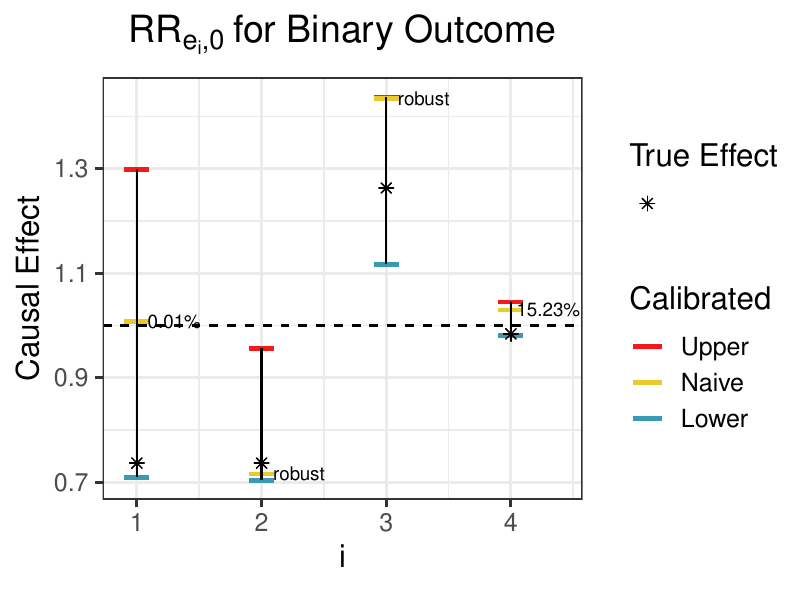}
		\caption{\label{fig:nonlinear_yt_binaryY} Binary Outcome} 
	\end{subfigure}
	\caption{\label{fig:nonlinear_yt} Estimated ignorance region for $e_i$ in case when $Z_Y$ is nonlinear in $T$.
    (a) $R_{Z_Y \sim U|T}^2 = 0$ denotes the treatment effects estimated based on the observed data only, i.e., under the assumption of no confoundedness.
    $R_{Z_Y \sim U|T}^2 = 1$ and $R_{Z_Y \sim U|T}^2= 1$ correspond to the case when all residual variation in $Y$ is due to the confounding, respectively denoting the upper and lower bounds of the ignorance region. 
    (b) In the binary setting, even though the estimand is a non-linear function of the latent Gaussian outcome, the width of the ignorance region and general robustness pattern is largely consistent with the implications of Corollary \ref{thm:anyt}.}
\end{figure}

For Gaussian outcomes, the width of the ignorance regions are larger for the treatments most correlated with confounders as characterized in Corollary \ref{thm:general_gaus} (see Figure \ref{fig:nonlinear_yt}). Since $B$ is a vector, the width of the ignorance region of $\text{PATE}_{t_1, t_2}$ can be examined by looking at the dot product between $B$ and the treatment contrasts. The larger the dot product, the wider the ignorance region. As expected, the ignorance region of the treatment effect is widest when $t_1 = e_1$ (RV $\approx 0\%$) and narrowest when $t_1 = e_4$, since $B^\top  e_1$ has the largest magnitude while $B^\top  e_4$ has the smallest.  Despite the fact that $t_1 = e_4$ has the smallest ignorance region, it is not robust to confounding because the naive effect is already close to zero (RV = $9 \%$). For the second and third treatment contrasts, estimates are robust to confounders, as their entire ignorance regions exclude 0.  These results require the Gaussian copula assumption (Assumption \ref{asm:copula}), but in Appendix D, we show via simulation that alternative choices for the copula yield results that lie within the worst-case Gaussian bounds for $R^2_{Y\sim U \mid T} = 1$.  In Appendix Figure \ref{fig:cop_misspec}, we include the causal effects implied by some Archimedean copulas as well as an example with a non-monotone copula (e.g. quadratic relationship between $U$ and $Y$).  Thus, while the Gaussian copula will not hold exactly in practice, it is likely a that the Gaussian bounds cover the true causal effect when the true copula is non-Gaussian. 

For the simulation with binary outcomes, we compute ignorance regions for the risk ratio.  Although we do not have a theoretical result about the ignorance regions of the risk ratio, the general trends in the size of the ignorance region and the robustness of effects are  comparable to the Gaussian.  Most notably, the treatments with the largest ignorance regions are still those that are most correlated with the confounder. On the other hand, because the outcome is non-linear in $U$, the naive estimate is not at the center of the ignorance region (Figure \ref{fig:nonlinear_yt_binaryY}). In fact, the ignorance region is also non-monotone in $R^2_{Z_Y\sim U |T}$ because the variance of the intervention distribution also depends on $\gamma$.  In this case, one of the endpoints of the ignorance region corresponds to $R^2_{Z_Y \sim  U |T} = 1$ but the other does not. We compute the endpoints of the ignorance region numerically (see Appendix \ref{sec:binary} for more details).  

\subsection{Example with Simulated Genome Wide Association Study}
\label{sec:sim_nongaussian_t}

We now explore a slightly more complex setting motivated by applications in biology, particularly in genome wide association studies (GWAS).
GWAS investigate the association between hundreds or thousands of genetic features (i.e., single nucleotide polymorphisms, or SNPs) and observable traits (i.e., phenotypes), such as disease status.
Despite having ``association'' in the name,
measures of association in GWAS are often adjusted to afford a causal interpretation in which conclusions speak to how a phenotype would change if the genome were intervened upon.
For example, most analyses adjust for ``population structure'', which correspond to broad genetic patterns induced by population dynamics that are often confounded with geography, ancestry, environment, and other lifestyle factors \citep{price2006principal,song2015testing}. 
\citet{wang2018blessings} cite this literature as motivation for their work.

Here, we construct a simulated GWAS to demonstrate two properties of our sensitivity analysis method.
First, we show that flexible latent variable models can be plugged into our sensitivity model.
Secondly, we demonstrate how minimizing multiple contrast criteria (MCC) can be used to select interesting candidate models that conform to broad hypotheses about the nature of genetic effects.

In this simulation, we generate data with high-dimensional binary treatments (SNPs), and set the true causal effects to be mostly small, with a small fraction of treatments having effects of larger magnitudes.
The simulation is then designed so that unobserved confounding biases estimates for each of these treatment effects, obscuring the difference between large and small effects.
To generate data, we follow the template in Equations~\ref{eq:sim_first}--\ref{eq:sim_last}.
We generate data with $m=3$ latent confounders and $k=500$ treatments, $T \in \{0,1\}^k$, where $T_j = 1$ if the the $j$th site shows a deviation from the baseline sequence (i.e., the presence of at least one minor allele).
We set the response function $g(T) = \tau^\top  T$ to be linear in the treatments (a common assumption in GWAS), and set the outcome $Y$ to be Gaussian by setting $h_Y(Z_Y) = Z_Y$.
We focus on estimating 
\begin{equation}
\frac{1}{n} \sum_{i = 1}^n PATE_{t_i^{j}, t_i^{-j}} \ \text{for all} \ j = 1, \cdots, k, 
\end{equation}
where $t_i^j$ and $t_i^{-j}$ correspond to the $i^{th}$ observed treatment vector with the $j^{th}$ SNP set to be 1 and 0 respectively.  Note that since $g(T)$ is linear in $T$, $\frac{1}{n} \sum_{i = 1}^n PATE_{t_i^{j}, t_i^{-j}} = \tau_j$, the $j^{th}$ element of $\tau$.  
We generate $\tau$ from a two component mixture with 90\% of the coefficients from a $\text{Uniform}(-0.1,\ 0.1)$ (small effects) and 10\% from a $\text{Uniform}(-2,\ 2)$ (large effects).
We assume that there are $m=3$ latent confounders.

\begin{figure}[htb!]
	\begin{subfigure}[t]{0.71\textwidth}
        \centering
        \includegraphics[width=\textwidth]{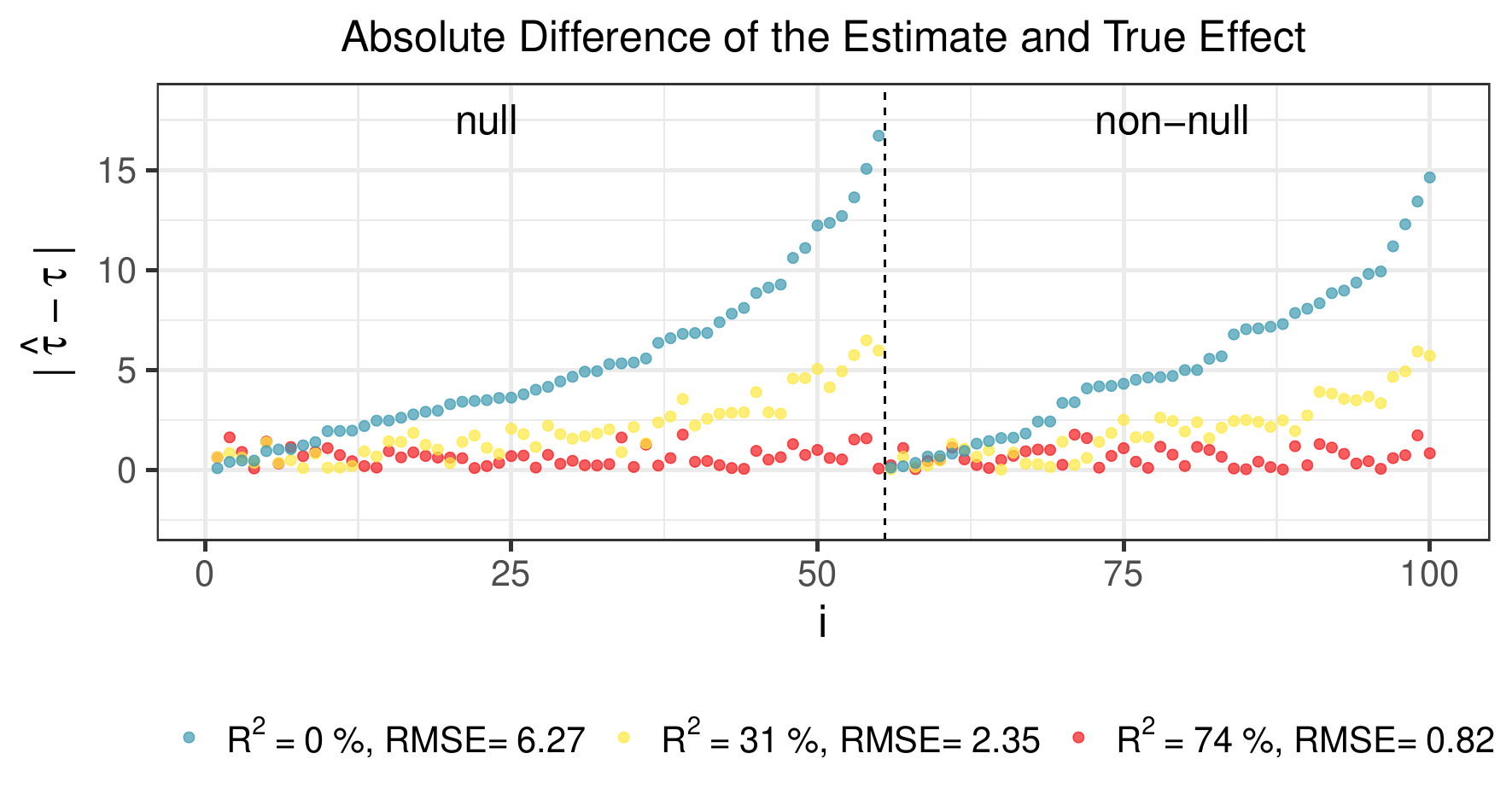}
         \caption{\label{fig:multi_robust_scatter} L1 minimizing effects}
	\end{subfigure}
    \begin{subfigure}[t]{0.28\textwidth}
        \centering
        \includegraphics[width=\textwidth]{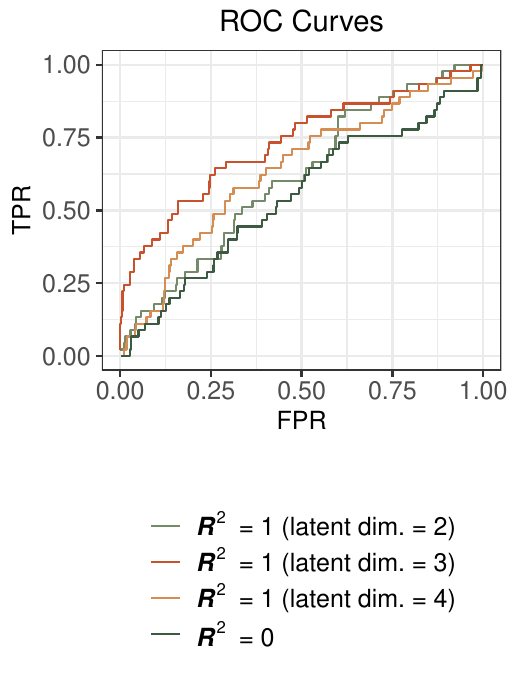}
        \caption{\label{fig:roc} Non-null classification}
	\end{subfigure}
		\caption{ Causal inference with 500 binary treatments with $k=3$ latent confounders. The omitted confounder bias of the naive estimates are large for both the null and non-null effects due to unmeasured confounding. 
    (a) The absolute difference between the true effects and the inferred minimum L1-norm treatment effects shown for fifty randomly chosen small effects (``null'' contrasts) and all large effects (``non-null'' contrasts) for three different limits on the magnitude of confounding, $\mathcal{R}^2 \in \{0, 0.3, 1.0\}$. When $\mathcal{R}^2=1$, the overall L1 minimizer of the treatment effects is achieved for the sensitivity vector which explains $R^2_{Y \sim U|T} = 74\%$ of the residual outcome variance.
    (b) We construct a simple non-null classifier from minimum L1 treatment effects with $\mathcal{R}^2 = 1$ and naive effects ($\mathcal{R}^2 = 0$).  The blue curve represents the ROC curves from the naive estimates and the green, yellow and red curves represents the L1 minimizer of the treatment effect estimates for inferred confounder models with dimensions $\hat k \in \{2, 3, 4\}$.
     The area under the curve (AUC) for the naive estimates is 0.54, whereas the AUC for the L1-minimized estimates are 0.61 ($\hat k=2$), 0.73 ($\hat k  =3$) and 0.64 ($\hat k=4$).}
\end{figure}

We consider a model for the observed data with two components, paying special attention to the latent confounder model.
In particular, we model the conditional distribution of confounders given treatment $f(u \mid t)$ using a variational autoencoder (VAE), which is a popular, flexible neural network--based approximate latent variable model.
This model is particularly appropriate because it yields an approximate Gaussian conditional distribution $f(u \mid t)$, even for discrete $T$ as we have here. 
(We discuss latent confounder inference with VAEs in more detail in Appendix \ref{sec:lv_inf}.)
We fit the observed outcome model $f(y\mid t)$ using a simple linear regression, ignoring confounding, which corresponds to the setting in which $R^2_{Y \sim U \mid T} =0$.

\paragraph{Worst-Case Ignorance Regions.}
With this simulation setup, we first examine whether the ignorance regions contain the true causal effects.
Importantly, because the VAE is an approximate latent variable model, and we are currently ignoring estimation uncertainty, it is not immediate that the ignorance regions should be valid.
We find that, even using our plug-in approach, the worst case ignorance regions cover 498 out of 500 of the true treatment effects.
In all cases, the worst case bounds communicate substantial fundamental uncertainty about the true treatment effects (See Appendix Figure~\ref{fig:gwas_worstcase}).


\paragraph{Finding Candidate Models with MCCs.}
Investigators often have strong hypotheses about the aggregate properties of SNP treatment effects.
For example, while some phenotypes can be predominantly explained by only a small number of SNPs, other phenotypes may be more plausibly described by the omnigenic hypothesis, which suggests that some observable effects must be explained by the sum of many small effects across many SNPs \citep{boyle2017expanded}. 
Here, we show that some of these aggregate hypotheses can be formalized in terms of MCCs, and in these cases, the MCC minimization procedure from Section~\ref{sec:joint_cali} can be used to find useful candidate causal models that align with these hypothesis while being fully consistent with the observed data.

To motivate candidate model selection, we consider the use case of estimating effect sizes from a single coherent model, under the hypothesis that the median effect size is small.
Specifically, we formalize this hypothesis by defining a MCC $\omega(\textbf{PATE}_{\mathcal T^2}(\gamma))$ to be the $L_1$ norm of the effects of each contrast $\mathcal{T}^2 = \{(t_i^j, t_i^{-j}) : i \in (1, ... , n)\}$ for all treatments $j=1, \cdots, k$.
We then select the model that minimizes this criterion by selecting $\gamma$ subject to different allowed levels of confounding $R^2_{Y\sim U\mid T}$.

In Figure \ref{fig:multi_robust_scatter}, we plot the the resulting coefficients estimates for three values of $\mathcal{R}^2$: $0$ (naive effects), $0.3$ and $1$.
Because the true effects are much smaller in magnitude than the na\"ive effects, the RMSE of the estimates decreases as we increase $R^2_{Y\sim U | T}$, although all effects are equally compatible with the observed data. In this simulation, the L1 norm of naive estimates is approximately 2525 and the norm of the true effects is drastically smaller at approximately 75.  

Models selected using this MCC minimization procedure are also useful for the coarser goal of separating small and large effects. From the naive regression, the coefficients are overdispersed to the true causal effects and the true small coefficients are practically indistinguishable from true large coefficients.  
Meanwhile, models chosen with the MCC minimization procedure provide more useful signal. 
To formalize this, we consider a classifier that separates large and small effects using the magnitude of the inferred coefficients as the classification score.
In Figure \ref{fig:roc} we plot the receiver operating characteristic (ROC) curves for the classifiers based on the naive estimates as well as the overall $L1$ minimizer of the treatment effects ($\mathcal{R}^2 = 1$, i.e. no limit on the value $R^2_{Y \sim U\mid T}$).



Importantly, the difference in conditional confounder means, $\mu_{u \mid t_i^j} - \mu_{u \mid t_i^{-j}}$, varies between non-null and null contrasts.  This leads to a larger reduction in the relative magnitude of the null effects for models chosen through MCC minimization, accentuating the differences between large and small treatment effects (See Appendix Figure \ref{fig:tau_classicifaction}).  For models selected by MCC minimization, the area under the ROC curve (AUC) increases from 0.54 (almost no ability to distinguish small and large treatments) to 0.72 ($\hat k=3$, red curve). The selected model achieves nearly 25\% true positive rate without accruing any false positives.  Naturally, the classifier performance is the best when we fit a latent variable model with the correct number of latent factors, although the classifier based on latent variable models of dimensions $\hat k = 2$ and $\hat k = 4$ still outperform classification from naive effects. In the Discussion, we note how this approach relates to, and complements recent identification results for a similar setting in \citet{miao2020identifying}.

\section{Analysis of Mouse Obesity Data}
\label{sec:mouse}
In this section, we apply our sensitivity analysis to  mice obesity data generated by \citet{wang2006genetic} and \citet{ghazalpour2006integrating}, and compiled into a single data set by \citet{lin2015regularization}. The data consists of body weight and gene expression levels for 17 genes in each of 227 mice, and the goal is to estimate the causal effect of the gene expression levels on mouse weight. In gene expression data sets like this one, batch effects can induce confounding when the batches are correlated with outcomes.  This problem has motivated several approaches for removing sources of potential unwanted variation prior to analysis \citep{gagnon2012using, Listgarten16465, leek2007capturing}.  \citet{miao2020identifying} analyze the mouse obesity data set in the context of the multiple treatment problem, under the assumption that at least half of the true treatments have no causal effect on the outcome.  Here, we provide a complementary analysis, and explore the broader set of causal effects that are compatible with the observed expression data.  

First, we use the linear treatment and outcome model, Equations \ref{eqn:confounder}-\ref{eqn:outcome}, to model the data. To represent the possible relationship between treatments and confounders we fit a linear factor model, which is commonly used to characterize the unmeasured confounding in gene expression studies \citep{gagnon2012using},  using the \texttt{factanal} method.  From the scree plot of the singular values of the gene expression matrix, we find that there are two singular values which exceed the rest, which suggests that an $m=2$ confounder model is a reasonable choice (Appendix Figure \ref{fig:mouse_scree}).  We then fit a Bayesian linear regression model of mouse weight on gene expression levels using the default prior distributions from the \texttt{rstanarm} package \citep{rstanarm}. In Appendix Table \ref{tab:rv_mouse}, we report the posterior mean for the observed regression coefficients, $\tau^\text{na\"ive}_\text{pm}$, as well as the endpoint of the 95\% posterior credible interval closest to zero, $\tau^\text{na\"ive}_\text{endpt}$, for genes whose 95\% posterior credible interval excludes zero.  We also report the robustness value, $RV$, in terms of the percentage of outcome variance explained by confounding needed for the true causal effect to change sign, using $\tau^\text{na\"ive}_\text{endpt}$ for a more conservative measure of robustness that accounts for estimation uncertainty. Only five genes are found to be significantly different from zero without confounding.  The significance of two genes, Sirpa and Avpr1a, are extremely sensitive to confounding in that confounders would only need to explain less than 2\% of the residual outcome variance to change the sign of the effect.  In contrast, while the gene Gstm2 does not have the largest magnitude of $\tau^\text{na\"ive}_\text{pm}$ among the significant genes, it is the most robust to confounding in the two-factor model (RV=80\%).

\begin{figure}[h!]
        \centering
        \includegraphics[width=0.8\textwidth]{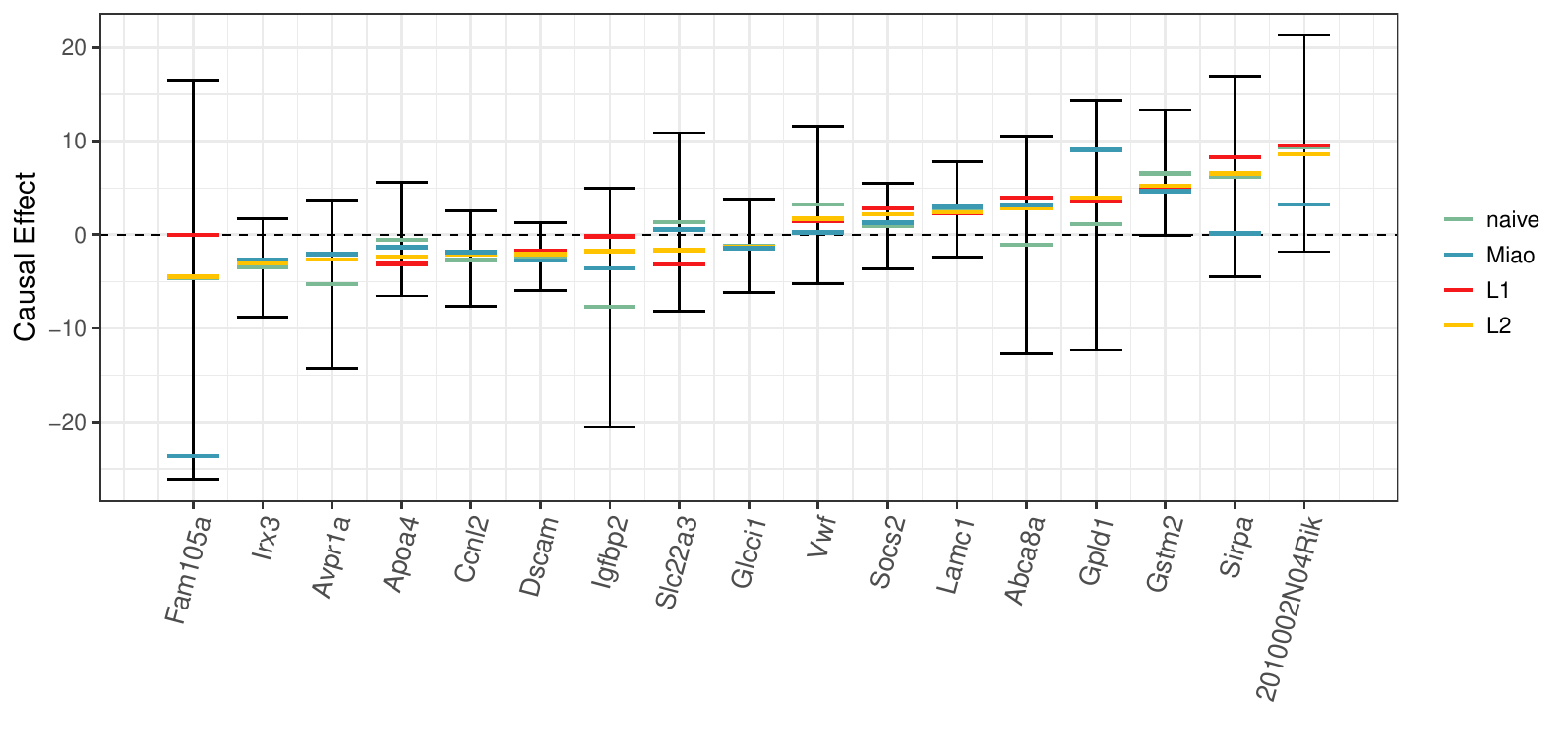}
		\caption{Possible causal effects for a one standard deviation change in expression level on mouse obesity.  Black bars define the support of causal effects consist with the observed data under the linear treatment and outcome model, Equations \ref{eqn:confounder}-\ref{eqn:outcome}. To accommodate both estimation uncertainty and uncertainty due to confounding, we construct this region from the endpoints of the 95\% posterior credible interval of the naive effects.	Inside the ignorance region we plot the naive treatment coefficients, the L1 and L2 minimizing MCC solutions, and the results of the null treatment approach from \citet{miao2020identifying}.   \label{fig:mice_mcc_miao}}
\end{figure}

We also use the MCC approach to examine the treatment effects with the smallest L1 and L2 norm, and compare these results to the results from \citet{miao2020identifying}, who use robust linear regression to infer multiple causal effects under a sparsity assumption.  We apply the MCC criteria to the causal effects associated with all 17 genes, to identify how small the causal effects can be in aggregate\footnote{See \citet{zheng2022bayesian} for an example in which prior knowledge is used to apply a similar shrinkage criteria to only a subset of the genes, which are  \emph{a priori} thought to have little to no effect on mouse obesity.}. In Figure \ref{fig:mice_mcc_miao}, we show how these additional identifying assumptions still lead to  solution vectors inside the worst-case ignorance regions.  To accommodate both estimation uncertainty and uncertainty due to confounding, we construct the ignorance region from the endpoints of the 95\% posterior credible interval of the naive effects.

While similar in spirit, the L1 and L2 MCC methods are distinct from the null treatments approach of \citet{miao2020identifying} in that the MCC solutions encourage small causal effects across \emph{all} treatments, and thus identify the solution for which the entire gene expression profile causes the smallest change in mouse weight.  This MCC approach tends to reduce the number of causal outliers.  In contrast, the null treatments assumption can accommodate some genes with significantly larger causal effects (e.g. Fam105a), as long as at least half of the treatments are true null genes.

Our sensitivity analysis can also be applied with more complex non-linear models.  We demonstrate this using Bayesian Additive Regression Trees (BART), a method that has previously been applied for estimating (heterogeneous) causal effects in the presence of observed confounders in the single-treatment settings \citep{hill2011bayesian, hahn2020bayesian}.  Here, we use BART to infer non-linearities in the causal effects across multiple treatments while characterizing robustness to unobserved confounding. As our estimand, we consider the population average treatment effect of changing gene $j$ from the median level to the $q$th quantile:
$$\tau_j^q = E[Y \mid do(t_j^q)] - E[Y \mid do(t_j^{0.5})]$$
where $t_j^q$ denotes the treatment vector with all treatments assigned to the median level in the observed population except for the $j$th treatment which is assigned to the $q$th quantile. This is a useful set of estimands when the outcome is nonlinear in the level of the exposure, precisely because it reveals such nonlinearities.

\begin{figure}[h!]
        \centering
        \includegraphics[width=0.8\textwidth]{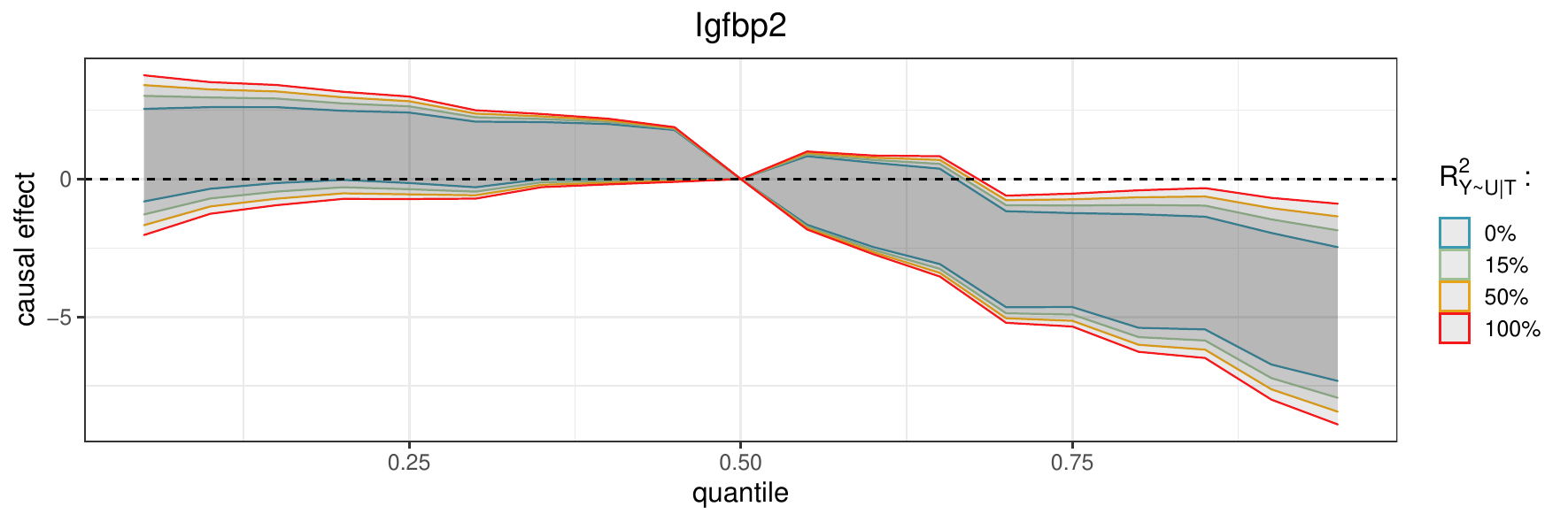}
		\caption{Causal effect of Igfbp2 expression level on mouse obesity, relative to the median level, inferred using BART.  We use an m=2 in the factor model to infer the distribution of latent confounders and find that there is no discernible effect of Ifgbp2 on mouse obesity when expressed at low levels, but for levels above the $0.75$ quantile, Igfbp2 reduces mouse weight and is robust to confounding. \label{fig:mice_igfbp2_bart}}
\end{figure}

Using BART, we found only one gene, Igfbp2, had 95\% posterior credible regions for $t_j^q$ which excluded zero for at least one $q$ under the no unobserved confounding assumption.  In Figure \ref{fig:mice_igfbp2_bart} we show the posterior 95\% region as a function of the expression quantile, $q$, for different values of $R^2_{Y \sim U|T}$.  With this particular data set, we can see that the estimation uncertainty is fairly large relative to the uncertainty induced by 2-factor shared confounding across the multiple-treatments.  For all values of $q$ < 0.7, $t_j^q$ is not significantly different from zero, even without confounding $(R^2_{Y\sim U \mid T} =0)$, whereas for $q \geq 0.75$, $t_j^q$ is significantly negative even if all the residual outcome variance was explained by shared confounding $(R^2_{Y\sim U \mid T}=1)$.  As such, we might conclude that high levels of Igfbp2 have an effect on mouse weight, but there is no robust difference in mouse weight for average and low levels of Igfbp2.


\section{Discussion}

In this paper, we introduced a framework for sensitivity analysis with multiple treatments which  provides further context to the growing literature on the challenges of inference in this setting.  Unlike previous work, we emphasize the importance of carefully defined estimands and show that bounds on the magnitude of confounding bias depend on the particular estimands of interest.  Our work also provides a practical solution to characterizing and calibrating the robustness of causal effects across multiple treatments in the presence of unobserved confounding.  Code to replicate all analyses is available \citep{replication_code} and an R package implementing our methodology is also available and in active development \citep{CopSens}.  In addition to the GWAS simulation and gene expression data set analyzed in this paper, we also include  a reanalysis of the TMDB 5000 Movie Data Set  \citep{moviedataset} in Appendix \ref{sec:movie}.  This data was extensively analyzed by \citet{wang2018blessings} and \citet{grimmer2020ive}, where the goal is to infer the causal effect of an actor's presence in a movie on revenue.


\ed{In this work, we propose a general strategy for sensitivity analysis involving copulas.  While we provide an algorithm for inference for any arbitrary copula, the examples and simulations assume the conditional distribution of latent confounders is Gaussian.  It would be useful to explore how best to handle multi-treatment sensitivity analysis with logistic or poisson factor analysis models for which the joint distribution of latent confounders given treatments is non-Gaussian. The main challenge with non-Gaussian confounder distributions relates to calibration, benchmarking and interpretation, as the mechanics of computing bounds given a copula are clearly described in this work.  Relatedly, we do not explore situations in which $U$ is discrete.  Follow up work which explores additional challenges for discrete confounders would also be interesting. }

In addition, we focused primarily on partial identification results and less on practical issues pertaining to estimation. In this regard, generalizations based on joint inference of the treatment and outcome models should be explored. Joint inference is essential for accounting for both estimation uncertainty and uncertainty due to unobserved confounding.  This is particularly important for the multiple contrast criteria which, as described, does not incorporate parameter uncertainty into the objective function.  Future work exploring frequentist strategies for constructing confidence intervals, perhaps leverage bootstrap methods.  In a follow up to this paper, \citet{zheng2022bayesian} consider uncertainty quantification in multi-treatment inference in the Bayesian paradigm.  Here, they view MCC criteria as Bayesian priors and consider how such shrinkage priors influence posterior estimates of treatment effects.  They also consider additional constraints on the calibration criteria, by considering the role of negative control exposures (NCE) \citep{shi2020selective}.  A set of NCEs are a subset of causes that were known a priori to have zero (or bounded) causal effect on the outcome, and can be considered a degenerate prior on some treatment effects.  Such additional constraints can further shrink the bounds on all causal contrasts.

Finally, it is worth further exploring the relationship between inference with multiple treatments and inference with multiple outcomes.  In follow-up work, \citet{zheng2022sensitivity} consider the related problem of causal inference with a scalar treatment and multiple outcomes, but do not consider settings with both multiple treatments and multiple outcomes.  The ideas in this work can be combined with the strategies applied to multi-outcome causal inference, to yield even more informative bounds on causal effects.  We leave exploration of these extensions to future work.

\singlespacing
\bibliographystyle{chicago}
\bibliography{references}

\newpage
\appendix




\clearpage
\section{Theory}

\subsection{General Contrast Estimation Algorithm}
\label{sec:algm2_derivation}

\begin{algorithm} 
     \caption{Marginal Contrast Estimation for Arbitrary Copulas}
     \label{algm:general}
     \begin{algorithmic}[1] 
     \Function{ComputeMean}{$t$, $\psi$}
        \For{$k = 1, 2, \ldots, M$}
            \State Sample $y_k$ from $f(y \mid t)$
            \For{$i = 1, 2, \ldots, n$}
                \For{$j = 1, 2, \ldots, N$}
                    \State Sample $u_{ij}$ from $f(u \mid t_i)$
                    \State Compute $c_{ij} \leftarrow c_{\psi}(y_k, u_{ij} \mid t)$
                \EndFor
            \EndFor
            \State Compute $w_k \leftarrow \frac{1}{nN} \sum_{ij} c_{ij}$
        \EndFor
        \State \Return $\frac{1}{M}\sum_{k} \nu(y_k) w_k$
     \EndFunction
     \State Return $\tau(\text{ComputeMean}(t_1, \psi), \text{ComputeMean}(t_2, \psi)$
     \end{algorithmic}
\end{algorithm}

\subsection{
Contrast Estimation Algorithm with Gaussian Copulas
}
Suppose that the observed data is generated by model in Equations \ref{eqn:u_mid_t} - \ref{eqn:y_ytilde_relationship_gauss}:
\begin{align*}
    U &= \mu_{u\mid t} + \Sigma_{u\mid t}^{1/2}\epsilon_U ,\\
    Z_Y &= \gamma_t^\top U + \sqrt{1 - \gamma_t^\top \Sigma_{u\mid t}\gamma_t}\epsilon_{Z_Y}, \\
     Y &= F^{-1}_{Y|T}(\Phi(Z_Y -\gamma_t^\top \mu_{u\mid t})),
\end{align*}
given sensitivity vectors $\gamma_t$, any marginal contrast estimand can be estimated by Algorithm \ref{algm:gaussian}.
  \begin{algorithm}
\caption{Marginal Contrast Estimation with Gaussian Copulas.}
\label{algm:gaussian}
\begin{algorithmic}[1]
\Function{ComputeMean}{$t$, $\gamma$}
        \For{$i = 1, 2, \ldots, n$}
        \State $\mu_{i} \leftarrow \gamma^\top  (\mu_{u|t_i} - \mu_{u|t})$
            \For{$j  =1, 2, \ldots, nSim$}
            \State Sample $z_{ij}$ from $N(\mu_{i}, 1)$ 
            \State $y_{ij} \leftarrow F_{Y|t}^{-1}(\Phi( z_{ij} ))$
            \EndFor
        \EndFor
        \State \Return $\frac{1}{n}\sum_{ij} v(y_{ij})$
\EndFunction
\State Return $\tau(\text{ComputeMean}(t_1, \gamma), \text{ComputeMean}(t_2, \gamma))$
\end{algorithmic}
\end{algorithm}

\subsubsection*{\textbf{Derivation of Algorithm \ref{algm:gaussian}}}
\noindent Since we have Equation \ref{eqn:interven_expect} and \ref{eqn:weight}, furthermore, we can write
\begin{equation}
    E[v(Y) \mid do(t)]\ =  \iint v(y) f(y \mid z_y) w_\psi(z_y, t) 
    f(z_y \mid t) dz_y d y,
\end{equation}
where $w_\psi(z_y, t) \approx \frac{1}{|\mathcal{T}|}\sum_{t_i \in \mathcal{T}} \left[ \int c_{\psi}(F_{Z_Y|t}(z_y), F_{U|t}(u) \mid t) f(u \mid t_i) du\right].$
To verify Algorithm \ref{algm:gaussian}, we only need to show that 
\begin{equation}
    \int f(z_y \mid t, u) f(u \mid t_i) du \sim N(\gamma^\top (\mu_{u|t_i} - \mu_{u|t}), \ 1),
\end{equation}
where $f(z_y \mid t, u) = f(z_y \mid t) c_{\psi}(F_{Z_Y|t}(z_y), F_{U|t}(u) \mid t).$
Based on model in Equations \ref{eqn:u_mid_t} - \ref{eqn:y_ytilde_relationship_gauss}, we have
\begin{align}
    f(u \mid t_i) &\sim N(\mu_{u|t_i}, \ \Sigma_{u\mid t}),\\
    f(z_y \mid t, u) &\sim N(\gamma_t^\top  (u - \mu_{u|t} ), \ 1 - \gamma_t^\top \Sigma_{u\mid t}\gamma_t). \label{eqn_tilde_complete}
\end{align}
By integrating out the $U$, 
\begin{equation}
    \int f(z_y \mid t, u) f(u \mid t_i) du\\
    = \frac{1}{\sqrt{2\pi}}
    \text{exp}\left\{
    - \frac{(z_y - \gamma_t^\top (\mu_{u|t_i} - \mu_{u|t}))^2 }{2}
    \right \}. \label{eqn_u_integrated}
\end{equation}

\subsection{Additional Figures Illustrating the Proposed Sensitivity Analysis}

\begin{figure}	
	\centering
	\begin{subfigure}[t]{0.20\textwidth}
		\centering
		\includegraphics[width=\textwidth]{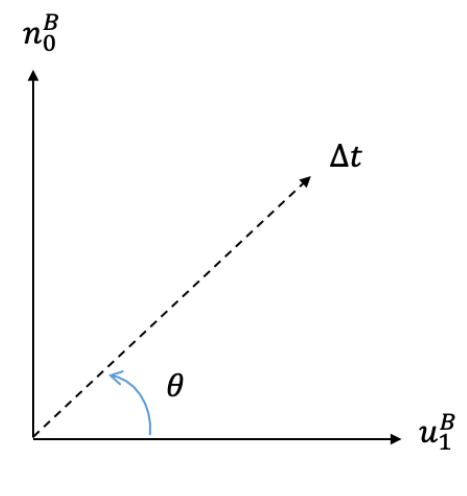}
		\caption{$\Delta \bm{t}:= \bm{t_1} - \bm{t_2}$} 
		\label{fig:rotation}
	\end{subfigure}
	\quad\quad
	\begin{subfigure}[t]{0.35\textwidth}
		\centering
		\includegraphics[width=\textwidth]{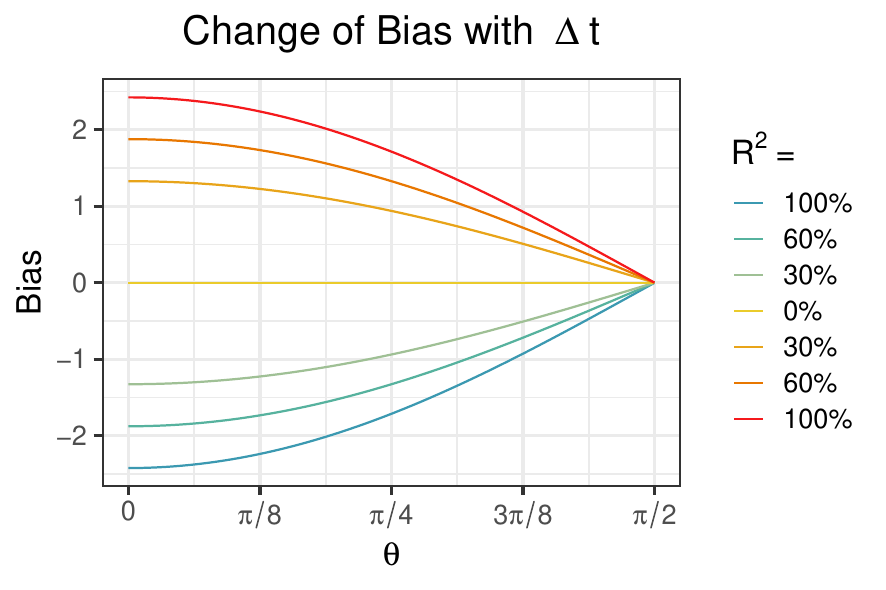}
		\caption{Estimation Bias of $\text{PATE}_{(t_1 - t_2)}$}
		\label{fig:ate_bias_bound}
	\end{subfigure}
	\quad\quad
	\begin{subfigure}[t]{0.35\textwidth}
	    \centering
	    \includegraphics[width=\textwidth]{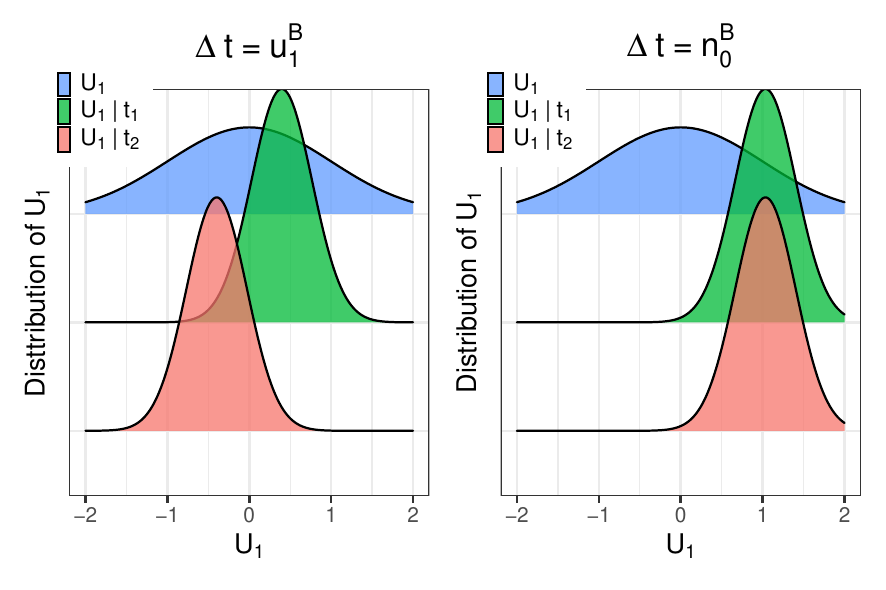}
	    \caption{Distribution of $U_1$}  \label{fig:udist}
    \end{subfigure}
    \caption{Illustration of Corollary \ref{thm:anyt}. (a) We parameterize $(t_1 - t_2)$ with $\theta$, the angle between $n_0^B$, a vector in the null space of $B$, and $u_1^B$, the first left singular vector of $B$. (b) The confounding bias of naive estimates of $\text{PATE}_{t_1, t_2}$ changes with $\theta$ and depends on $R_{Y \sim U|T}^2$. (c) Confounder densities in different populations. The blue, green, red densities denote distributions of $U_1$ in the observed population, the subpopulation receiving $t_1$ and the subpopulation receiving treatment $t_2$ respectively.  Observed data estimates of $\text{PATE}_{t_1, t_2}$ are unbiased when $(t_1 - t_2) = n_0^B$, since the confounder distributions are the same in two treatment arms.  However, observed data estimates of $\text{PATE}_{t_1, \sbullet}$ and $\text{PATE}_{t_2, \sbullet}$ are biased since in general the superpopulation distribution of the confounder is different.
    \label{fig:gaussian_example}}
\end{figure}

We illustrate some key insights from Corollary \ref{thm:anyt} in Figure \ref{fig:gaussian_example} where we display the worst-case bias as a function of the treatment contrasts, $(t_1 - t_2)$.  In this illustration, we assume that $(t_1 - t_2)$ lies on a plane spanned by  $u^B_1$ and $n_0^B$, an arbitrary vector in the null space of $B$. We let $\theta = \text{arccos}((t_1 - t_2)^\top  u_1^B)$ be the angle of $(t_1 - t_2)$ relative to $u_1^B$, Figure \ref{fig:rotation}.  Figure \ref{fig:ate_bias_bound} depicts the bias as function of $\theta$ for different values of $R_{Y \sim U|T}^2$.  When $(t_1 - t_2)$ is in the null space of $B$, $\text{PATE}_{t_1, t_2}$ is identified because the confounder distributions are identical in the two treatment arms, i.e.\ there is no confounding for this particular contrast.  When $(t_1 - t_2)$ is colinear with $u_1^B$ the scaled difference in means of $u$ is largest, which implies the largest worst-case bias for the treatment effect, Figure \ref{fig:udist} (left).    Even when $\text{PATE}_{t_1, t_2}$ is identified, we emphasize that $\text{PATE}_{t_1, \sbullet}$ and $\text{PATE}_{t_2, \sbullet}$ are both biased, since the distribution of confounders in the treatment arm differs from the distribution of confounder in the superpopulation, Figure \ref{fig:udist} (right). As noted by others, identification of $\text{PATE}_{t_1, t_2}$ for $(t_1 - t_2)$ in the null space of $B$ arises due to bias cancellation in intervention means of the two treatment arms \citep{grimmer2020ive}.

\begin{figure}	
	\centering
	\begin{subfigure}[t]{0.31\textwidth}
		\centering
		\includegraphics[width=\textwidth]{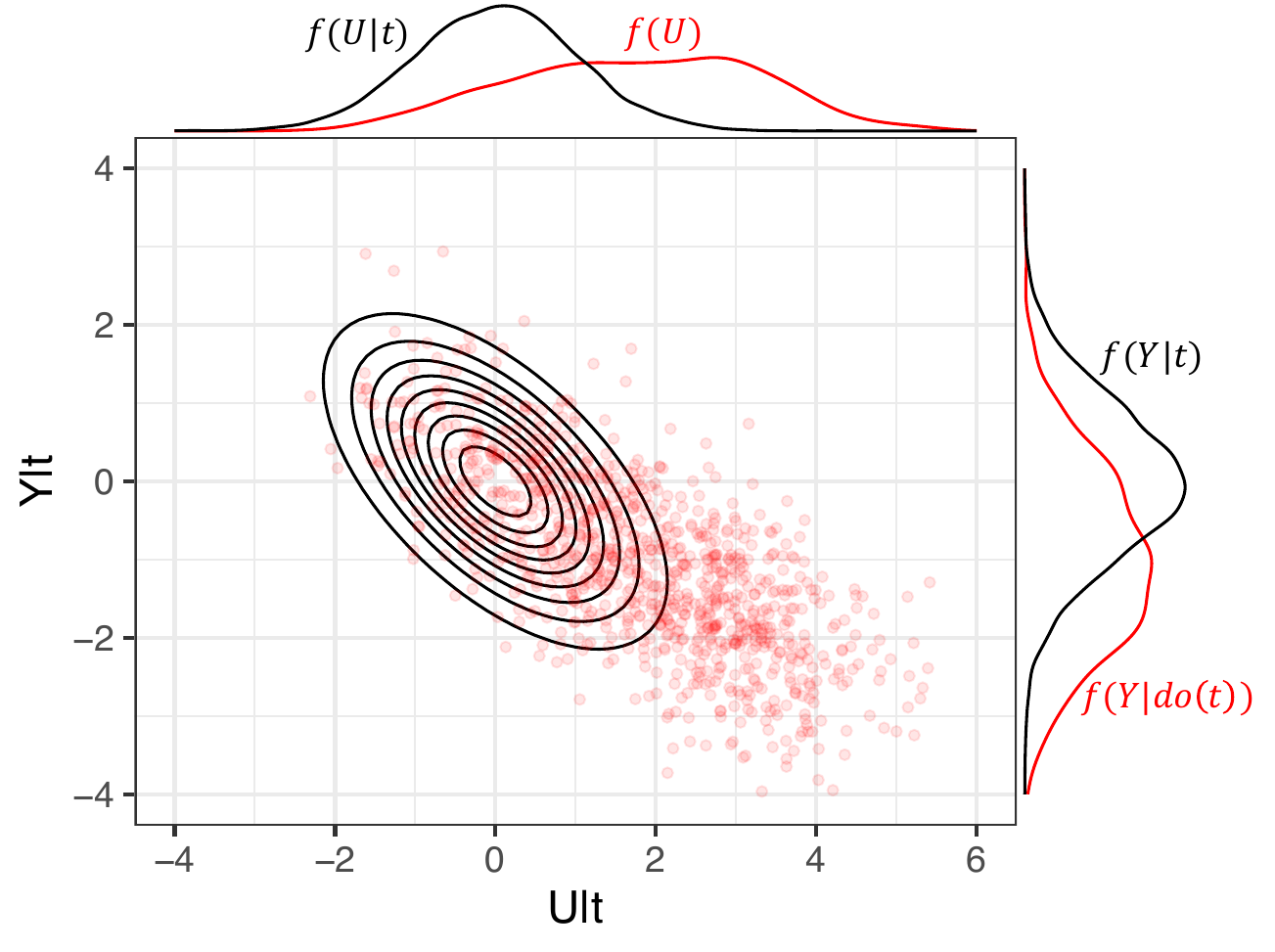}
		\caption{$\sqrt{R^2_{Y\sim U \mid T}} = 0.6$}
		\label{fig:cali_neg}
	\end{subfigure}
	\quad
	\begin{subfigure}[t]{0.31\textwidth}
		\centering
		\includegraphics[width=\textwidth]{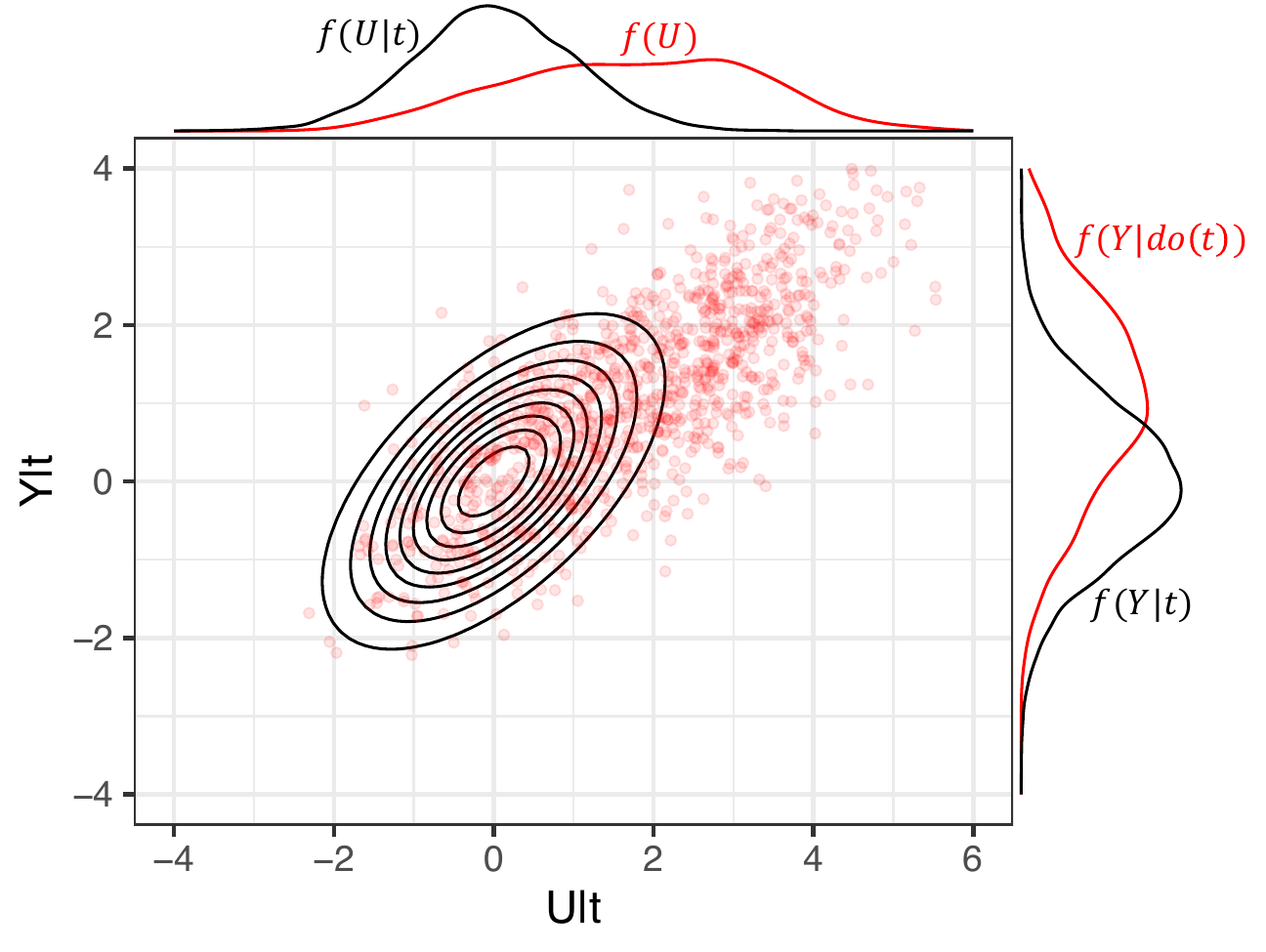}
		\caption{$\sqrt{R^2_{Y\sim U \mid T}} = 0.6$\newline (opposite sign)}
		\label{fig:cali_pos_l}
	\end{subfigure}
	\quad
	\begin{subfigure}[t]{0.31\textwidth}
	    \centering
	    \includegraphics[width=\textwidth]{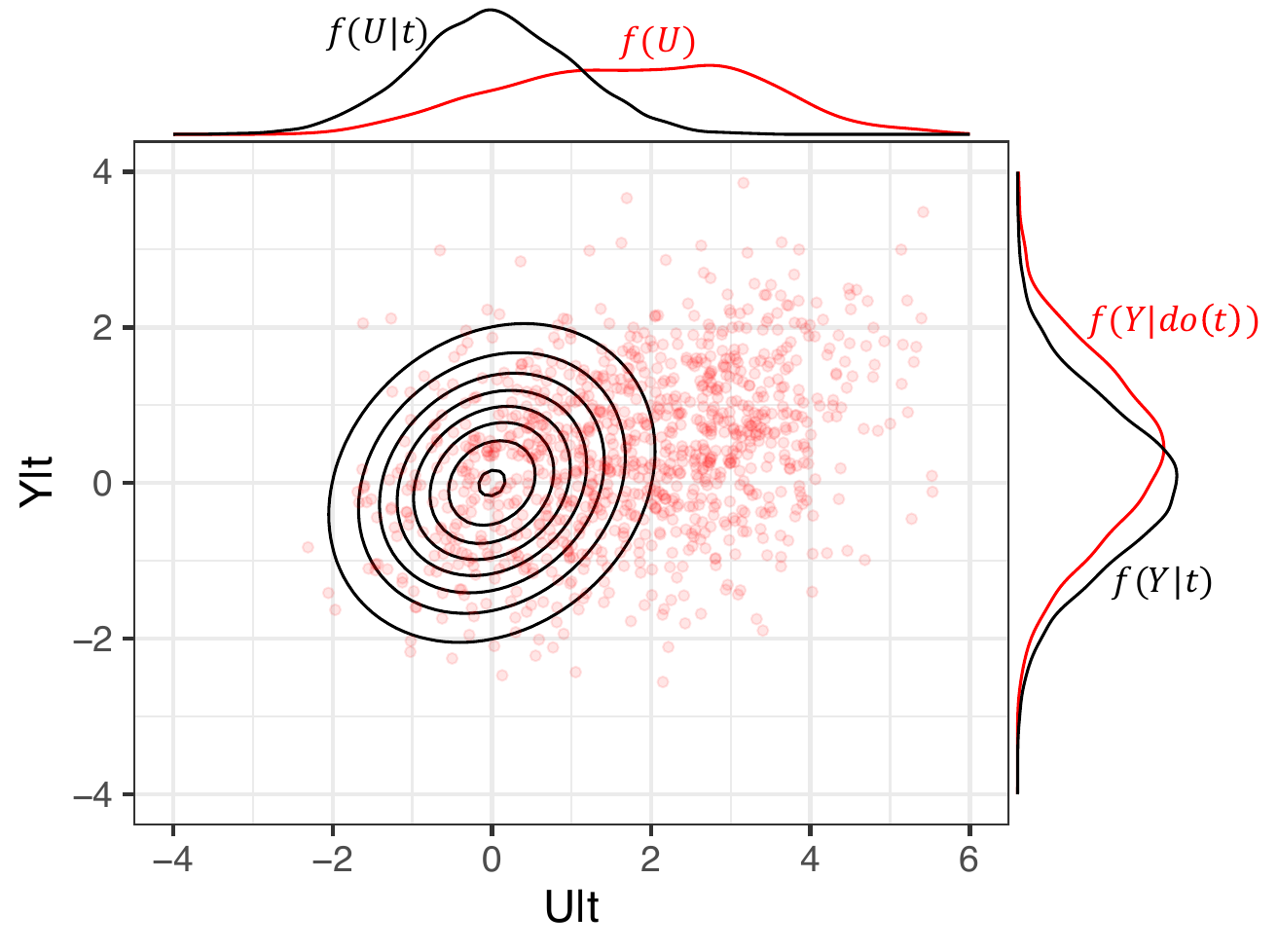}
	    \caption{$\sqrt{R^2_{Y\sim U \mid T}} = 0.2$}  
	    \label{fig:cali_pos_s}
    \end{subfigure}
    \caption{Differences between observed and intervention densities as a function of the fraction of outcome variance explained by a single confounder.  The black contours depict the conditional Gaussian copula, $c_{\gamma}(F_{Y|t}(y), F_{U|t}(u) \mid t)$ whereas red points represent samples from the joint distribution, $f(y, u \mid do(t)) \propto f(y \mid t) c_{\gamma}(F_{Y|t}(y), F_{U|t}(u) \mid t) 
    f(u)$.  We visualize the shift in the outcome density for different conditional correlations and note that smaller values of $R^2_{ Y\sim U \mid T}$ imply smaller biases in the outcome despite large imbalance in the distribution of U.
    \label{fig:calibration}}
\end{figure}




\subsection{Proof of Theorem \ref{thm:linear_bound}}
\label{sec:proof_thm_linear_bound}
\textbf{Theorem 1.} 
\begin{itshape}
Suppose that the observed data is generated by model in Equations \ref{eqn:confounder}-\ref{eqn:outcome}:
\begin{align*} 
    U & = \epsilon_u \\
    T &= BU + \epsilon_{t} \\
    Y &= \tau^\top  T + \gamma^\top  U + \epsilon_{y} 
\end{align*}
with $\Lambda_{t\mid u} > 0$. 
Then,  $\forall \gamma$ satisfying Assumptions 1 and 2, 
    \begin{equation} 
    {\gamma}^\top  \Sigma_{u\mid t} {\gamma} \leq \sigma_{y\mid t}^2
    \end{equation}

\noindent For any given $t_1$, $t_2$, we have
    \begin{equation}
    \text{Bias}_{t_1, t_2}^2 \leq \sigma^2_{y\mid t}R_{Y \sim U\mid T}^2 \parallel  \Sigma_{u\mid t}^{-1/2} (\mu_{u\mid t_1} - \mu_{u\mid t_2}) \parallel_2^2.
    \end{equation}
The bound is achieved when ${\gamma}$ is colinear with $\Sigma_{u\mid t}^{-1}(\mu_{u\mid t_1} - \mu_{u\mid t_2})$ and is maximized when all the residual outcome variance is due to unmeasured confounders, e.g. $R^2_{Y\sim U\mid T} = 1$.\\
\end{itshape}

\proof{ Under model in Equations \ref{eqn:confounder}-\ref{eqn:outcome}, the variance of the observed outcome equals
\begin{equation}
    \begin{split}
    \label{eqn:var(Y|T)}
         \sigma_{y\mid t}^2 &:= Var(Y \mid T) \\
         & = \sigma^2 + \gamma^\top (I - B^\top (BB^\top  + \Lambda_{t})^{-1}B)\gamma\\
         &  = \sigma^2 + \gamma^\top  \Sigma_{u\mid t} \gamma,
    \end{split}
\end{equation}
\sloppy
where $\gamma^\top  \Sigma_{u\mid t} \gamma$ corresponds to the confounding variation, and $\sigma^2$ stands for the non-confounding variation in the residual of observed outcome. Hence, the fraction of confounding variation in the residual of $Y$, $R_{Y \sim U\mid T}^2$, can be expressed in terms of Equation \ref{eqn:partial_r2}, which produces a constrain for $\gamma$ (Equation \ref{eqn:ellipse_gamma}) that the confounding variation in the residual of $Y$, $\gamma^\top  \Sigma_{u\mid t} \gamma$, should not be larger than $\sigma_{y\mid t}^2 R_{Y \sim U\mid T}^2$ for a given level of $R_{Y \sim U\mid T}^2$.

Let
\begin{equation} \label{eqn:defZ}
    W := \Sigma_{u\mid t}^{1/2} \gamma, 
\end{equation}
then the omitted variable bias in Equation \ref{eqn:pate_bias} can be written as 
$$
    Bias_{t_1, t_2} = W^\top \Sigma_{u\mid t}^{-1/2}(\mu_{u\mid t_1} - \mu_{u\mid t_2}),
$$
where $W^\top  W \leq \sigma_{y\mid t}^2R_{Y \sim U\mid T}^2$, implied by inequality in Equation \ref{eqn:ellipse_gamma}.

Therefore,
\begin{align}
\label{eqn:bias2}
    \text{Bias}_{t_1, t_2}^2  &= W^\top \Sigma_{u\mid t}^{-1/2}(\mu_{u\mid t_1} - \mu_{u\mid t_2}) (\mu_{u\mid t_1} - \mu_{u\mid t_2})^\top  \Sigma_{u\mid t}^{-1/2} W\\
    & \leq \sigma_{y\mid t}^2R_{Y \sim U\mid T}^2 \parallel \Sigma_{u\mid t}^{-1/2} (\mu_{u\mid t_1} - \mu_{u\mid t_2}) \parallel_2^2,
\end{align}
where the bounds are reached when $W =  \frac{\sqrt{\sigma^2_{y\mid t}R^2_{Y \sim U \mid  T}}\Sigma_{u\mid t}^{-1/2}(\mu_{u\mid t_1} - \mu_{u\mid t_2})}{\parallel \Sigma_{u\mid t}^{-1/2}(\mu_{u\mid t_1} - \mu_{u\mid t_2})\parallel_2}$, i.e., ${\gamma}$ is colinear with the $\Sigma_{u\mid t}^{-1}(\mu_{u\mid t_1} - \mu_{u\mid t_2})$ inferred by the relationship defined in Equation \ref{eqn:defZ}.
}

\subsection{Proof of Corollary \ref{thm:anyt}}

\noindent\textbf{Corollary \ref{thm:anyt}}
\label{sec:proof_thm_anyt}
\begin{itshape}
Assume $\Lambda_{t} = \sigma^2_{t} I$ and let $d_1$ be the largest singular value of $B$.  For all $t_1, t_2$ with $\parallel (t_1 - t_2) \parallel_2 = 1$, the squared bias  is bounded by
\begin{equation}
\text{Bias}_{t_1, t_2}^2
\leq \frac{ d_1^2}{ (d_1^2 + \sigma_{t}^2)}\frac{\sigma_{y|t}^2}{\sigma_{t}^2} R_{Y \sim U|T}^2,
\end{equation}
with equality when $(t_1 - t_2) = u_1^{B}$, the first left singular vector of $B$. When $(t_1 - t_2) \in  Null(B^\top )$, the naive estimate is unbiased, that is, $PATE_{t_1, t_2} = \tau_{\text{naive}}^\top  (t_1 - t_2)$. \\

\end{itshape}

\proof{ Suppose that the matrix $B$ has the singular value decomposition, 
$$B = U D V^\top ,$$ 
where the diagonal entries of $D$ are the singular values of $B$ in descending order. Then, we can write
\begin{equation} 
\label{eqn:mu_u_t_decomp}
    (\mu_{u|t_1} - \mu_{u|t_2}) = VD(D^2 + \sigma_{t}^2 I)^{-1} U^\top  (t_1 - t_2),
\end{equation}
and 
\begin{equation}
\label{eqn:sigma_u_t_decomp}
    \Sigma_{u|t}^{-1} = V[I + \frac{1}{\sigma_{t}^2} D^2] V^\top .
\end{equation}
\noindent By plugging Equation \ref{eqn:mu_u_t_decomp} and \ref{eqn:sigma_u_t_decomp} into the result of theorem \ref{thm:linear_bound}, we have 
\begin{align}
    \text{Bias}_{t_1, t_2}^2  
    & \leq \frac{\sigma_{y|t}^2}{\sigma_{t}^2}R_{Y \sim U|T}^2 \parallel V D ( \sigma_{t}^2 I + D^2)^{-1/2} U^\top  (t_1 - t_2) \parallel_2^2,
\end{align}
where, according to Rayleigh quotient \citep{horn1985matrix}, the squared L2 norm reaches its maximum, $\frac{ d_1^2}{ (d_1^2 + \sigma_{t}^2)}$, when $(t_1 - t_2)$ equals the first column of $U$, i.e., the first left singular vector of $B$. 

Therefore, we have
\begin{align}
    \text{Bias}_{t_1, t_2}^2  
    & \leq \frac{ d_1^2}{ (d_1^2 + \sigma_{t}^2)}\frac{\sigma_{y|t}^2}{\sigma_{t}^2} R_{Y \sim U|T}^2.
\end{align}
}

\subsection{Proof of Proposition \ref{prop:equivalence}}
\label{sec:proof_prop_equivalence}
\textbf{Proposition \ref{prop:equivalence}} 
\begin{itshape}
Suppose that the observed data is generated by model in Equations \ref{eqn:confounder}-\ref{eqn:outcome}:
\begin{align*} 
    U & = \epsilon_u, \\
    T &= BU + \epsilon_{t}, \\
    Y &= \tau^\top  T + \gamma^\top  U + \epsilon_{y}. 
\end{align*}
If B is rank $m$ and there remain two disjoint matrices of rank $m$ after deleting any row of $B$, then $\| \Sigma_{u|t}^{-1/2} (\mu_{u|t_1} - \mu_{u|t_2}) \|_2^2$ and $\frac{ d_1^2}{ (d_1^2 + \sigma_{t}^2)}$ are both identified. \\
\end{itshape}

\proof{
With model in Equations \ref{eqn:confounder}-\ref{eqn:outcome}, the conditional distribution of confounder $U$ 
\begin{equation}
    f_{B}(u \mid t) \sim N(\mu_{u\mid t}, \ \Sigma_{u\mid t}),
\end{equation}
where $\mu_{u\mid t} :=  B^\top (BB^\top  + \Lambda_{t})^{-1}t$, $\Sigma_{u\mid t} := I - B^\top (BB^\top  + \Lambda_{t})^{-1}B$. 

From \cite{anderson1956statistical}, we know that, if B is rank $m$ and there remain two disjoint matrices of rank $m$ after deleting any row of $B$, then we $B$ is identifiable up to rotations from the right. Let $\tilde B = BA$ for an arbitrary positive matrix $A$ such that the observed treatments are consistent with $Cov(T) = BB^\top  + \Lambda_t = \tilde B \tilde B^\top  + \Lambda_t$, implying that $AA^\top  = I$ and $A$ is an orthogonal matrix. With $\tilde B$, we have the conditional distribution of confounder $U$ as
\begin{equation}
    f_{\tilde B}(u \mid t) \sim N(\tilde \mu_{u\mid t}, \ \tilde \Sigma_{u\mid t}),
\end{equation}
where 
\begin{align}
 \tilde \mu_{u\mid t} &:= \tilde B^\top (\tilde B \tilde B^\top  + \Lambda_{t})^{-1}t = A^\top   \mu_{u\mid t} \label{eqn:mu_u_t_tildeB},\\
 \tilde \Sigma_{u\mid t} &:= I - \tilde B^\top (\tilde B \tilde B^\top  + \Lambda_{t})^{-1} \tilde B = A^\top  \Sigma_{u|t} A \label{eqn:Sigma_u_t_tildeB}.
\end{align}

With Equation \ref{eqn:mu_u_t_tildeB} and \ref{eqn:Sigma_u_t_tildeB}, we have
\begin{align}
     \| \tilde \Sigma_{u|t}^{-1/2} (\tilde \mu_{u|t_1} - \tilde \mu_{u|t_2}) \|_2^2
    &= (\tilde \mu_{u|t_1} - \tilde \mu_{u|t_2})^\top  \tilde \Sigma_{u|t}^{-1/2}  \tilde \Sigma_{u|t}^{-1/2} (\tilde \mu_{u|t_1} - \tilde \mu_{u|t_2}) \\
    &= (A^\top  \mu_{u|t_1} - A^\top  \mu_{u|t_2})^\top \tilde \Sigma_{u|t}^{-1} (A^\top  \mu_{u|t_1} - A^\top  \mu_{u|t_2})\\
    &= (\mu_{u|t_1} - \mu_{u|t_2})A(A^\top \Sigma_{u|t}A)^{-1}
    A^\top  (\mu_{u|t_1} - \mu_{u|t_2})\\
    &= (\mu_{u|t_1} - \mu_{u|t_2}) \Sigma_{u|t}^{-1}(\mu_{u|t_1} - \mu_{u|t_2})^\top  \\
    &= \| \Sigma_{u|t}^{-1/2} (\mu_{u|t_1} - \mu_{u|t_2}) \|_2^2,
\end{align}
Therefore, $\| \Sigma_{u|t}^{-1/2} (\mu_{u|t_1} - \mu_{u|t_2}) \|_2^2$ is identified.

For $\frac{ d_1^2}{ (d_1^2 + \sigma_{t}^2)}$, it depends on the largest singular value of $B$, which is rotation-invariant and therefore identified.

}

\subsection{Proof of Proposition \ref{prop:vine_cop}}
\label{prop:vine_cop_proof}
\noindent\textbf{Proposition \ref{prop:vine_cop}}
\begin{itshape}
The joint (m+1) dimensional copula between the scalar outcome and $m$ latent confounders can be re-expressed
\begin{align}
&c_{\psi}(F_{Y \mid t}(y), F_{U_1\mid t}^{\psi_T}(u_1), \cdots, F^{\psi_T}_{U_m\mid t}(u_m)\mid t) =\\ 
&c_{\psi_T}(F_{U_1\mid t}^{\psi_T}(u_1), \cdots, F^{\psi_T}_{U_m\mid t}(u_m)\mid t)\prod_{i=1}^{m}
c_{\psi}(y,u_i; u_1, \cdots, u_{i-1}, t)
\end{align}
where $c_{\psi_T}$ is the $m$-dimensional copula governing the dependence amongst latent variables, $U$ given $T=t$ and we define $c_{\psi}({y,u_i; u_1, \cdots, u_{i-1}}, t) := c_{\psi}(F_{Y\mid u_1, \cdots u_{i-1}, t}(y),F_{U_i\mid u_1, \cdots u_{i-1}, t} (u_i) \mid u_1, \cdots, u_{i-1}, t)$ is a bivariate (conditional) copula characterizing the dependence between $Y$ and $U_i$ given $i-1$ of the ``preceding'' variables. Given Assumption \ref{asm:latent_identification}, $c_{\psi_T}$ and $F^{\psi_T}_{U_i\mid t}$ are fully identified.  The conditional density $f_{\psi_Y}$ is unidentified without further assumptions.
\end{itshape}

\proof{ Based on Theorem 4.7 of D-vine density from \cite{czado2019analyzing}, we have 
\begin{align}
&c_{\psi}(F_{Y \mid t}(y), F_{U_1\mid t}^{\psi_T}(u_1), \cdots, F^{\psi_T}_{U_m\mid t}(u_m)\mid t)\\
= & \prod_{i=1}^{m} c^i_{\psi_Y}(F_{Y|t}(y), F^{\psi_T}_{U_i|t}(u_i) \mid u_1, \cdots, u_{i-1}, t) \times \\
& 
\prod_{j=1}^{m-1} \prod_{i=1}^{m-j} 
c_{\psi_T}( F^{\psi_T}_{U_i|t}(u_i), F^{\psi_T}_{U_{i+j}|t}(u_{i+j}) \mid u_{i+1}, .., u_{i+j-1}), \label{eqn:dvine_decomp_2nd}
\end{align}
where the first part of the decomposition involves the bivariate copulas $c^i_{\psi_Y}$ describing the dependence between the outcome $Y$ and each latent confounder $U_i$, and the second part in Equation \ref{eqn:dvine_decomp_2nd} involves the joint copula of the latent confounders.

Again, by applying the D-vine Theorem to the joint distribution of $U$ given $T$, we have
\begin{equation}
    c_{\psi_T}(F_{U_1\mid t}^{\psi_T}(u_1), \cdots, F^{\psi_T}_{U_m\mid t}(u_m)\mid t) = 
\prod_{j=1}^{m-1} \prod_{i=1}^{m-j} 
c_{\psi_T}( F^{\psi_T}_{U_i|t}(u_i), F^{\psi_T}_{U_{i+j}|t}(u_{i+j}) \mid u_{i+1}, .., u_{i+j-1})
\label{eqn:dvine_decomp_u}
\end{equation}

Plugging Equation \ref{eqn:dvine_decomp_u} into \ref{eqn:dvine_decomp_2nd}, then we have 
\begin{align}
&c_{\psi}(F_{Y \mid t}(y), F_{U_1\mid t}^{\psi_T}(u_1), \cdots, F^{\psi_T}_{U_m\mid t}(u_m)\mid t) =\\ 
&c_{\psi_T}(F_{U_1\mid t}^{\psi_T}(u_1), \cdots, F^{\psi_T}_{U_m\mid t}(u_m)\mid t)\prod_{i=1}^{m} c^i_{\psi_Y}(F_{Y|t}(y), F^{\psi_T}_{U_i|t}(u_i)\mid u_1, \cdots, u_{i-1}, t) 
\end{align}

}

\subsection{Proof of Proposition \ref{prop:model_gcop}}
\label{sec:proof_prop_model_gcop}
\noindent\textbf{Proposition \ref{prop:model_gcop}}
\begin{itshape}
Given Assumptions \ref{asm:gaussian_latent} and \ref{asm:copula}, the full $(m+1)$-dimensional conditional copula between the outcome and all $m$ latent confounders given treatments,\\ $c_{\psi}(F_{Y \mid t}(y), F_{U_1\mid t}^{\psi_T}(u_1), ..., F^{\psi_T}_{U_m\mid t}(u_m)\mid t)$, is a Gaussian copula. Further, any data generating process satisfying Assumptions 5 and 6 can be represented using the following structural equation model, which links $Y$  to $U$ through a scalar latent Gaussian variable $Z_Y$ with unit variance,
 \begin{align}
    U &= \mu_{u\mid t} + \Sigma_{u\mid t}^{1/2}\epsilon_U \\
    Z_Y &= \gamma_t^\top U + \sqrt{1 - \gamma_t^\top \Sigma_{u\mid t}\gamma_t}\epsilon_{Z_Y} \\
     Y &= F^{-1}_{Y|T}(\Phi(Z_Y -\gamma_t^\top \mu_{u\mid t}))
\end{align}
where $\epsilon_U \sim N(0, I_m)$, and $\epsilon_{Z_Y} \sim N(0, 1)$. $\psi_Y = \{\gamma_t\}$ is a reparameterization of the $\rho_i$ parameterization.

\end{itshape}

\proof{ 

\noindent \textbf{1) The full \((m + 1)\)-dimensional conditional copula is a Gaussian copula.}

\noindent By Assumption \ref{asm:gaussian_latent}, we have $f_{\psi_T}(u\mid t) \sim N(\mu_{u\mid t}, \Sigma_{u\mid t})$ with $\psi_T = \{\mu_{u\mid t}, \Sigma_{u\mid t} : t \in \mathcal{T}\}$. The Gaussian copula density can be derived by transforming marginal variables $U$ into a standard normal distributions. Define the transformed variables as follows: 
$$
Z_{U_1} := \Phi^{-1}(F_{U_1\mid t}^{\psi_T}(U_1))), \cdots, Z_{U_m}:=  \Phi^{-1}(F_{U_m\mid t}^{\psi_T}(U_m))
$$
Given $\Sigma_{u|t}$ and utilizing the properties of the Gaussian distribution, the correlation matrix of $[Z_{U_1} , \cdots, Z_{U_k} ]$ conditional on $T = t$ for any $k \leq m$ can be written as:
\begin{equation}
    C_{u_{1:k}|t}:= \text{Corr}(Z_{U_1} , \cdots, Z_{U_k} \mid t) = D_{u_{1:k}|t}^{-1/2} \Sigma_{u_{1:k}|t} D_{u_{1:k}|t}^{-1/2},
\end{equation}
\sloppy
where $D_{u_{1:k}|t} = \text{diag}(\sigma_{11}, \cdots, \sigma_{kk})$ is a diagonal matrix containing the variances of $U_1, \cdots, U_k$ along the diagonal, and $\Sigma_{u_{1:k}|t}$ is the top $k \times k$ submatrix of $\Sigma_{u|t}$. Therefore, the density of the joint conditional copula of $U$ given $T$ can be written as:

\begin{align}
& c_{\psi_T}(F_{U_1\mid t}^{\psi_T}(u_1), \cdots, F^{\psi_T}_{U_m\mid t}(u_m)\mid t) \notag \\
=&
\frac{1}{\sqrt{\det C_{u_{1:m}|t}}} \exp \left( -\frac{1}{2} 
\left( 
\begin{array}{c}
z_{u_1} \\
\vdots \\
z_{u_m}
\end{array} 
\right)^\top  
\cdot \left( C_{u_{1:m}|t}^{-1} - I \right) 
\cdot \left( 
\begin{array}{c}
z_{u_1} \\
\vdots \\
z_{u_m}
\end{array} 
\right) 
\right).
\label{eqn:copula_density_u_t}
\end{align}


\noindent From Assumption \ref{asm:copula}, we have the density of the conditional bivariate copula, describing the conditional dependence of $Y$ on $U_i$ given ``previous'' $U$'s and $T$, as
\begin{align}
    & c^i_{\psi_Y}(F_{Y|t}(y), F^{\psi_T}_{U_i|t}(u_i)\mid u_1, \cdots, u_{i-1}, t) \\
    = & \frac{1}{\sqrt{ 1- \rho_i^2  }}
    \exp \left(
    -
    \frac{
    \rho_i^2 \big((z_y - \gamma_t^\top \mu_{u\mid t})^2 + z_{u_i}\big) - 2\rho_i (z_y - \gamma_t^\top \mu_{u\mid t}) z_{u_i}
    }{
    2(1 - \rho_i^2)
    }
    \right)
    \label{eqn:copula_density_yui}
\end{align}

\noindent From Proposition \ref{prop:vine_cop}, the density of the joint $(m+1)$ dimensional copula, $c_{\psi}($ $F_{Y \mid t}(y)$,$F_{U_1\mid t}^{\psi_T}(u_1)$, $\dots$, $F^{\psi_T}_{U_m\mid t}(u_m)\mid t)$, is the product of product of the copula density $c_{\psi_T}(F_{U_1\mid t}^{\psi_T}(u_1), \cdots, F^{\psi_T}_{U_m\mid t}(u_m)\mid t)$ in Equation \ref{eqn:copula_density_u_t}, and the conditional bivariate copula densities $c_{\psi}(y,u_i; u_1, \cdots, u_{i-1}, t)$ in Equation \ref{eqn:copula_density_yui}. By plugging these densities, with algebra we derive the density of the joint $(m+1)$-dimensional copula, which follows the Gaussian copula as:
\begin{align}
& c_{\psi}(F_{Y \mid t}(y), F_{U_1\mid t}^{\psi_T}(u_1), \cdots, F^{\psi_T}_{U_m\mid t}(u_m)\mid t) \notag \\
= & \frac{1}{\sqrt{\det C}} \exp \left( -\frac{1}{2} 
\left( 
\begin{array}{c}
z_y - \gamma_t^\top \mu_{u\mid t} \\
z_{u_1}\\
\vdots \\
z_{u_m}
\end{array} 
\right)^\top  
\cdot \left( C^{-1} - I \right) 
\cdot \left( 
\begin{array}{c}
z_y - \gamma_t^\top \mu_{u\mid t} \\
z_{u_1}\\
\vdots \\
z_{u_m}
\end{array} 
\right) 
\right),
\end{align}
where the correlation matrix $C$ is given by:
\begin{equation}
    C = 
    \begin{bmatrix}
        1 & c_{yu_{1:m}|t}\\
        c_{yu_{1:m}|t}^\top  & D_{u_{1:m}|t}^{-1/2} \Sigma_{u_{1:m}|t} D_{u_{1:m}|t}^{-1/2}
    \end{bmatrix}.
    \label{eqn:copula_corrmat}
\end{equation}
In this matrix, $c_{yu_{1:m}|t}$ is defined as $c_{yu_{1:k}|t} := [c_{yu_1|t},$ $ .., c_{yu_k|t}]$ for any $k \leq m$ with $c_{yu_i|t}$ representing the correlation between $Z_Y$ and each $Z_{U_i}$ given $T = t$. 
The elements of $c_{yu_{1:m}|t}$ depend on the unknown parameters $\rho_i$ and the identifiable components of the covariance matrix $\Sigma_{u|t}$.
To simplify, we define $c_{u_iu_{1:(i-1)|t}}$ as $\big[\text{Corr}(Z_{u_i}, Z_{u_1} \mid t), \cdots, \text{Corr}(Z_{u_i}, Z_{u_{(i-1)}} \mid t) \big]$, which can be written in terms of the covariance elements $\sigma_{ij}$ of $\Sigma_{u|t}$ as $c_{u_iu_{1:(i-1)|t}} = \big[\frac{\sigma_{i1}}{\sqrt{\sigma_{ii}\sigma_{11}}}, \cdots,  \frac{\sigma_{i(i-1)}}{\sqrt{\sigma_{ii}\sigma_{(i-1)(i-1)}}} \big]$ with $\sigma_{ij}$ representing the covariance between $U_i$ and $U_j$ given $t$. With these definitions, the elements of the vector $c_{yu_{1:m}|t}$ can be expressed as 
\begin{align}
    c_{yu_1|t} =& \rho_{1}, \notag\\
    c_{yu_i|t} =& \rho_i 
    \sqrt{(1-c_{yu_{1:(i-1)}|t}^\top  C_{u_{1:(i-1)}|t}^{-1}c_{yu_{1:(i-1)}|t})(1 - c_{u_iu_{1:(i-1)|t}}^\top  C_{u_{1:(i-1)}|t}^{-1} c_{u_iu_{1:(i-1)|t}})}  \label{eqn:c_yui_rho} \\
    &+ 
    c_{yu_{1:(i-1)}|t}^\top  C_{u_{1:(i-1)}|t}^{-1}c_{u_iu_{1:(i-1)|t}}) \text{\hspace{4pt} for \hspace{4pt}} i = 2, \cdots, m.
    \notag
\end{align}
Note that, in the above, the only unidentifiable parameters are $\rho_i$'s, while all other terms can be fully determined by $\Sigma_{u|t}$.

\noindent \textbf{2) Structural equation model }

\noindent Now, we prove that the structural Equations \ref{eqn:u_mid_t} - \ref{eqn:y_ytilde_relationship_gauss} is implied by the previous assumptions and Gaussian copula structure.
First, note that Equation \ref{eqn:u_mid_t} is a direct consequence of Assumption \ref{asm:gaussian_latent}. Obviously, $Z_Y$ and $U$ are multivariate normal, and since $Y$ is a simply a transformation of $Z_Y$ then by definition $Y$ and $U$ jointly have a Gaussian copula. Equation \ref{eqn:u_mid_t} determines the identifiable part of the correlation matrix in Equation \ref{eqn:copula_corrmat}, specifically $D_{u_{1:m}|t}^{-1/2} \Sigma_{u_{1:m}|t} D_{u_{1:m}|t}^{-1/2}$. 
The correlations between $Z_y$ and $[Z_{u_1}, \cdots, Z_{u_m}]$ conditional on $T = t$, determine the Y-U subcopula, where $Cor(Z_Y, U_1, ..., U_m) = \gamma_t^\top  \Sigma_{u|t} D_{u_{1:m}|t}^{-1/2}$.  By equating  $c_{yu_{1:m}|t}$ from the correlation matrix in \ref{eqn:copula_corrmat} to $\gamma_t^\top  \Sigma_{u|t} D_{u_{1:m}|t}^{-1/2}$, we can solve for $\gamma_t$ as a function of the copula paramters $\{  
\rho_i \}_{i = 1}^m$:
\begin{equation} 
    \gamma_t^\top = c_{yu_{1:m}|t}D_{u_{1:m}|t}^{1/2}\Sigma_{u|t}^{-1}.
\end{equation}
}

\subsection{Proof of Theorem \ref{prop:non_gaussian_bound}}
\label{sec:prop_non_gaussian_bound}
\noindent\textbf{Theorem \ref{prop:non_gaussian_bound}}
\begin{itshape}
Assume model  \ref{eqn:u_mid_t} - \ref{eqn:y_ytilde_relationship_gauss}:
 \begin{align*}
    U &= \mu_{u\mid t} + \Sigma_{u\mid t}^{1/2}\epsilon_U \\
    Z_Y &= \gamma_t^\top U + \sqrt{1 - \gamma_t^\top \Sigma_{u\mid t}\gamma_t}\epsilon_{Z_Y} \\
     Y &= F^{-1}_{Y|T}(\Phi(Z_Y -\gamma_t^\top \mu_{u\mid t}))
\end{align*}
and that $\sigma_{y|t}$, $\Sigma_{u|t}$, and $\gamma_t$ can vary with $t$ and assume $\mu_{u|t_1}$ and $\mu_{u|t_2}$ are in the row space of $\Sigma_{u|t_1}$ and $\Sigma_{u|t_2}$ respectively, and $Y$ is continuous.  Then the omitted variable bias for all quantile treatment effects are bounded. If $U_i$ is marginally symmetric for all $i=1, \cdots, m$ then the median treatment effect, $MTE_{t_1, t_2} = \text{med}\big(Y \mid do(t_1)\big) - \text{med}\big(Y \mid do(t_2)\big)$  is in the interval $m_l \leq MTE_{t_1, t_2} \leq m_u$ where
\begin{align}
m_l &= F^{-1}_{Y|T=t_1}(\Phi(-\parallel(\Sigma_{u|t_1}^{\dagger})^{1/2}\mu_{u|t_1}\parallel_2)) - F^{-1}_{Y|T=t_2}(\Phi(\parallel(\Sigma_{u|t_2}^{\dagger})^{1/2}\mu_{u|t_2}\parallel_2))\\
m_u &= F^{-1}_{Y|T=t_1}(\Phi(\parallel(\Sigma_{u|t_1}^{\dagger})^{1/2}\mu_{u|t_1}\parallel_2)) - F^{-1}_{Y|T=t_2}(\Phi(-\parallel(\Sigma_{u|t_2}^{\dagger})^{1/2}\mu_{u|t_2}\parallel_2))
\end{align}
\noindent where $\Sigma_{u\mid t_1}^{\dagger}$ and $\Sigma_{u\mid t_2}^{\dagger}$ denote the pseudo-inverses of $\Sigma_{u \mid t_1}$ and $\Sigma_{u \mid t_2}$ respectively and $m_l$ and $m_u$ are identifiable under Assumptions \ref{asm:latent_identification} and \ref{asm:copula}.\\
\end{itshape}

\proof{
In Equation \ref{eqn:y_ytilde_relationship_gauss},
\begin{equation*}
    Y = F^{-1}_{Y|T}(\Phi(Z_Y -\gamma_t^\top \mu_{u\mid t})),
\end{equation*}
where $F^{-1}_{Y|T}$ and $\Phi$ are both non-decreasing monotone functions, so the composition of these functions is also monotone non-decreasing.  The quantile of a monotone function of a random variable is same as the monotone function of the quantile of the random variable.  As such, we can first consider the quantiles in the space of $Z_Y$, i.e., $(Z_Y - \gamma_t^\top  \mu_{u|t}) = \gamma_t^\top (U - \mu_{u|t}) + \sqrt{1 - \gamma_t^\top \Sigma_{u\mid t}\gamma_t}\epsilon_{Z_Y}$, then the observed and intervention means are
\begin{align}
    E\big[ Z_Y - \gamma_t^\top  \mu_{u|t} \mid T = t \big] &= 0\\
    E\big[ Z_Y - \gamma_t^\top  \mu_{u|t} \mid do(T = t)\big] & = - \gamma_t^\top \mu_{u|t}
\end{align}
To show that the omitted variable bias for all
quantile treatment effects are bounded, it is suffice to show that $\text{E}\big[Z_Y - \gamma_t^\top  \mu_{u|t} \mid do(T = t)\big]$ is bounded when $\mu_{u|t}$ is in the row space of $\Sigma_{u|t}$. 

Let $W:= \Sigma_{u|t}^{1/2} \gamma_t$, with $W^\top W \leq 1$ as a consequence of the constraint on $\gamma_t$. Then, we have 
\begin{align}
    \parallel \text{E}\big( (Z_Y - \gamma_t^\top  \mu_{u|t}) \mid do(T = t)\big) \parallel_2^2 &= \parallel \gamma_t^\top \mu_{u|t} \parallel_2^2\\
    & = W^\top (\Sigma_{u|t}^{\dagger})^{1/2}\mu_{u|t} \mu_{u|t}^\top  (\Sigma_{u|t}^{\dagger})^{1/2}W\\
    &\leq \parallel (\Sigma_{u|t}^{\dagger})^{1/2}\mu_{u|t} \parallel_2^2. \label{eqn:bound_Ew_dot}
\end{align}
where the bounds are reached when $W$ is colinear with $(\Sigma_{u|t}^{\dagger})^{1/2}\mu_{u|t}$, i.e., ${\gamma}$ is colinear with the $\Sigma_{u|t}^{\dagger}\mu_{u|t}$. 

Suppose that $\Sigma_{u|t}$ has the eigendecomposition,
\begin{equation}
    \Sigma_{u|t} = Q \Lambda Q^\top ,
\end{equation}
where Q is the square $s \times s $ matrix whose $j$th column is the eigenvector $q_j$ of $\Sigma_{u|t}$, and $\Lambda$ is the diagonal matrix whose diagonal elements are the corresponding eigenvalues, $\Lambda_{jj} = \lambda_j$, in descending order. If $\Sigma_{u|t}$ is non-invertible and has rank $p$ ($p \leq s$), we have $\lambda_j = 0$ for $j = p+1, \cdots, s$. 
when $\mu_{u|t}$ is in the row space of $\Sigma_{u|t}$, it can be expressed as a linear combination of $q_j$, $\sum_{j = 1}^p a_jq_j$, $a_j \in \mathbb{R}$, then we have $\text{E}\big( (Z_Y - \gamma_t^\top  \mu_{u|t}) \mid do(T = t)\big)$ is bounded as 
\begin{align}
    \parallel \text{E}\big( (Z_Y - \gamma_t^\top  \mu_{u|t}) \mid do(T = t)\big) \parallel_2^2
    &\leq \parallel (\Sigma_{u|t}^{\dagger})^{1/2}\mu_{u|t} \parallel_2^2\\
    & = \parallel Q (\Lambda^{\dagger})^{1/2} Q^\top  \sum_{j = 1}^p a_j q_j  \parallel_2^2,\\
    & = \sum_{i=1}^s(\sum_{j=1}^p a_j \lambda_j^{-\frac{1}{2}} Q_{ij} )^2.
\end{align}

Since the mean of $\text{E}\big( (Z_Y - \gamma_t^\top  \mu_{u|t}) \mid do(T = t)\big)$ is bounded, all quantiles of the intervention density of $Z_Y$ are also bounded.  Since $Y$ is a monotone increasing function of $Z_Y$, we also have and hence any quantile treatment effect is bounded.

Next we consider the special case when $U_i$ is symmetric for all $i = 1, \cdots, m$.  In this case, it is also true that the intervention distribution for $Z_Y - \gamma_t^\top \mu_{u\mid t}$ is symmetric because a linear combination of symmetric random variables is symmetric.  Thus,  $\text{med}\big(  (Z_Y - \gamma_t^\top  \mu_{u|t}) \mid do(T = t) \big) = \text{E}\big(  (Z_Y - \gamma_t^\top  \mu_{u|t}) \mid do(T = t) \big) = -\gamma_t^\top \mu_{u|t}$.

Plugging $t_1$ and $t_2$ into inequality \ref{eqn:bound_Ew_dot}, we have 
\begin{equation}
    - \parallel (\Sigma_{u|t_1}^{\dagger})^{1/2}\mu_{u|t_1} \parallel_2
   \leq 
   \text{med} \big(Z_Y - \gamma_t^\top  \mu_{u|t_1} \mid do(T = t_1) \big)
   \leq 
   \parallel (\Sigma_{u|t_1}^{\dagger})^{1/2}\mu_{u|t_1} \parallel_2,
\end{equation}
\begin{equation}
    - \parallel (\Sigma_{u|t_2}^{\dagger})^{1/2}\mu_{u|t_2} \parallel_2
   \leq 
   \text{med} \big(Z_Y - \gamma_t^\top  \mu_{u|t_2} \mid do(T = t_2) \big)
   \leq 
   \parallel (\Sigma_{u|t_2}^{\dagger})^{1/2}\mu_{u|t_2} \parallel_2.
\end{equation}

\noindent Again since both $F^{-1}_{Y|T}$ and $\Phi$ are monotonously non-decreasing functions, $MTE_{t_1, t_2}$ would reach its smallest when $\text{med} \big(Z_Y - \gamma_t^\top  \mu_{u|t_1} \mid do(T = t_1) \big)$ being its smallest and $\text{med} \big(Z_Y - \gamma_t^\top  \mu_{u|t_2} \mid do(T = t_2) \big)$ being its largest, i.e.,
\begin{equation}
      MTE_{t_1, t_2} \geq m_l:= F^{-1}_{Y|T=t_1}(\Phi(-\parallel (\Sigma_{u|t_1}^{\dagger})^{1/2}\mu_{u|t_1}\parallel_2)) - F^{-1}_{Y|T=t_2}(\Phi(\parallel (\Sigma_{u|t_2}^{\dagger})^{1/2}\mu_{u|t_2}\parallel_2)), 
\end{equation}
Conversely, $MTE_{t_1, t_2}$ would reach its largest when $\text{med}\big(Z_Y - \gamma_t^\top  \mu_{u|t_1} \mid do(T = t_1) \big)$ being its largest and $\text{med}\big(Z_Y - \gamma_t^\top  \mu_{u|t_2} \mid do(T = t_2) \big)$ being its smallest, i.e.,
\begin{equation}
    MTE_{t_1, t_2} \leq m_u:= F^{-1}_{Y|T=t_1}(\Phi(\parallel (\Sigma_{u|t_1}^{\dagger})^{1/2}\mu_{u|t_1}\parallel_2)) - F^{-1}_{Y|T=t_2}(\Phi(-\parallel (\Sigma_{u|t_2}^{\dagger})^{1/2}\mu_{u|t_2}\parallel_2)).
\end{equation}
}

\subsection{Proof of Corollary \ref{thm:general_gaus}}
\label{sec:proof_thm_general_gaus}

\noindent\textbf{Corollary \ref{thm:general_gaus}}
\begin{itshape}
Assume the model in Equations \ref{eqn:u_mid_t} - \ref{eqn:y_ytilde_relationship_gauss}:
 \begin{align*}
    U &= \mu_{u\mid t} + \Sigma_{u\mid t}^{1/2}\epsilon_U \\
    Z_Y &= \gamma_t^\top U + \sqrt{1 - \gamma_t^\top \Sigma_{u\mid t}\gamma_t}\epsilon_{Z_Y} \\
     Y &= F^{-1}_{Y|T}(\Phi(Z_Y -\gamma_t^\top \mu_{u\mid t}))
\end{align*}
where $Y$ is conditionally Gaussian given treatments, and where $\sigma_{y|t}$, $\Sigma_{u|t}$, and $\gamma_t$ can vary with $t$. If $\Sigma_{u|t_1}$ and $\Sigma_{u|t_2}$ are non-invertible, then $\text{Bias}_{t_1, t_2}$ is bounded if and only if $\mu_{u|t_1}$ and $\mu_{u|t_2}$ are in the row space of $\Sigma_{u|t_1}$ and $\Sigma_{u|t_2}$ respectively.  When bounded, 
\begin{equation}
\label{eqn:gaus_homoskedastic_app}
\text{Bias}_{t_1, t_2}^2 \leq 
    \bigg(
     \sigma_{y|t_1}\sqrt{R_{ Y \sim U|t_1}^2} \| (\Sigma_{u|t_1}^{\dagger})^{1/2}\mu_{u|t_1}\|_2 + 
     \sigma_{y|t_2}\sqrt{R_{ Y \sim U|t_2}^2} \| (\Sigma_{u|t_2}^{\dagger})^{1/2}\mu_{u|t_2}\|_2 
    \bigg)^2,
\end{equation}
with equality when $\gamma_{t_1} \propto \Sigma_{u\mid t_1}^{\dagger}\mu_{u \mid t_1}$ and $\gamma_{t_2} \propto \Sigma_{u\mid t_2}^{\dagger}\mu_{u \mid t_1}$ and where $\Sigma_{u\mid t_1}^{\dagger}$ and $\Sigma_{u\mid t_2}^{\dagger}$ are the pseudo-inverses of $\Sigma_{u \mid t_1}$ and $\Sigma_{u \mid t_2}$.  If $\Sigma_{u\mid t_1} = \Sigma_{u\mid t_2} = \Sigma_{u|t}$ and $\gamma_t = \gamma$ is invariant to $t$ (i.e. there are no treatment-confounder interactions), and $\sigma^2_{y\mid t_1} = \sigma^2_{y \mid t_2} = \sigma^2_{y|t}$ (homoskedastic outcome model) then $\text{Bias}_{t_1, t_2}$ is bounded if and only if $\mu_{u|t_1} - \mu_{u|t_2}$ is in the row space of $\Sigma_{u|t}$ and when bounded, 
\begin{equation}
  \text{Bias}_{t_1, t_2}^2 \leq \sigma_{y \mid t}^2 R_{ Y \sim U | T}^2\|(\Sigma_{u|t}^{\dagger})^{1/2}(\mu_{u|t_1} - \mu_{u|t_2}) \|_2^2.
\end{equation}\\

\end{itshape}

\proof{
Under model in Equations \ref{eqn:u_mid_t} - \ref{eqn:y_ytilde_relationship_gauss} with Gaussian outcomes, and $\sigma_{y|t}$, $\Sigma_{u|t}$, $\gamma_t$ varying with $t$, we have
\begin{equation}
    \text{PATE}_{t_1, t_2} = (\mu_{y|t_1} - \mu_{y|t_2}) -(\sigma_{y \mid t_1}\gamma_{t_1}^\top  \mu_{u|t_1} - \sigma_{y \mid t_2}\gamma_{t_2}^\top  \mu_{u|t_1}),
\end{equation}
and 
\begin{equation}
    \text{Bias}_{t_1, t_2} = \sigma_{y \mid t_1}\gamma_{t_1}^\top  \mu_{u|t_1} -\sigma_{y \mid t_2} \gamma_{t_2}^\top  \mu_{u|t_1}
\end{equation}
with sensitivity parameter $\gamma_{t_1}$ and $\gamma_{t_2}$ satisfying constraints $\gamma_{t_1}^\top  \Sigma_{u|t_1} \gamma_{t_1} \leq R^2_{Y \sim U \mid t_1}$ and $\gamma_{t_2}^\top  \Sigma_{u|t_2} \gamma_{t_2} \leq R^2_{Y \sim U \mid t_2}$ respectively, since $\gamma_{t_1}^\top  \Sigma_{u|t_1} \gamma_{t_1}$ and $\gamma_{t_2}^\top  \Sigma_{u|t_2} \gamma_{t_2}$ correspond to the confounding variations and should not be larger than a given level of $R^2_{Y \sim U \mid T}$.

Let
\begin{align}
    W_1 := \Sigma_{u|t_1}^{1/2}\gamma_{t_1},\\
    W_2 := \Sigma_{u|t_2}^{1/2}\gamma_{t_2},
\end{align}
then the ommitted variable bias can be written as
\begin{equation}
    \text{Bias}_{t_1, t_2} = \sigma_{y \mid t_1}W_1^\top  (\Sigma_{u|t_1}^{\dagger})^{1/2}\mu_{u|t_1} - \sigma_{y \mid t_2} W_2^\top  (\Sigma_{u|t_2}^{\dagger})^{1/2}\mu_{u|t_2}
\end{equation}
with $W_1^\top  W_1 \leq R_{ Y \sim U|t_1}^2$ and $W_2^\top  W_2 \leq R_{ Y \sim U|t_2}^2$.
Then, 
\begin{align}
    \mid \text{Bias}_{t_1, t_2}\mid  &\leq \sigma_{y \mid t_1}
    \mid W_1^\top  (\Sigma_{u|t_1}^{\dagger})^{1/2}\mu_{u|t_1} \mid  + 
    \sigma_{y \mid t_2} \mid W_2^\top  (\Sigma_{u|t_2}^{\dagger})^{1/2}\mu_{u|t_2} \mid \\
    & \leq
    \sigma_{y|t_1}\sqrt{ R_{ Y \sim U|t_1}^2}
    \parallel (\Sigma_{u|t_1}^{\dagger})^{1/2}\mu_{u|t_1}\parallel_2 + 
    \sigma_{y \mid t_2}\sqrt{ R_{ Y \sim U|t_2}^2}\parallel(\Sigma_{u|t_2}^{\dagger})^{1/2}\mu_{u|t_2}\parallel_2 \label{eqn:biast1t2_hetero_bound}
\end{align}
where the bounds are reached when $W_1$ and $W_2$ are respectively colinear with $(\Sigma_{u|t_1}^{\dagger})^{1/2}   \mu_{u|t_1}$ and $(\Sigma_{u|t_2}^{\dagger})^{1/2} \allowbreak \mu_{u|t_2}$, i.e., ${\gamma_{t_1}}$ and $\gamma_{t_2}$ are respectively colinear with the $\Sigma_{u|t_1}^{\dagger} \allowbreak \mu_{u|t_1}$ and $\Sigma_{u|t_2}^{\dagger} \allowbreak \mu_{u|t_2}$. 
}

To show that $\text{Bias}_{t_1, t_2}$ is bounded if and only if $\mu_{u|t_1}$ and $\mu_{u|t_2}$ are in the row space of $\Sigma_{u|t_1}$ and $\Sigma_{u|t_2}$ respectively, we can first show that the first term in Equation \ref{eqn:biast1t2_hetero_bound}, $\sigma_{y|t_1}\sqrt{ R_{ Y \sim U|t_1}^2}\parallel (\Sigma_{u|t_1}^{\dagger})^{1/2}\mu_{u|t_1}\parallel_2$, is bounded if and only if $\mu_{u|t_1}$ is in the row space of $\Sigma_{u|t_1}$.
Suppose that $\Sigma_{u|t_1}$ has the eigendecomposition,
\begin{equation}
    \Sigma_{u|t_1} = Q \Lambda Q^\top , \label{eqn:eigenD_Sigma_u_t}
\end{equation}
where Q is the square $s \times s $ matrix whose $j$th column is the eigenvector $q_j$ of $\Sigma_{u|t_1}$, and $\Lambda$ is the diagonal matrix whose diagonal elements are the corresponding eigenvalues, $\Lambda_{jj} = \lambda_j$, in descending order. If $\Sigma_{u|t_1}$ is non-invertible and has rank $p$ ($p \leq s$), we have $\lambda_j = 0$ for $j = p+1, \cdots, s$. 

On the one hand, when $\mu_{u|t_1}$ is in the row space of $\Sigma_{u|t_1}$, it can be expressed as a linear combination of $q_j$, $\sum_{j = 1}^p a_jq_j$, $a_j \in \mathbb{R}$. Then, we have the squared first term in Equation \ref{eqn:biast1t2_hetero_bound} as
\begin{align}
    &\sigma_{y|t_1}^2 R_{ Y \sim U|t_1}^2\parallel (\Sigma_{u|t_1}^{\dagger})^{1/2}\mu_{u|t_1}\parallel_2^2, \\
    = &\sigma_{y|t_1}^2R_{ Y \sim U|t_1}^2 
        \parallel Q (\Lambda^{\dagger})^{1/2} Q^\top  \sum_{j = 1}^p a_j q_j  \parallel_2^2,\\
    = &\sigma_{y|t_1}^2R_{ Y \sim U|t_1}^2 
    \sum_{i=1}^s(\sum_{j=1}^p a_j \lambda_j^{-\frac{1}{2}} Q_{ij} )^2,
\end{align}
where $Q_{ij}$ denotes the element at the $i$th row and $j$th column of matrix $Q$, and $\Lambda^{\dagger}$ is the pseudo-inverse of $\Lambda$ by taking the reciprocal of each its non-zero element on the diagonal, leaving the zeros in place.

On the other hand, when $\sigma_{y|t_1}\sqrt{ R_{ Y \sim U|t_1}^2}\parallel (\Sigma_{u|t_1}^{\dagger})^{1/2}\mu_{u|t_1}\parallel_2$ is bounded, let's assume that $\mu_{u|t_1}$ is not in the row space of $\Sigma_{u|t_1}$, say $\mu_{u|t_1} = q_s$. Since $\lambda_s = 0$, $\lambda_s^{-1/2} = \infty$, resulting in $\parallel (\Sigma_{u|t_1}^{\dagger})^{1/2}\mu_{u|t_1}\parallel_2$ equaling $\infty$, which contradicts the condition that $\sigma_{y|t_1}\sqrt{ R_{ Y \sim U|t_1}^2}\parallel (\Sigma_{u|t_1}^{\dagger})^{1/2}\mu_{u|t_1}\parallel_2$ is bounded.

Similarly, we can show that the second term in Equation \ref{eqn:biast1t2_hetero_bound}, $\sigma_{y \mid t_2}\sqrt{ R_{ Y \sim U|t_2}^2} \parallel(\Sigma_{u|t_2}^{\dagger})^{1/2}$ \ $\mu_{u|t_2}\parallel_2$, is bounded if and only if $\mu_{u|t_2}$ is in the row space of $\Sigma_{u|t_2}$ by expressing $\Sigma_{u|t_2}$ in its eigendecomposition. Altogether, we can see that $\text{Bias}_{t_1, t_2}$ is bounded if and only if $\mu_{u|t_1}$ and $\mu_{u|t_2}$ are in the row space of $\Sigma_{u|t_1}$ and $\Sigma_{u|t_2}$ respectively.

Lastly,  let's consider the case when $\Sigma_{u|t_1} = \Sigma_{u|t_2}$ and $\gamma_t = \gamma$, the constraints on the sensitivity parameters can be unified as
\begin{equation}
\label{eqn:ellipse_gamma_general}
    \gamma^\top \Sigma_{u \mid t}\gamma \leq R_{ Y \sim U \mid T}^2
\end{equation}
where $R_{ Y \sim U \mid T}^2$ denotes the fraction of confounding variation in residual variance of $Y$ conditional on $T$ \footnote{$R_{ Y \sim U \mid T}^2$ coincides with $R_{ Y \sim U}^2$ here, but we use notation $R_{ Y \sim U \mid T}^2$ for consistency. }.

Let
\begin{equation}
    W := \Sigma_{u|t}^{1/2} \gamma, 
\end{equation}
then the omitted variable bias can be written as 
\begin{equation}
    \text{Bias}_{t_1, t_2} =  \sigma_{y \mid t} W^\top  (\Sigma_{u|t}^{\dagger})^{1/2} (\mu_{u|t_1} - \mu_{u|t_2}),   
\end{equation}
where $W^\top  W \leq R_{ Y \sim U|T}^2$, implied by inequality in Equation \ref{eqn:ellipse_gamma_general}.

Therefore,
\begin{align}
    \text{Bias}_{t_1, t_2}^2  &= \sigma_{y \mid t}^2 W^\top (\Sigma_{u|t}^{\dagger})^{1/2}(\mu_{u|t_1} - \mu_{u|t_2}) 
    (\mu_{u|t_1} - \mu_{u|t_2})^\top  (\Sigma_{u|t}^{\dagger})^{1/2} W \label{eqn:biast1t2_homo} \\
    & \leq \sigma_{y|t}^2 R_{ Y \sim U|T}^2 \parallel (\Sigma_{u|t}^{\dagger})^{1/2}(\mu_{u|t_1} - \mu_{u|t_2}) \parallel_2^2,
\end{align}
where the bounds are reached when $W$ is colinear with $(\Sigma_{u|t}^{\dagger})^{1/2}(\mu_{u|t_1} - \mu_{u|t_2})$, i.e., ${\gamma}$ is colinear with the $\Sigma_{u|t}^{\dagger}(\mu_{u|t_1} - \mu_{u|t_2})$. 

Similarly as above in showing that $\sigma_{y|t_1}\sqrt{ R_{ Y \sim U|t_1}^2}\parallel (\Sigma_{u|t_1}^{\dagger})^{1/2}\mu_{u|t_1}\parallel_2$ is bounded if and only if $\mu_{u|t_1}$ is in the row space of $\Sigma_{u|t_1}$, we can show that $\text{Bias}_{t_1, t_2}$ in Equation \ref{eqn:biast1t2_homo} is bounded if and only if $\mu_{u|t_1} - \mu_{u|t_2}$ is in the row space of $\Sigma_{u|t}$. We leave out the details here to avoid redundancy.

\section{Causal Equivalence}
\label{sec:causal_equiv}
\subsection{Multiple Treatments and Causal Equivalence Classes}

In this section, we formally clarify why the latent variable identification is only required up to invertible linear transformations (Assumption \ref{asm:latent_identification}.  We say that $\psi_T$ is identified up to a \emph{causal equivalence class} when, for any value of $\psi_T$ that is compatible with the observed data distribution, the set of possible causal effects, as indexed by $\psi_Y$, is invariant to the particular value of $\psi_T$ in the equivalence class.  

\begin{Def}[Causal equivalence class]
$[\psi_T]$ is a causal equivalence class of $\psi_T$ if and only if for any $\tilde \psi_T$ in $[\psi_T]$, then, for every $\psi_Y$ there exists a $\tilde \psi_{Y}$ such that $f_{\psi_Y, \psi_T}(y \mid do(T = t)) = f_{\tilde \psi_{Y}, \tilde \psi_{T}}(y \mid do(T = t))$ for all $y, t$.
\end{Def}

For the purposes of sensitivity analysis, when $\psi_T$ is identified up to a causal equivalence class, we can assume that $\psi_T$ is point-identified at a particular value within the class $[\psi_T]$ without loss of generality. Crucially, the copula-based formulation enables valid sensitivity analysis without observable implications, even in these cases where $\psi_T$ is restricted by the observed data. In this case, the outcome-confounder copula $c_{\psi_Y}$ remains the lone degree of freedom in the sensitivity model.
As we will show, this restriction can induce qualitatively different sensitivity regions compared two cases where $\psi_T$ is unrestricted.
For example, sensitivity regions can be bounded, even without additional restrictions on $\psi_Y$.

We consider causal equivalence under the Gaussian copula model in Equations \ref{eqn:u_mid_t} - \ref{eqn:y_ytilde_relationship_gauss}:
 \begin{align*}
    U &= \mu_{u\mid t} + \Sigma_{u\mid t}^{1/2}\epsilon_U \\
    Z_Y &= \gamma_t^\top U + \sqrt{1 - \gamma_t^\top \Sigma_{u\mid t}\gamma_t}\epsilon_{Z_Y} \\
     Y &= F^{-1}_{Y|T}(\Phi(Z_Y -\gamma_t^\top \mu_{u\mid t}))
\end{align*}
The Gaussian copula under this model is fully determined by the covariance matrix
\begin{equation} 
    \text{Cov}([Z_Y, U] \mid T=t) = 
    \begin{bmatrix}
        1 & \gamma_t^\top \Sigma_{u|t}\\
    \ \Sigma_{u|t}\gamma_t & \ \Sigma_{u|t}
    \end{bmatrix}
\end{equation}
with parameters $\psi_T = \{\mu_{u\mid t}, \Sigma_{u\mid t} : t \in \mathcal{T}\}$ and $\psi_Y = \{\gamma_t : \mathcal{T}\}$ where $\mathcal{T}$ is the space of all treatments.  Given Assumption \ref{asm:latent_identification}, $\psi_Y = \{\gamma_t\}$ is the sole $m-$dimensional sensitivity vector governing the magnitude of the omitted variable bias. The following theorem establishes that the class of $\psi_T$ defined by all invertible linear transformations of $U$ is a causal equivalence class.  

\begin{thm}
\label{thm:equivalence_general}
Assume model in Equations \ref{eqn:u_mid_t} - \ref{eqn:y_ytilde_relationship_gauss}. Let $[\psi_T] = \{ \tilde \psi_T = \{A\mu_{u\mid t}, A\Sigma_{u\mid t}A : t \in \mathcal{T}\} : A \in \mathcal{S}^+\}$ where $\mathcal{S}^+$ is the space of symmetric positive definite matrices.  Then $[\psi_T]$ is a causal equivalence class.
\end{thm}

\proof{
The intervention distribution for $Z_Y$ is defined as 
\begin{equation}
    f_{\psi}(z_y \mid do(t)) = \int \left[\int f_{\psi_Y}(z_y \mid t, u) f_{\psi_T}(u \mid \tilde t) du \right] f(\tilde t) d\tilde t
\end{equation}
where $\psi_Y = \gamma$ and $\psi_T = \{\mu_{u\mid t}, \Sigma_{u\mid t}\}$.  Then,  $\int f_{\gamma_t}(z_y \mid t, u) f_{\psi_T}(u \mid \tilde t) du \sim N(\gamma_t^\top \mu_{u|\tilde t}, \ 1)$ for any $\gamma_t$ such that $\gamma_t^\top  \Sigma_{u|t}\gamma_t \le 1$ (see Equation \ref{eqn_u_integrated}).  Let $\tilde \psi_T = \{A\mu_{u\mid t}, A\Sigma_{u\mid t}A\} \in [\psi_T]$ where $A \in \mathcal{S}^+$ is a positive definite matrix and assume $\tilde \psi_Y = \tilde \gamma_t$.  Then,
\begin{equation}
\int f_{\tilde \gamma_t}(z_y \mid t, u) f_{\tilde \psi_T}(u \mid \tilde t) du \sim N(\tilde \gamma_t^\top  A\mu_{u|\tilde t}, \ 1).
\end{equation}
Let $\tilde \gamma_t = A^{-1} \gamma_t$ be a bijective mapping from $\gamma_t$ to $\tilde \gamma_t$.  For any $\gamma_t$ and positive definite $A$, we have $\tilde \gamma_t^\top  A \Sigma_{u|t}A \tilde \gamma_t = \gamma_t^\top \Sigma_{u|t}\gamma_t \le 1$ so that $\tilde \gamma_t$ is a valid copula parameter.  In addition, $\int f_{\gamma_t}(z_y \mid t, u) f_{\psi_T}(u \mid \tilde t) du = \int f_{\tilde \gamma_t}(z_y \mid t, u) f_{\tilde \psi_T}(u \mid \tilde t) du$, which implies $f_{\gamma_t, \psi_T}(z_y \mid do(t)) = f_{\tilde \gamma_t, \tilde \psi_T}(z_y \mid do(t))$.  Since $Y$ is a deterministic function of $Z_Y $, this implies $f_{\gamma_t, \psi_T}(y \mid do(t)) = f_{\tilde \gamma_t, \tilde \psi_T}(y \mid do(t))$. Therefore, $[\psi_T]$ is a causal equivalence class.\\
}

\noindent The gist of the proof is that for any invertible linear transformation, $A$, of $U$, the copula parameterized by $\tilde \gamma_t = A^{-1} \gamma_t$ yields equivalent causal effects in the reparameterized coordinates of $U$ as $\gamma$ does in the original confounder coordinates.  See \citet{miao2020identifying} for more general theory about causal equivalence in larger class of latent variable models.

\section{Modeling Choice Details}
\subsection{Identification and Inference in the Factor Model}
\label{sec:factor_identification}
Here, we briefly elaborate on identifiability of the probabilistic principal components model, which is a prerequisite for our multi-cause sensitivity analysis.  Identifiability under various factor model assumptions is well studied and has a long history in the literature \citep{Mardia1980, everett2013introduction}.  In the specific probabilistic principal components model  \ref{eqn:ppca}, \citet{tipping1999probabilistic} provide a maximum likelihood solution for inferring the latent confounder parameters conditional on $m$.  Many  procedures are available for selecting the appropriate value of $m$, using for example Bayesian model selection techniques \citep{minka2001automatic} or large $p$, small $n$ asymptotics \citet{Gavish2014}.  


When the observed outcome distribution is non-Gaussian, we cannot necessarily express $\text{PATE}_{t_1, t_2}$ analytically. 
In particular, for non-Gaussian $Y$, when $f(u \mid t_1) \sim f(u \mid t_2)$ the average treatment effect among the $t_1$- and $t_2$-treated units is unconfounded, but the bias of $\text{PATE}_{t_1, t_2}$ may be nonzero since $f(u \mid t) \nsim f(u)$.  The causal effects, however, can still be  calculated using Algorithm \ref{algm:general}.

In the following, we would like to further elucidates
important situations in which we cannot bound the omitted variable bias due to non-identifiability of the factor model.  Again, we focus on the rotated treatments $\tilde T \sim N(0, \Delta + \sigma_{t}^2I)$ and highlight two simple situations in which we cannot bound the the causal effects.  First, when $B$ is rank $k$, i.e. there exist $m=k$ independent confounders,  $\Delta$ has no non-zero entries on the diagonal and thus we cannot identify either $\Delta$ nor $\text{Cov}(\epsilon_{t}) = \sigma_{t}^2 I_k$, only their sum.  Second, if $\text{Cov}(\epsilon_{t})$ is an unknown arbitrary diagonal matrix (as opposed to a matrix proportional to the identity), then $\text{Cov}(\epsilon_{t})$ is not distinguishable from $\Delta$.  In both of these cases, the worst-case bias is unbounded since the non-confounding variation of the treatment assignment, $\text{Cov}(\epsilon_{t})$, can be arbitrarily small.  In such settings, we can still apply approaches used in single cause sensitivity analysis, by specifying both $\Psi_Y$ and $\Psi_T$;  when the factor model is not identifiable, $\Psi_T$ must be chosen as a true parameter, e.g. by bounding the fraction of treatment variation due to confounding, $R^2_{T\sim U}$.

\subsection{Confounder Inference with Variational Autoencoders}

\label{sec:lv_inf}

Probabilistic Principal Component Analysis should only be used when the treatments are approximately Gaussian treatments.  For binary and other general treatment distributions, more sophisticated probabilistic latent variables models are required. Examples of such latent variable models include models for count data like the logistic factor analysis \citep{hao2015probabilistic} and Poisson factor analysis methods \citep{gopalan2013scalable}. Unfortunately, these models imply posteriors which are non-Gaussian, violating Assumption \ref{asm:copula}.  

As such, for general treatment distributions, our approach is to infer a conditional Gaussian latent variable model using a variational autoencoder (VAE).  VAEs have been extremely popular in machine learning, in particular for generating low dimensional representations of complex inputs like images \citep{pu2016variational} but more recently have been used in scientific and decision-making applications \citep{lopez2020decision} and in applications to causal inference \citep{louizos2017causal}.  A VAE consists of a prior distribution, $f(u)$, typically for the low-dimensional latent variables, a stochastic encoder, and a stochastic decoder. In our application, the inferred stochastic decoder, $\hat f_\theta(t \mid u)$, is a non-linear map from latent confounders to a distribution over causes.  Together, the prior distribution for $u$ and the decoder imply a posterior confounder distribution, $\hat f(u \mid t)$.  

In practice, inference for the true posterior is intractable and so a variational approximation, called the encoder, $q_\phi(u \mid t)$, is used in place of the true posterior.  Typically the encoder is chosen to be a normal distribution with mean and variance which are non-linear functions of the input, $q_\phi = N(\mu_\phi(t), \sigma^2_\phi(t))$.  A crucial question is that how well the Gaussian encoder approximates the true posterior; improving the variational approximation to the true latent variable posterior is an area of active research.  In this work, we follow a common strategy of using the encoder learned by the VAE as the proposal distribution in an importance sampler \citep{lopez2020decision}.

Specifically, we apply a variant of the Constant-Variance Variational Autoencoder (CV-VAE) \citep{ghosh2019variational} to infer the conditional confounder distribution, $f(u \mid t) \sim N(\mu_{u|t}, \Sigma_{u|t})$, in which $\Sigma_{u|t}$ does not depend on the level of $t$.  We use the importance sampling to improve estimates of the conditional mean $\mu_{u\mid t}$, and posterior variance, $\Sigma_{u\mid t}$.  While this approach only yields an approximation to the true posterior, we demonstrate the practical effectiveness of this approach in Section \ref{sec:simulation} and in Appendix \ref{sec:movie}.





\subsection{Binary Outcomes}
\label{sec:binary}

\begin{figure}[ht!]
    \begin{subfigure}{0.24\textwidth}
        \centering
        \includegraphics[width=\textwidth]{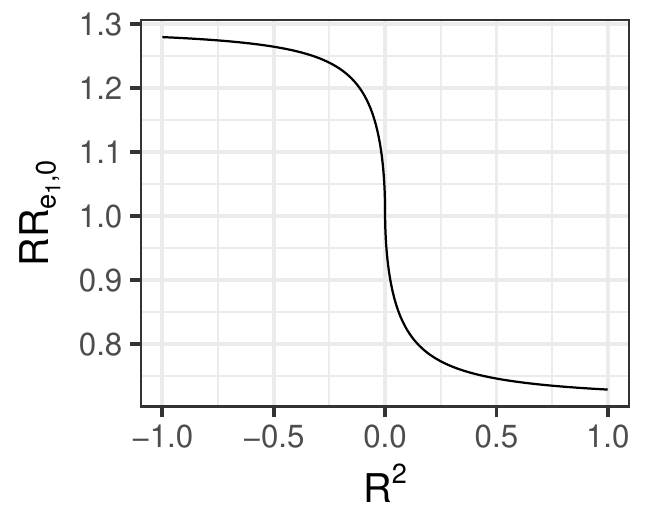}
        \caption{$RR_{e_1,0}$ vs $R^2$}
	\end{subfigure}
	\begin{subfigure}{0.24\textwidth}
        \centering
        \includegraphics[width=\textwidth]{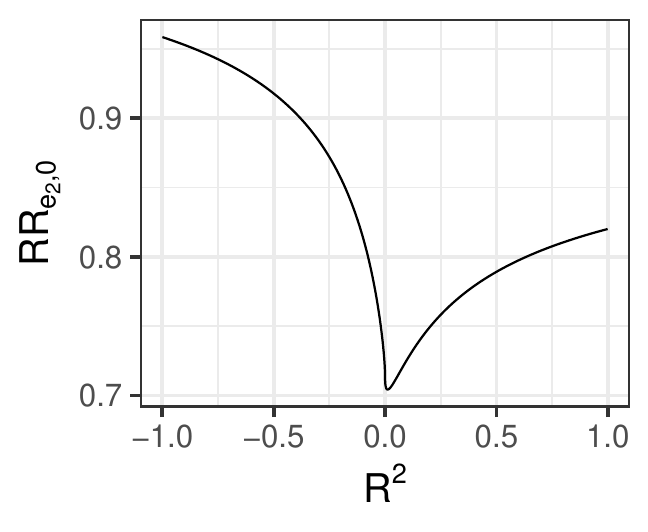}
        \caption{$RR_{e_2,0}$ vs $R^2$}
	\end{subfigure}
	\begin{subfigure}{0.24\textwidth}
        \centering
        \includegraphics[width=\textwidth]{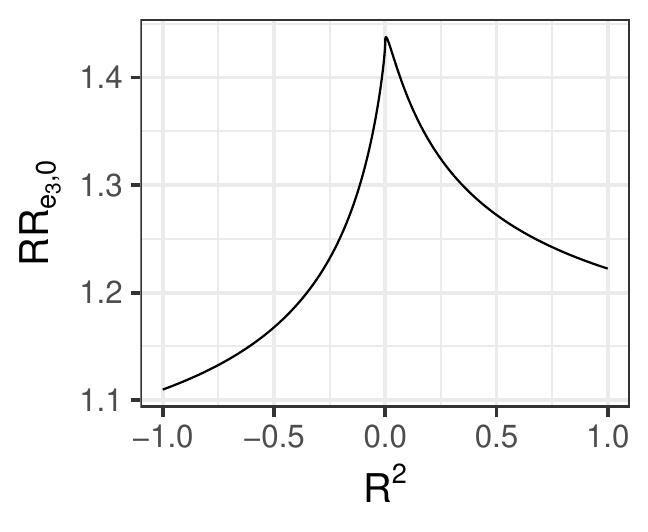}
        \caption{$RR_{e_3,0}$ vs $R^2$}
	\end{subfigure}
	\begin{subfigure}{0.24\textwidth}
        \centering
        \includegraphics[width=\textwidth]{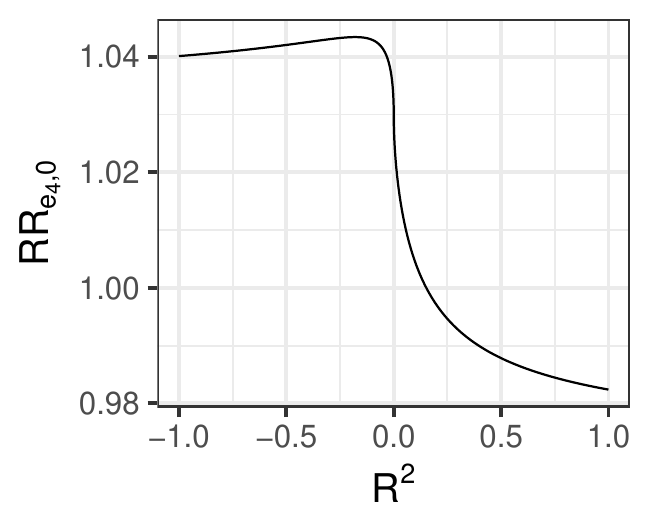}
        \caption{$RR_{e_4,0}$ vs $R^2$}
	\end{subfigure}
		\caption{\label{fig:r2_rr} $RR_{t_1, t_2}$ is non-monotone in $R_{Z_Y \sim U | T}^2$. Positive values of $R^2$ indicates that $U$ is positively correlated with $Z_Y$, and negative values of $R^2$ means that $U$ is negatively correlated with $Z_Y$. }
\label{fig:binary-non_monotone}		
\end{figure}

Under model in Equations \ref{eqn:u_mid_t} - \ref{eqn:y_ytilde_relationship_gauss}, the binary outcomes with the risk ratio estimand:
\begin{equation} \label{eqn:rr_t}
    \begin{split}
    RR_{t, \sbullet}
    &= \sum_{t_i \in \mathcal{T}} \Phi \left( \Phi^{-1}(\mu_{y|t})  + \gamma^\top  (\mu_{u|t_i} - \mu_{u|t})
    \right)
    \bigg/ Pr(Y=1),
    \end{split}
\end{equation}
%
\noindent which implies that
%
\begin{equation} 
\label{eqn:rr_t1t2}
    \begin{split}
    RR_{t_1, t_2} \!
    &= \! \sum_{t_i \in \mathcal{T}} \! \Phi \! \left( \Phi^{-1}(\mu_{y|t_1}) \! + \! \gamma^\top \! (\mu_{u|t_i} - \mu_{u|t_1})
    \right)  \! \bigg/ \!
    \sum_{t_i \in \mathcal{T}} \! \Phi \! \left( \Phi^{-1}(\mu_{y|t_2})  \! + \! \gamma^\top  \! (\mu_{u|t_i} - \mu_{u|t_2})
    \right),
    \end{split}
\end{equation}
where $\sigma_{z_y |t}^2$ is the conditional variance of the transformed outcome $Z_Y$ given $T$, and $R_{Z_Y \sim U | T}^2$ denotes the partial R-squared of potential unobserved confounders within the $Z_Y$ space, constrained by $\gamma^\top  \Sigma_{u|t} \gamma \leq \sigma_{z_y |t}^2 R_{Z_Y \sim U | T}^2$. We can numerically explore values of $RR_{t_1, t_2}$ within the valid domain of $\gamma$, and calculate the corresponding implicit partial R-squared by $R_{Z_Y \sim U | T}^2 = \frac{\gamma^\top  \Sigma_{u|t} \gamma}{\sigma_{z_y |t}^2}$.  To calculate the robustness value, we only need to find the value of $R_{Z_Y \sim U | T}^2$ for which the corresponding $RR_{t_1, t_2} = 1$. Noticeably, $RR_{t_1, t_2}$ is not monotone in $R_{Z_Y \sim U | T}^2$, since the variance of intervention distribution also depends on $\gamma$. This is evident in the simulation in Section \ref{sec:simulation} where we fit the observed outcome model by probit regression and the valid range for scalar $\gamma$ is $[- \frac{1}{\sigma_{u|t}}, \frac{1}{\sigma_{u|t}}]$. We visualize the non-monotone relationship between $RR_{t_1, t_2}$ and $R_{Z_Y \sim U | T}^2$ in Figure \ref{fig:binary-non_monotone}.

\pagebreak

\section{Additional Simulation Results}

\subsection{Robustness of the Gaussian Copula Assumption}
\label{sec:copula_misspecification}
Here, we explore how alternative copula specifications affect the results.  We use the observed data generating process defined in Section \ref{sec:sim_nongaussian}.  

We explore three classes of copulas:  Clayton copulas, a quadratic dependence copula and a Gaussian mixture copula.  The Clayton copula is one of the well-studied Archimedian copulas \citep{hofert2008sampling}.  For the quadratic dependence, we consider the copula implied by the relationship $f(y \mid t) = F_{Y|T}^{-1}(a(u-b)^2)$.  Finally, we consider a Gaussian mixture (G.M.) based copula, $f(y \mid t) = F_{Y|T}^{-1}(h_\epsilon(u))$ where $h_\epsilon(u) = a_1 u + b_1$ with probability 1/2 and $h_\epsilon(u) = a_2 u + b_2$ with probability 1/2. Examples of these copula densities are presented in Figure \ref{fig:cop_densities}.  

\begin{figure}[ht!]
        \centering
        \includegraphics[width=\textwidth]{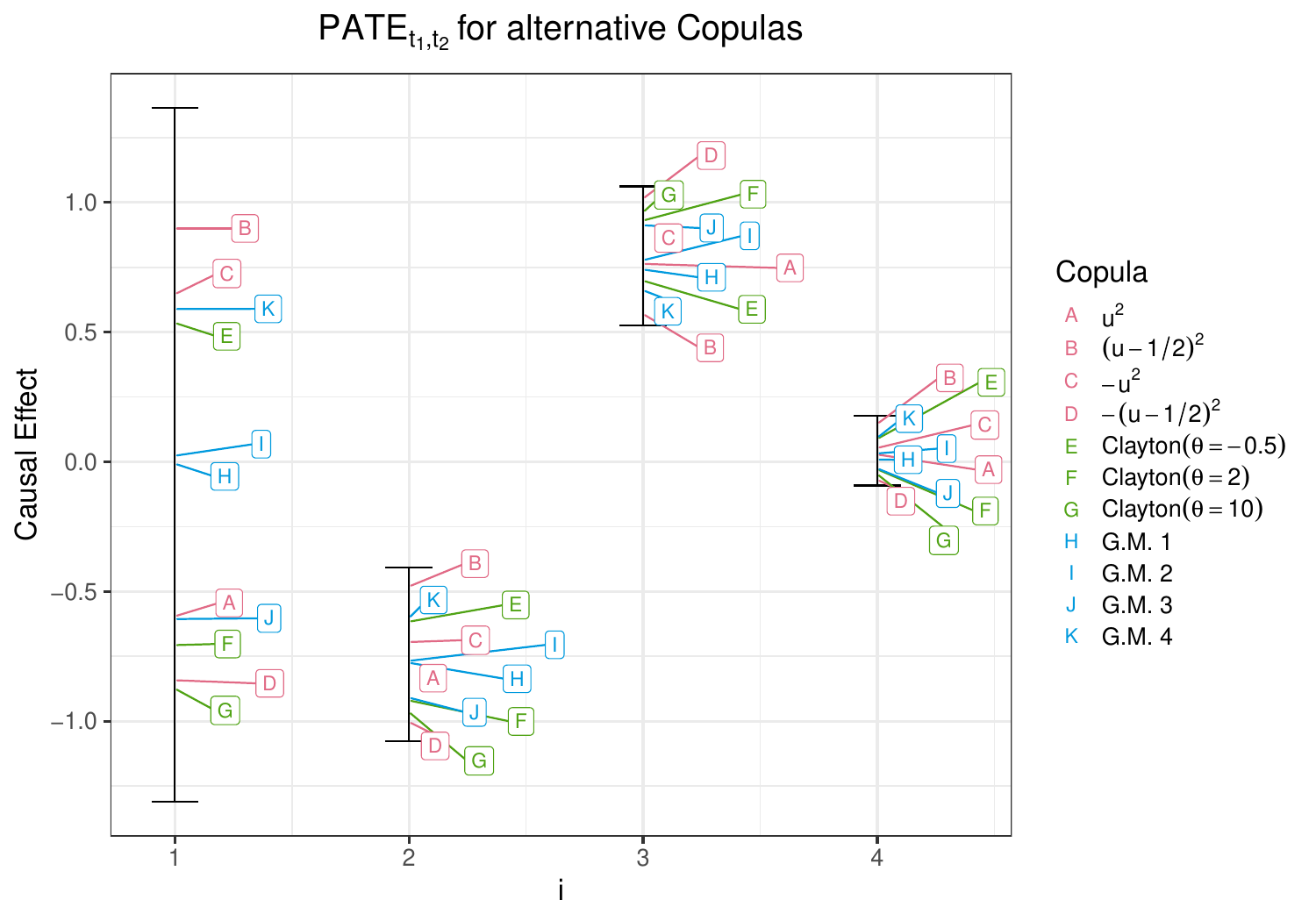}
		\caption{Robustness to alternative copulas.  Black intervals are the ignorance regions for the Gaussian copula ($R^2_{Y \sim U | T} = 1$).  We consider quadratic non-monotone copulas (pink), Clayton copulas (green) \citep{hofert2008sampling}, and four Gaussian mixture-based copulas. All alternative specifications lie within the bounds defined by the Gaussian copula. \label{fig:cop_misspec}}
\end{figure}

\begin{figure}[ht!]
        \centering
        \includegraphics[width=\textwidth]{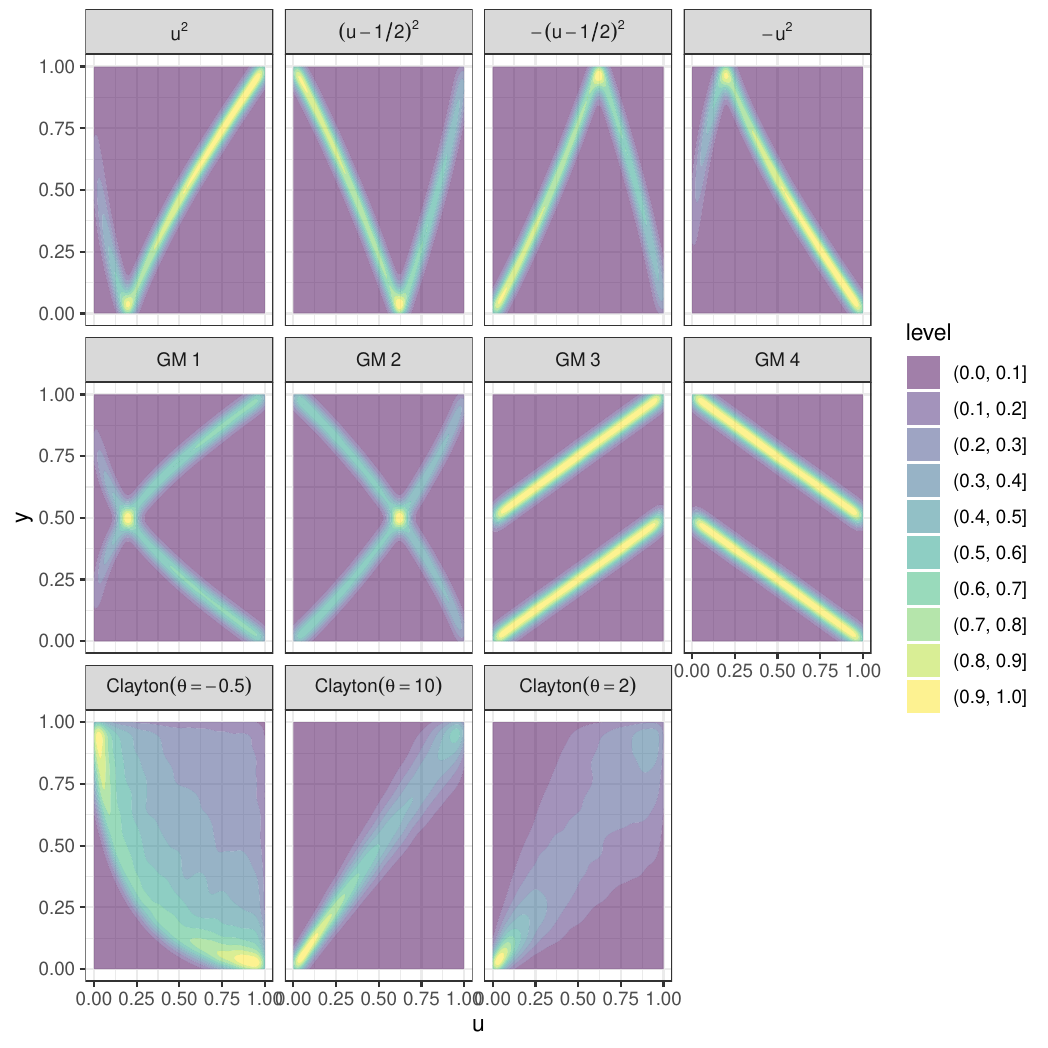}
		\caption{Conditional copula densities associated with effects in Figure \ref{fig:cop_misspec} for $t=(0,0,0,0)$ \label{fig:cop_densities}}
\end{figure}

Most importantly, for all of the alternative copulas we explored, the PATE was inside the bounds implied by the Gaussian copula (Figure \ref{fig:cop_misspec}).  This suggest that the Gaussian bounds are quite robust to a range of different types of dependence.

\clearpage

\subsection{Additional Results from Simulation in Sparse Effects Setting}
\begin{figure}[ht!]
        \centering
        \includegraphics[width=\textwidth]{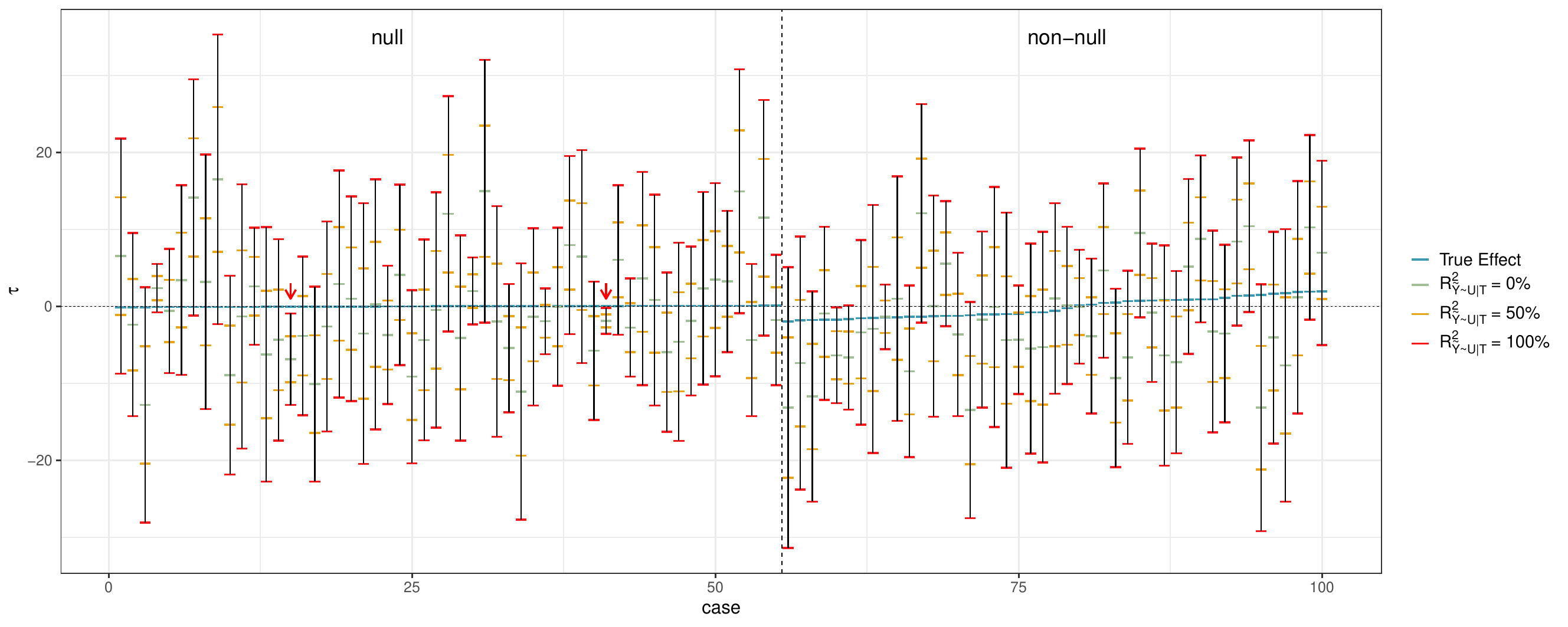}
		\caption{Worst-case ignorance regions for 55 randomly chosen null effects (left) and all 
		45 non-null effects (right) ordered by the magnitude of true effects in each group.  Two red arrows indicate non-null treatments for which the worst-case ignorance region does not cover the true effect.  This appears to be due to estimation error in the outcome model, more so than with the VAE. \label{fig:gwas_worstcase}}

\end{figure}

\begin{figure}[ht!]
    \begin{subfigure}[t]{0.48\textwidth}
        \centering
        \includegraphics[width=\textwidth]{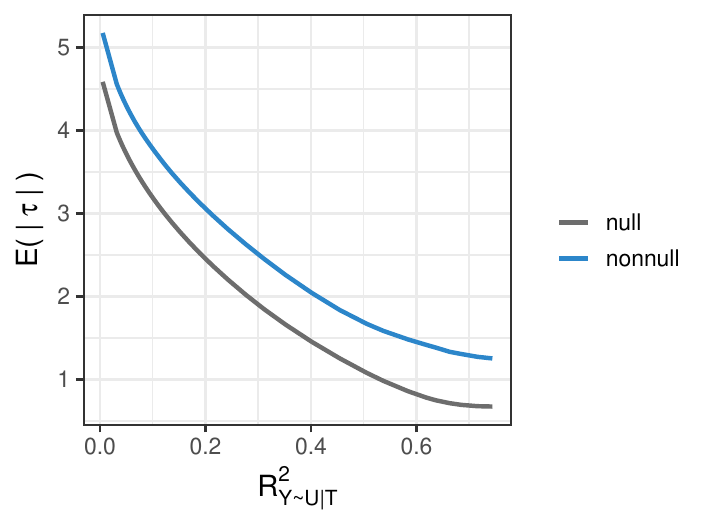}
        \caption{ \label{fig:r2_norm}}
	\end{subfigure}
	\quad
	\begin{subfigure}[t]{0.4\textwidth}
        \centering
        \includegraphics[width=\textwidth]{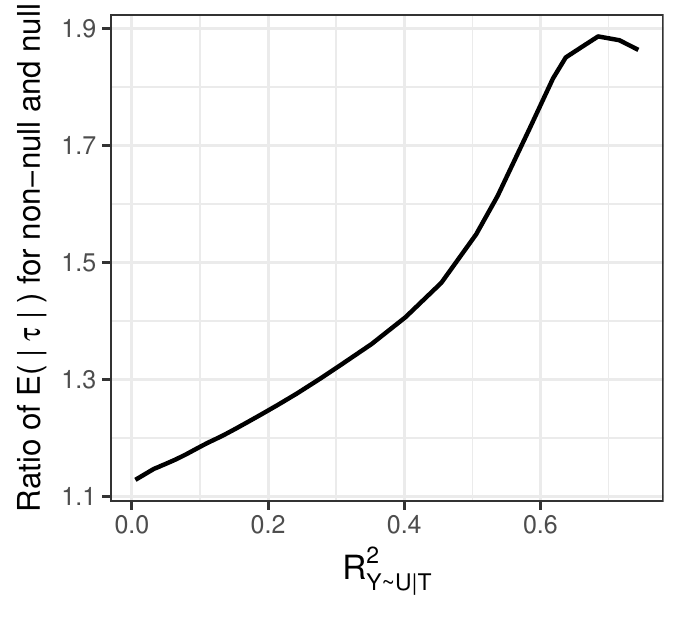}
         \caption{\label{fig:r2_normratio}}
	\end{subfigure}
		\caption{Change in $E(| \tau |)$ for the L1-minimized estimates as a function of $R_{Y \sim U | T}^2$, separated by null and non-null effects.
		(a) The magnitude of effects decreases with $\mathcal{R}^2$, with a larger relative decrease for null contrasts.
		(b) The relative magnitude of non-null and null effects increases with $\mathcal{R}^2$ in general. The magnitude of non-null effects can be as large as 1.9 times the null effects when $\mathcal{R}^2$ is large.
    }
    \label{fig:tau_classicifaction}
\end{figure}

\pagebreak

\section{Addition Results From the Mouse Obesity Analysis}

\begin{table}[ht!]
\caption{Point estimates and robustness values for the effect of gene expression on mouse obesity.  Only genes which have 95\% posterior credible intervals which exclude zero under the no unobserved confounding assumption are included in the table.  The first two columns correspond to the posterior mean estimate of the regression coefficients under no unobserved confounding, $\tau^\text{na\"ive}_\text{pm}$, as well as the endpoint of the 95\% posterior credible interval closest to zero for these coefficients.  The third column is the robustness value based on the interval endpoint.}
\centering
\begin{tabular}[t]{lccc}
\toprule
Gene & $\tau^\text{na\"ive}_\text{pm}$ & $\tau^\text{na\"ive}_\text{endpt}$ & $RV (\%)$\\
 \midrule
2010002N04Rik & 9.35 & 1.98 & 29\\
Gstm2 & 6.54 & 2.02 & 80\\
Sirpa & 6.17 & 0.30 & 1\\
Avpr1a & -5.30 & -0.48 & 2\\
Igfbp2 & -7.67 & -3.95 & 17\\
\bottomrule
\end{tabular}
\label{tab:rv_mouse}
\end{table}

\begin{figure}[ht!]
        \centering
        \includegraphics[width=0.5\textwidth]{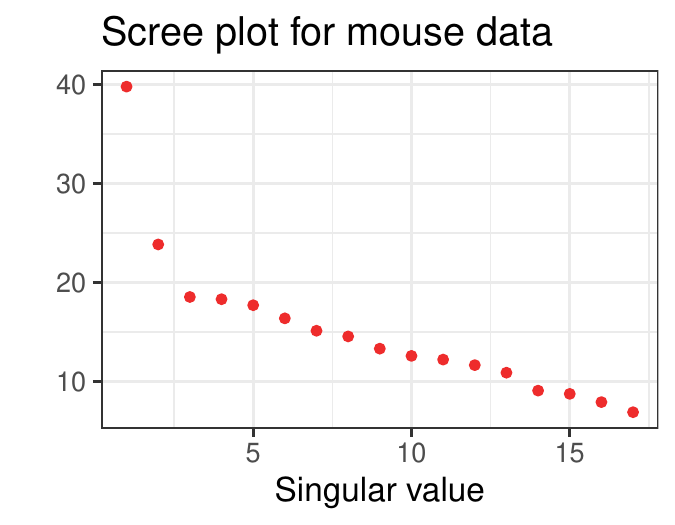}
		\caption{Scree plot for gene expression values in the mouse obesity data set. Three singular values are significantly larger than the rest, so we use a use a 3 factor model to estimate conditional confounder distributions.\label{fig:mouse_scree}}
\end{figure}

\pagebreak
\section{A Reanalysis of the Actor Case Study}
\label{sec:movie}

In this section, we compare our approach to other recent analyses of the TMDB 5000 Movie Data Set \citep{moviedataset} which was analyzed extensively by \citet{wang2018blessings} and \citet{grimmer2020ive}. The data set consists of 5000 movies and their corresponding revenue, budget, genre and the identities of the lead cast members.  Following Wang and Blei, we focus on estimating the causal effect of an actor's presence on the movie's log revenue. We let $Y$ denote the log revenue and $T_i = (T_{i1}, \cdots, T_{ik})$ encode the movie cast, where the binary random variable $T_{ij} \in \{ 0, 1 \}$ indicates whether actor $j$ appeared in the movie $i$ and $T_i \in \mathcal{T} = \{T_1, \cdots, T_n\}$.  We also let $\mathcal{T}^{j}$ denote the set of all movies $T_i$ for which $T_{ij} = 1$.  We define the estimand of interest, $\eta_j$, as the the total log revenue contributed by actor $j$:  

\begin{equation}
\eta_j := \sum_{t_i \in \mathcal{T}^{j}} \text{PATE}_{t_i, t_i^{\-j}} \label{eqn:pate_actor}
\end{equation} 
where $t_i^{\-j}$ corresponds to the observed treatment vector for movie $i$ excluding actor $j$.  
This estimand is a non-parametric generalization of the regression coefficient $\tau_j$, which was targeted in the analysis in \citet{wang2018blessings}.
Specifically, under the assumption that log-revenue is linear in the cast indicators, $\eta_j$ reduces to $n_j \tau_j$, the effect of actor $j$ scaled by the number of movies they appeared in, where $\tau_j$ are the regression coefficients for actor $j$.
Our estimand is well-defined without this linearity assumption.


We regress the log revenue on cast indicators to estimate actor effects, $\tau_j^{\text{naive}}$, under an assumption of no unobserved confounding. In order to demonstrate the applicability of our sensitivity analysis, we explicitly induce unobserved confounding by excluding observed confounders.  We validate our analysis, by comparing calibrated effect estimates when the confounders are excluded to estimates when the confounder is included.  Most importantly we exclude the movie's budget which we estimate to be the largest known source of confounding (computed using Equation \ref{eqn:partial_R2_tildeY}, see Appendix Figure \ref{fig:movie_pR2})\footnote{For illustrative purposes, we can assume that the budget is pre-treatment, meaning that the budget is decided prior to selecting the cast, which may be a dubious assumption in actuality.}.  

For simplicity, we model the observed outcome distribution with a linear regression, although other more flexible outcome models (e.g. \texttt{BART}) can also be used.  As in the previous section, we use a VAE to infer a Gaussian conditional confounder distribution, $f(u \mid t) \sim N(\mu_{u|t}, \Sigma_{u|t} )$ (See Appendix, Section \ref{sec:lv_inf}).   

\begin{figure}[htb!]
\centering
\includegraphics[width=\textwidth]{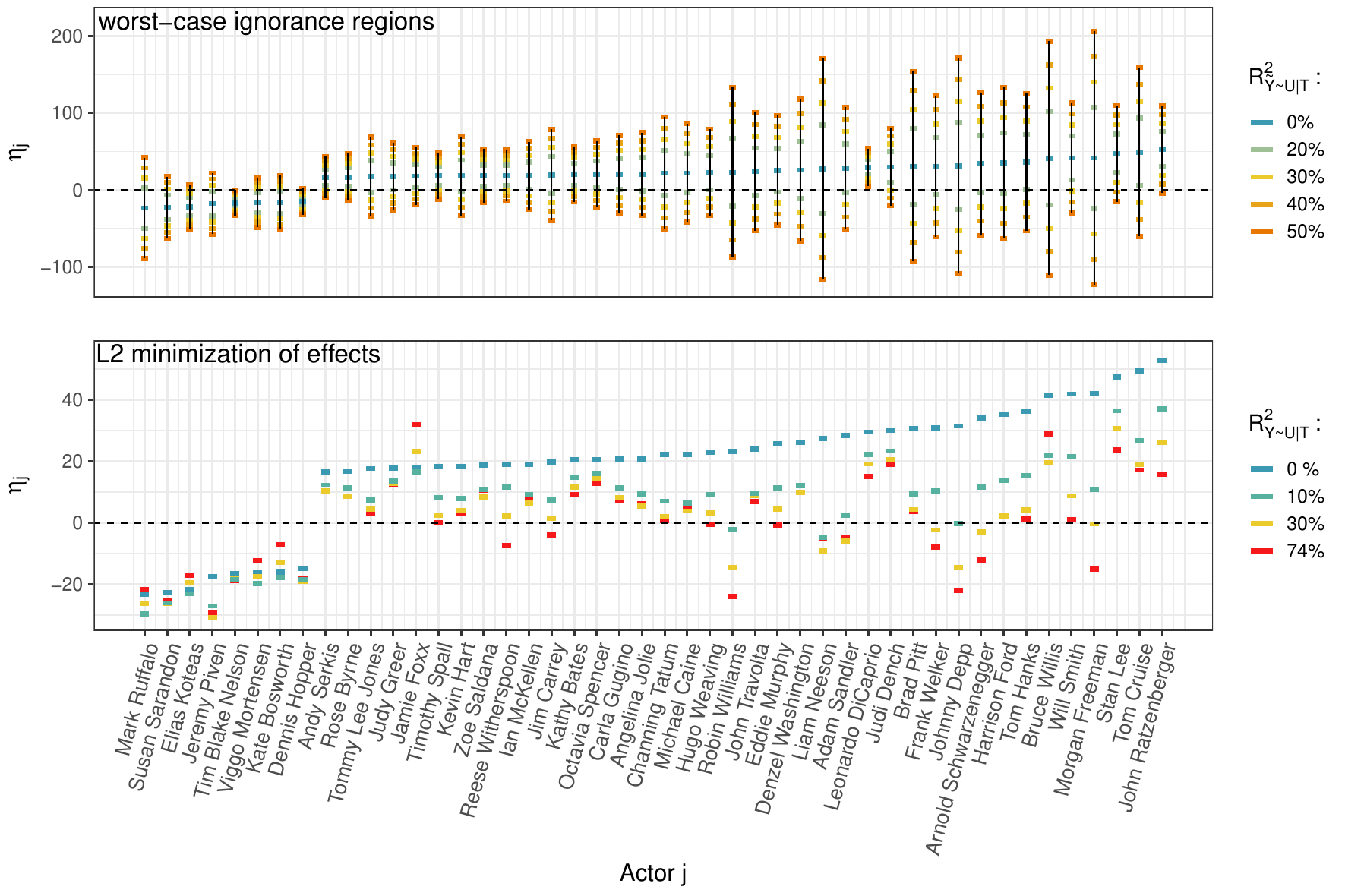}
 \caption{ Estimated total log revenue contributed by a given actor.  Top: worst-case ignorance region for each actor on a case by case basis.  The blue points correspond to $R_{Y \sim U | T}^2=0$, i.e. the naive estimates.  Robustness values can be found in Appendix Table \ref{tab:movie_rv}.  Bottom: Treatment effects for candidate models chosen with the L2 minimizing multiple contrast criterion (MCC). The color correspond to $\mathcal{R}^2$, the limit on the fraction of outcome variance explained by confounding.}
    \label{fig:movie_effect_combined}
\end{figure}

\vspace*{12pt}
\noindent \textbf{Results.} Since our focus is on confounding not estimation, in order to limit the influence of estimation uncertainty we subset the data to the $k=327$ actors who participated in at least twenty movies.  This reduces the total number of movies to 2439.   We fit the VAE to the treatments and use cross-validation to identify the appropriate latent confounder dimension, which we inferred to be $\hat m = 20$ (See Appendix Figure \ref{fig:latent_dim_select}).  We then plot the worst-case ignorance region for the causal effect on log revenue as a function of $R^2_{Y \sim U |T}$ for the 46 actors with significant regression coefficients in the naive regression (Figure \ref{fig:movie_effect_combined}, top). Eight actors in the observed data regression have significantly negative coefficients, whereas 38 actors have significant positive coefficients.  However, the worst-case ignorance regions for each actor are all very wide and include zero, which suggests that none of the effects are robust to confounding. In Table \ref{tab:movie_rv} of the Appendix we include robustness values for these actors.  Leonardo DiCaprio has the largest robustness value at $36\%$, with the majority of the other actors well below $20\%$.  For reference, the log budget, which was explicitly excluded from our causal analysis, explains about 30\% of the variance in log revenue (Appendix Figure \ref{fig:movie_pR2}).  In other words, none of the causal effects are robust at a level which matches the variance explained by the most important excluded confounder.

The worst case ignorance regions depicted in top panel of Figure \ref{fig:movie_effect_combined} correspond to a different choice of $\gamma$ for each actor.  We can also explore the robustness of causal effects under a single model by applying an appropriate MCC. Specifically, we search for a ``worst case'' candidate model by finding the sensitivity vector, $\gamma_*$, that implies the smallest L2 norm of the regression coefficients, $\tau$. In this conservative model, the minimum L2 norm of the treatment coefficients is 4.4, down from 7.6 for the naive coefficients. In addition, 40 out of 46 actors have coefficients that are smaller in magnitude than the magnitudes of the naive coefficients (Figure \ref{fig:movie_effect_combined}, bottom). For this candidate model, it turns out that $\gamma_*^\top  E[U | T=t]$ is significantly correlated, albeit weakly, with budget (Spearman's rank correlation = 0.2, p-value $< 2e\-16 $).  Thus, the conservative model correctly attributes part of the outcome variation induced by the known excluded confounder to unobserved confounding.

\begin{figure}[ht!]
    \begin{subfigure}[t]{0.56\textwidth}
        \centering
        \includegraphics[width=\textwidth]{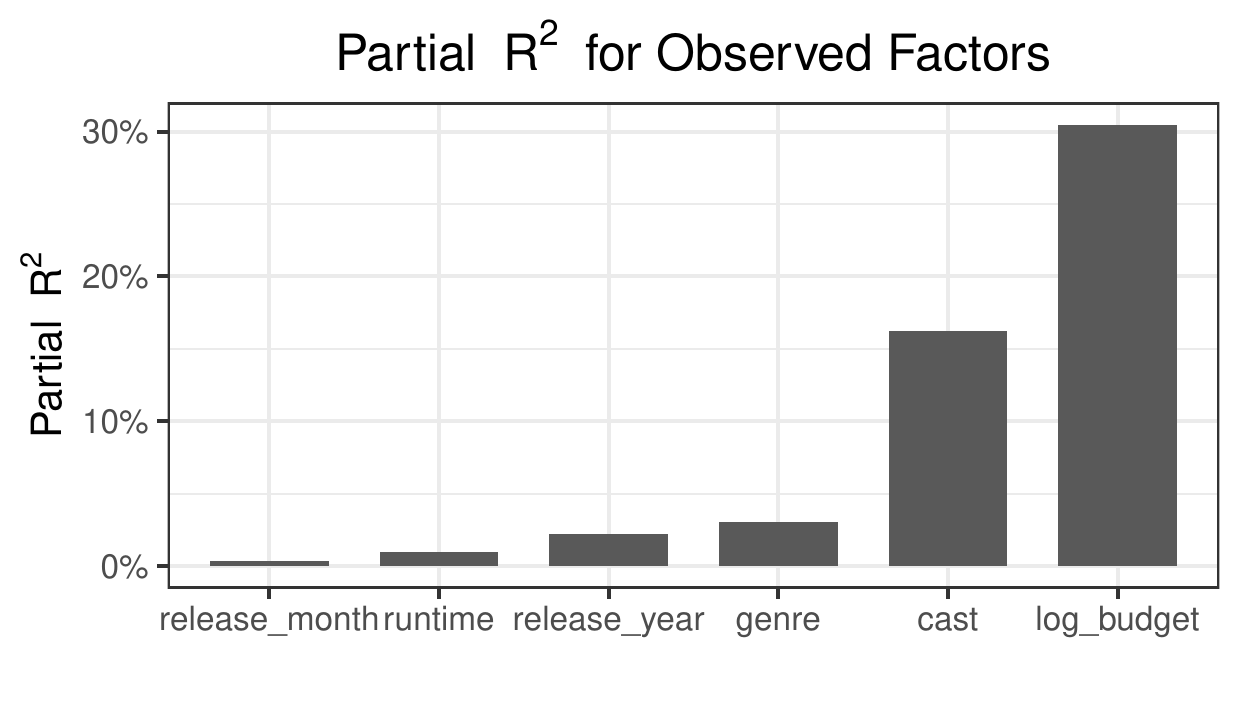}
        \caption{\label{fig:movie_pR2}}
	\end{subfigure}
	\quad
	\begin{subfigure}[t]{0.38\textwidth}
        \centering
        \includegraphics[width=\textwidth]{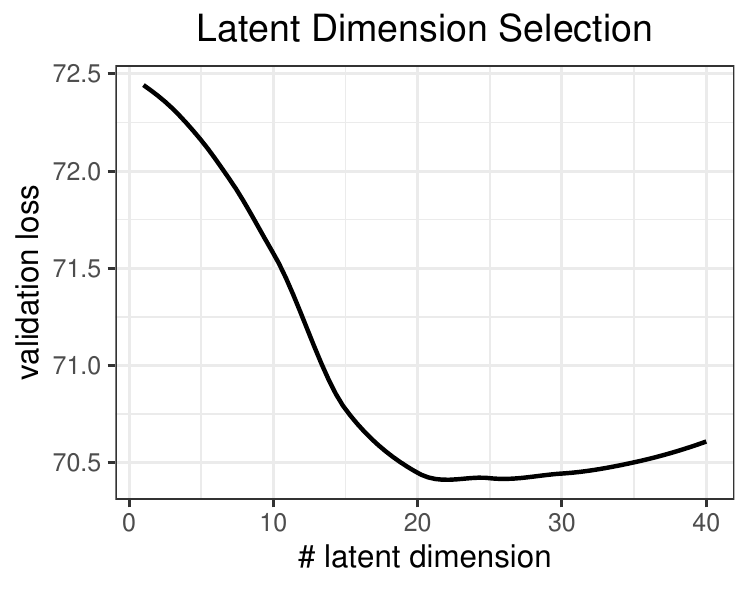}
         \caption{\label{fig:latent_dim_select}}
	\end{subfigure}
		\caption{
    (a) Estimated partial $R^2$ for observed confounders using method described in section \ref{sec:calibration_single}. Budget is the most dominant variable, which can explain significantly higher variation in outcome $Y$.
    (b) Latent confounder dimension selection, based on the reconstruction loss on the validation set.
    }
\end{figure}

\begin{table}
\caption{\label{tab:movie_rv}Robustness Value for Significant Actors}
\centering
\resizebox{0.62\textwidth}{!}{
\begin{tabular}[t]{lccc}
\toprule
  & Effect & $RV_{mean}$(\%) & $RV_{limit}$(\%)\\
\midrule
John Ratzenberger & 52.91 & 21.79 & 9.36\\
Tom Cruise & 49.48 & 5.09 & 1.59\\
Stan Lee & 47.53 & 14.43 & 4.16\\
Morgan Freeman & 41.95 & 1.63 & 0.26\\
Will Smith & 41.87 & 8.61 & 2.64\\
Bruce Willis & 41.44 & 1.86 & 0.3\\
Tom Hanks & 36.40 & 4.16 & 0.79\\
Harrison Ford & 35.26 & 3.25 & 0.6\\
Arnold Schwarzenegger & 34.13 & 3.39 & 0.55\\
Johnny Depp & 31.53 & 1.27 & 0.08\\
Frank Welker & 30.99 & 2.86 & 0.34\\
Brad Pitt & 30.61 & 1.54 & 0.09\\
Judi Dench & 30.01 & 8.87 & 1.41\\
Leonardo DiCaprio & 29.52 & 36.47 & 7.35\\
Adam Sandler & 28.39 & 3.22 & 0.19\\
Liam Neeson & 27.40 & 0.91 & 0.03\\
Denzel Washington & 26.09 & 2.01 & 0.11\\
Eddie Murphy & 25.82 & 3.30 & 0.27\\
John Travolta & 23.96 & 2.45 & 0.13\\
Robin Williams & 23.26 & 1.12 & 0.03\\
Hugo Weaving & 23.02 & 4.20 & 0.12\\
Michael Caine & 22.23 & 3.03 & 0.11\\
Channing Tatum & 22.22 & 2.33 & 0.06\\
Angelina Jolie & 20.81 & 3.72 & 0.09\\
Carla Gugino & 20.74 & 4.24 & 0.13\\
Octavia Spencer & 20.68 & 5.78 & 0.13\\
Kathy Bates & 20.58 & 8.25 & 0.17\\
Jim Carrey & 19.79 & 2.79 & 0\\
Ian McKellen & 19.02 & 4.70 & 0\\
Reese Witherspoon & 18.97 & 8.17 & 0.27\\
Zoe Saldana & 18.79 & 7.33 & 0.12\\
Kevin Hart & 18.41 & 3.21 & 0.05\\
Timothy Spall & 18.40 & 9.39 & 0.25\\
Jamie Foxx & 18.15 & 6.02 & 0.01\\
Judy Greer & 17.76 & 4.17 & 0.01\\
Tommy Lee Jones & 17.67 & 2.96 & 0\\
Rose Byrne & 16.85 & 7.62 & 0.05\\
Andy Serkis & 16.53 & 9.48 & 0.05\\
Dennis Hopper & -14.79 & 19.32 & 0\\
Kate Bosworth & -16.01 & 5.20 & 0.04\\
Viggo Mortensen & -16.24 & 6.50 & 0.02\\
Tim Blake Nelson & -16.58 & 24.78 & 0.09\\
Jeremy Piven & -17.45 & 4.86 & 0.06\\
Elias Koteas & -21.64 & 13.87 & 1.06\\
Susan Sarandon & -22.57 & 7.90 & 0.24\\
Mark Ruffalo & -23.29 & 3.17 & 0.11\\
\bottomrule
\end{tabular}
}
\end{table}

\end{document}